\documentclass[journal]{IEEEtran}

\ifCLASSINFOpdf
\else
\fi

\usepackage{mathrsfs}
\usepackage{bbm}
\usepackage{amsmath, amsthm}
\usepackage{tabularx}
\usepackage{cite}
\usepackage{listings}
\usepackage{graphicx}
\usepackage{float}
\usepackage{amssymb}
\usepackage{array}
\usepackage{subfigure}
\usepackage{fancyhdr}
\usepackage{cases}
 \usepackage{supertabular}
\usepackage{url}
\usepackage{psfrag}
\usepackage{color}
\usepackage{bbding}

\newcommand {\fo} {it follows that }

\newcommand {\C} {{\rm I\kern-5.5pt C}}

\newcommand{\bP}[1]{{\mathbb{P}}\left[{#1}\right]}
\newcommand{\bE}[1]{{\mathbb{E}}\left[{#1}\right]}

\newcommand{\1}[1]{{\bf 1}\left[#1\right]}       

\newcommand{\fsquare}{\vrule height6pt width7pt depth1pt}   
\newcommand{\myproof}{{\hfill \\ \bf Proof. \ }}           
\newcommand{\myendpf}{\hfill\fsquare \\[0.1in]}             

\newtheorem{fact}{Fact}
\newtheorem{lem}{Lemma}
\newtheorem{thm}{Theorem}

\newtheorem{cor}{Corollary}
\newtheorem{proposition}{Proposition}
\newtheorem*{proposition1.1}{Proposition 1.1}
\newtheorem*{proposition1.2}{Proposition 1.2}
\newtheorem*{proposition1.3}{Proposition 1.3}
\newtheorem*{proposition2.1}{Proposition 2.1}
\newtheorem*{proposition2.2}{Proposition 2.2}
\begin{document}
%
\title{$k$-Connectivity in Random Key Graphs with Unreliable Links}

%

%
%
%

\author{Jun~Zhao,~\IEEEmembership{Student Member,~IEEE,}
        Osman~Ya\u{g}an,~\IEEEmembership{Member,~IEEE,}
        and~Virgil~Gligor,~\IEEEmembership{Senior Member,~IEEE}
        \thanks{Manuscript received May 22, 2012; revised August 12, 2013; accepted
February 1, 2015. The
material in this paper was presented in part at the 2013 IEEE International
Symposium on Information Theory \cite{ISIT}. The full version of this paper is available on arXiv \cite{ZhaoYaganGligorArxiv}.

The authors are with CyLab and the Department of Electrical and Computer Engineering,
Carnegie Mellon University, Pittsburgh, PA 15213. (Emails: junzhao@alumni.cmu.edu, oyagan@ece.cmu.edu, gligor@cmu.edu).

Copyright (c) 2015 IEEE. Personal use of this material is permitted.
However, permission to use this material for any other purposes must be
obtained from the IEEE by sending a request to pubs-permissions@ieee.org.}} 

\maketitle

%
%

\thispagestyle{fancy} \pagestyle{fancy}

\markboth{}{}

\fancyhead[C]{{IEEE Transactions on Information Theory 2015}}

%




\begin{abstract}

Random key graphs form a class of random intersection graphs and are naturally induced by
the random key predistribution scheme of Eschenauer and Gligor for securing wireless sensor network (WSN)
communications. Random key graphs have received much interest recently, owing in part to their wide applicability in
various domains including recommender systems, social networks, secure sensor networks, clustering and classification analysis, and cryptanalysis to name a few.
In this paper, we study connectivity properties of random key graphs in the presence of
unreliable links. Unreliability of the edges are captured by independent Bernoulli random variables,
rendering edges of the graph to be {\em on}  or {\em off}  independently
from each other. The resulting model is an {\em intersection} of a random key graph and an Erd\H{o}s--R\'enyi graph, and
is expected to be useful in capturing various real-world networks; e.g., with secure WSN applications in mind, link unreliability can be
attributed to harsh environmental conditions severely impairing transmissions. We present conditions on how to scale this model's parameters so that
i) the minimum node degree in the graph is at least $k$, and
ii) the graph is $k$-connected, both with high probability as the number of nodes becomes large.
The results are given in the form of zero-one laws with critical thresholds identified and shown to coincide
for both graph properties.
These findings improve
the previous results by Rybarczyk on
the $k$-connectivity of random key graphs (with reliable links),
as well as the zero-one laws by Ya\u{g}an on the $1$-connectivity of random key graphs with unreliable links.
\end{abstract}


\section{Introduction}\label{sec:intro}

Random key graphs have received significant interest recently with applications spanning
key predistribution in secure wireless sensor networks (WSNs) \cite{r1,Rybarczyk,DiPietroTissec,adrian,virgil}, social networks \cite{virgillncs,ZhaoYaganGligorArxiv,DBLP:journals/corr/abs-1301-7320}, recommender systems \cite{r4},
clustering and classification analysis \cite{GodehardtJaworski,Assortativity}, cryptanalysis of hash functions \cite{r10},
circuit design \cite{SingerThesis}, and
the modeling of epidemics \cite{ball2014} and
``small-world'' networks \cite{5383986}.
They belong to a larger class of random graphs known
as {\em random intersection graphs} \cite{r1,r10,Rybarczyk,Assortativity,Perfectmatchings,EfthymiouaHM,NikoletseasHM,YShang,FarrellGLRVS14,DBLP:journals/corr/abs-1301-7320,SingerThesis}; in fact, they are referred to as {\em uniform} random intersection graphs
by some authors \cite{r1,Perfectmatchings,r10,NikoletseasHM,ryb3,zz,YShang,ZhaoCDC,ITAin}.

To fix the terminology, we will describe random key graphs in the context of secure WSNs, where they have originated from.
Security is expected to be a key challenge in resource constrained sensor networks. A widely accepted
solution for securing WSN communications is the random predistribution of cryptographic keys to sensor nodes, and
utilization of symmetric-key encryption modes~\cite{GD,cryptoeprint:2000:039,Rogaway01ocb} to ensure message secrecy and
authenticity. Among various key predistribution algorithms proposed to date, the original
scheme by Eschenauer and Gligor (EG) \cite{virgil} is still the most widely recognized one. According to the EG
scheme, each of the $n$ sensors is assigned $K_n$ distinct keys that are selected uniformly at random from a key pool of size $P_n$.
Two sensors can then {\em securely} communicate over an existing communication link if they have at least one key in common; i.e.,
if they share a common key. This notion of adjacency defines the random key graph, hereafter denoted
by $G(n,K_n,P_n)$. For generality, $K_n$ and $P_n$ are assumed to
scale with the number of nodes $n$, with the natural condition $1 \leq K_n \leq P_n$ always imposed.



In this paper, we study connectivity properties of random key graphs in the presence of
unreliable links. Unreliability of the edges are captured by independent Bernoulli random variables,
rendering each edge of $G(n;K_n, P_n)$ to be {\em on} (with probability $p_n$) or {\em off} (with probability $1-p_n$) independently
from all other edges. Put differently, we consider an Erd\H{o}s--R\'enyi (ER) graph $G(n;p_n)$ \cite{citeulike:4012374} on the same set of
$n$ vertices, with edges appearing between any pair of vertices independently with probability $p_n$. A random key graph with unreliable links
thus corresponds to the {\em intersection} of a random key graph and an ER graph. Hereafter, we denote this graph by
$\mathbb{G}_{on} = G(n;K_n, p_n) \cap G(n;p_n)$;
see Section \ref{sec:SystemModel} for precise definitions.

Just like the random key graph, the $\mathbb{G}_{on}$ model can be used in various applications, particularly when
links are expected to be unreliable. For example, in a secure WSN application,
links might be unreliable due to wireless media of the communication, or due to
physical obstacles and altering environmental conditions severally impairing the transmission. We refer the reader to \cite{yagan} and \cite{ZhaoYaganGligorArxiv}
for two other applications of $\mathbb{G}_{on}$: i) secure connectivity of WSNs
under an on-off channel model, and ii) large scale, distributed publish-subscribe services in online social networks, respectively.


The main goal of this paper to study $k$-connectivity of $\mathbb{G}_{on}$.
A network (or graph) is said to be $k$-connected if for
each pair of nodes there exist at least $k$ mutually
disjoint paths connecting them. An equivalent
definition of $k$-connectivity is that a network is $k$-connected if
the network remains connected despite the failure
of any $(k-1)$ nodes \cite{penrose}; a network is
said to be simply connected if it is $1$-connected.
$k$-connectivity is a fundamental graph property and is
important for various applications of random key graphs. For example,
in a WSN application where sensor nodes operate autonomously and
physically unprotected, $k$-connectivity provides communication security against an adversary
that is able to {\em compromise} up to $k-1$ links by launching a sensor capture attack
\cite{adrian}; i.e., two sensors can communicate securely as long as at least one of the $k$ disjoint paths
connecting them consists of links that are not compromised by the adversary. Also,
$k$-connectivity improves resiliency against network disconnection due to battery depletion,
in both normal mode of operation and under battery-depletion attacks \cite{LiWanWangYi}.
Furthermore, it enables flexible communication-load balancing across multiple paths so
that network energy consumption is distributed without penalizing
any access path \cite{Ganesan:2001:HEM:509506.509514}.

Our main contributions are {\em zero-one laws} for two related graph properties for $\mathbb{G}_{on}$:
i) the minimum node degree being at least $k$, and
ii) $k$-connectivity.
Namely, we present conditions on how to scale the model parameters $K_n$, $P_n$, $p_n$ such that these properties
hold with probability approaching to one and zero, respectively, as the number of nodes $n$ becomes large.
Our main results also imply a
zero-one law for $k$-connectivity in random key graph $G(n, K_n, P_n)$ (see Corollary \ref{cor:k_con_rkg}),
and the established result is shown to improve that given previously by
Rybarczyk \cite{ryb3}; see Section \ref{secref} for details.
Moreover, for the $1$-connectivity of $\mathbb{G}_{on}$, we provide a stronger form
of the zero-one law as compared to that given by Ya\u{g}an \cite{yagan_onoff}; see Section \ref{secref}.

We organize the rest of the paper as follows:
In Section \ref{related}, we
survey the relevant results from the literature, while
in Section \ref{sec:SystemModel} we give a detailed description
of the system model
$\mathbb{G}_{on}$. The main results of the paper are presented (see Theorem \ref{thm:unreliableq1})
in Section \ref{sec:MainResult}, with a detailed discussion and comparisons with the existing
results given in Section \ref{secref}; also, in Section \ref{subsec:Numerical} we provide numerical
results that confirm Theorem  \ref{thm:unreliableq1}.
 The basic ideas that pave the way
in establishing Theorem \ref{thm:unreliableq1} are given in
Section \ref{sec:BasicIdeas}.
Sections \ref{sec:proof_of_zero_law_oy} through \ref{sec:proof_Prop2}
are devoted to establishing the zero-law part of Theorem \ref{thm:unreliableq1},
whereas the one-law of  Theorem \ref{thm:unreliableq1} is established in
Sections \ref{sec:Proof_One_Law} through \ref{sec:Last_step_One_law}. The paper is concluded in
Section \ref{sec:Conclusion}, and some of the technical details are given in Appendix \ref{sec:addi:f:l}-\ref{secprflem}.


%

\section{Related Work} \label{related}

%
Erd\H{o}s and R\'{e}nyi \cite{citeulike:4012374} and
Gilbert \cite{Gilbert} introduces the random graph $G(n,p)$,
which is defined on $n$ nodes and there exists an edge between
any two nodes with probability $p$  \emph{independently} of all other
edges. The probability $p$ can also be a function of $n$, in which case
we refer to it as $p_n$. Throughout the paper, we refer to the
random graph $G(n, p_n)$ as an Erd\H{o}s-R\'{e}nyi (ER) graph following the
convention in the literature.

Erd\H{o}s and R\'{e}nyi \cite{citeulike:4012374} prove that when $p_n$
is $\frac{\ln  n + {\alpha_n}}{n}$, graph $G(n,p_n)$ is \emph{asymptotically almost
surely}\footnote{We say that an
event takes place \textit{asymptotically almost surely} if its probability
approaches to 1 as $n\to \infty$. Also, we use \lq\lq resp.'' as a shorthand for
``respectively''.} (a.a.s.)  connected (resp., not connected) if
$\lim_{n\to \infty}{\alpha_n}=+\infty$ (resp.,
$\lim_{n\to \infty}{\alpha_n}=-\infty$). In later work \cite{erdos61conn}, they further
explore $k$-connectivity \cite{citeulike:505396} in $G(n,p_n)$ and
show that if $p_n = \frac{\ln n + {(k-1)} \ln  \ln  n +
{\alpha_n}}{n}$, $G(n,p_n)$ is a.a.s. $k$-connected (resp.,
not $k$-connected) if $\lim_{n\to \infty}{\alpha_n}=+\infty$ (resp.,
$\lim_{n\to \infty}{\alpha_n}=-\infty$).

Previous work \cite{r1,ryb3,yagan} investigates the zero-one law for
connectivity in random key graph $G(n,K_n,P_n)$, where $P_n$ and
$K_n$ are the key pool size and the key ring size, respectively.
Blackburn and Gerke \cite{r1} prove that if $K_n \geq 2$ and
$P_n=\lfloor n ^{\xi}\rfloor$, where $\xi $ is a positive constant,
$G(n,K_n,P_n)$ is a.a.s. connected (resp., not connected) if
$\liminf_{n\to+\infty}\frac{K_n^2 n }{P_n \ln n} > 1$ (resp.,
$\limsup_{n\to+\infty}\frac{K_n^2 n }{P_n \ln n} < 1$). Ya\u{g}an
and Makowski \cite{yagan} demonstrate that if\footnote{We use the
standard asymptotic notation $o(\cdot), O(\cdot), \Theta(\cdot),
\Omega(\cdot), \sim$. That is, given two positive
sequences $f_n$ and $g_n$,
\begin{enumerate}
  \item $f_n = o \left(g_n\right)$ means $\lim_{n \to
  \infty}\frac{f_n}{g_n}=0$.
  \item $f_n = O \left(g_n\right)$ means that there exist positive
  constants $c$ and $N$ such that $f_n \leq c g_n$ for all $n \geq
  N$.
  \item $f_n = \Omega \left(g_n\right)$ means that there exist positive
  constants $c$ and $N$ such that $f_n \geq c g_n$ for all $n \geq
  N$. 
  \item $f_n = \Theta \left(g_n\right)$ means that there exist positive
  constants $c_1, c_2$ and $N$ such that
  $c_1 g_n \leq f_n \leq c_2 g_n$ for all $n \geq
  N$. 
  \item $f_n \sim g_n$ means that $\lim_{n \to
  \infty}\frac{f_n}{g_n}=1$; i.e., $f_n$
  and $g_n$ are asymptotically equivalent.%
\end{enumerate}
} $K_n \geq 2$, $P_n
= \Omega(n)$ and $\frac{K_n^2}{P_n}=\frac{\ln n + {\alpha_n}}{n}$,
then $G(n,K_n,P_n)$ is a.a.s. connected (resp., not
connected) if $\lim_{n\to \infty}{\alpha_n}=+\infty$ (resp.,
$\lim_{n\to \infty}{\alpha_n}=-\infty$). Rybarczyk \cite{ryb3}
obtains a stronger result without requiring $P_n = \Omega(n)$. In particular, she derives
the asymptotically exact probability of connectivity in $G(n,K_n,P_n)$ as follows: under $K_n \geq 2$, if the sequence
$\alpha_n$ defined through $\frac{K_n^2}{P_n}=\frac{\ln n + {\alpha_n}}{n}$ has a limit $\alpha^* \in [-\infty, \infty]$, then the probability of
$G(n,K_n,P_n)$ being connected approaches to $e^{-e^{- \alpha^{*}}}$ as $n\to \infty$. This asymptotically exact probability result is stronger than
a zero--one law since the latter can be obtained by setting $\alpha^{*}$ as $\infty$ and $-\infty$ in the former. Rybarczyk
also establishes \cite[Remark 1, p. 5]{zz} a zero-one law for
$k$-connectivity in $G(n,K_n,P_n)$ by showing
the similarity between $G(n,K_n,P_n)$ and a random
intersection graph \cite{Rybarczyk} via a coupling argument.
Specifically, she proves
that if $P_n = \Theta(n ^{\xi})$ for some $\xi >1$ and
$\frac{K_n^2}{P_n}=\frac{\ln n + (k-1) \ln \ln n + {\alpha_n}}{n}$,
then the $G(n,K_n,P_n)$ is
a.a.s. $k$-connected (resp., not $k$-connected) if
$\lim_{n\to \infty}{\alpha_n}=+\infty$ (resp.,
$\lim_{n\to \infty}{\alpha_n}=-\infty$).


Recently Ya\u{g}an \cite{yagan_onoff} gives a zero-one law for
connectivity (i.e., 1-connectivity) in graph $G(n,K_n,P_n) \cap
G(n,{p_n\iffalse_{on}\fi})$, which is the intersection of random key
graph $G(n,K_n,P_n)$ and random graph $G(n,{p_n\iffalse_{on}\fi})$,
and clearly is equivalent to our key graph $\mathbb{G}_{on}$;
see Section \ref{sec:SystemModel}. Specifically,
he proves that if $K_n \geq 2$, $P_n = \Omega(n)$ and
${p_n\iffalse_{on}\fi}\cdot \left[ 1- \frac{\binom{P_n- K_n}{K_n} }
{\binom{P_n}{K_n}} \right] \sim \frac{c\ln n}{n}$ hold, and
$\lim_{n\to \infty}({p_n\iffalse_{on}\fi}\ln n)$ exists, then graph
$G(n,K_n,P_n) \cap G(n,{p_n\iffalse_{on}\fi})$ is asymptotically almost
surely connected (resp., not connected) if $c>1$ (resp., $c<1$).
A comparison of our results with the related work is given in
Section \ref{secref}.

After the submission of this paper, we have derived the asymptotically exact probability of $k$-connectivity in $G(n,K_n,P_n)$ \cite{ZhaoCDC} (resp., $\mathbb{G}_{on}$ \cite{WiOpt15}). Based on the proofs in this paper, we show i) that \cite{ZhaoCDC}
under $P_n = \Omega(n)$, if the sequence
$\alpha_n$ defined through $\frac{K_n^2}{P_n}=\frac{\ln n + {(k-1)} \ln \ln n + {\alpha_n}}{n}$ has a limit $\alpha^* \in [-\infty, \infty]$, then the probability of
$G(n,K_n,P_n)$ being $k$-connected converges to $e^{- \frac{e^{-\alpha^{*}}}{(k-1)!}}$ as $n\to \infty$,
 and ii) that \cite{WiOpt15} under $P_n = \Omega(n)$ and $\frac{K_n}{P_n} = o(1)$, if the sequence
$\alpha_n$ defined through ${p_n\iffalse_{on}\fi}\cdot \left[ 1- \frac{\binom{P_n- K_n}{K_n} }
{\binom{P_n}{K_n}} \right] =\frac{\ln n + {(k-1)} \ln \ln n + {\alpha_n}}{n}$ has a limit $\alpha^* \in [-\infty, \infty]$, then the probability of
 $\mathbb{G}_{on}$ being $k$-connected converges to $e^{- \frac{e^{-\alpha^{*}}}{(k-1)!}}$ as $n\to \infty$.
%

%

\section{System Model  $\mathbb{G}_{on}$}
\label{sec:SystemModel}


Consider a vertex set $\mathcal {V} = \{v_1, v_2, \ldots, v_n
\}$. Each node $v_i \in \mathcal {V}$ is assigned a key ring $S_i$ that consists of
 $K_n$ distinct keys selected uniformly at random from a key pool
$\mathcal{P}$ of size $P_n$. The random key
graph $G(n, K_n, P_n)$ is defined on the vertex set
$\mathcal{V}$ such that two distinct nodes
$v_i$ and $v_j$ are adjacent, denoted $K_{ij}$, if their key rings have at least one
key in common; i.e.,
\[
K_{ij} = [S_i \cap S_j \neq \emptyset].
\]
For distinct nodes $v_x$ and $v_y$, we let $S_{xy}$ denote the intersection
of their key rings $S_x$ and $S_y$; i.e., $S_{xy} = S_x \cap S_y$.

Our main interest is to study random key graphs whose links are unreliable.
In particular, we assume that each link is {\em on} with probability $p_n$, or {\em off} with probability $1-p_n$, independently from
any other link. Namely, with $C_{ij}$ denoting the event that link between $v_i$ and $v_j$ is on,
$\{C_{ij}, ~ 1\leq i < j \leq n \}$  are mutually independent such
that
\begin{equation}
\bP{C_{ij}} = p_n, \quad 1\leq i < j \leq n.
\label{eq_prob_p}
\end{equation}
This unreliable link model can be represented \cite{citeulike:4012374}
by an Erd\H{o}s-R\'enyi (ER) graph $G(n, p_n)$ on the vertices
$\mathcal{V}$ such that there exists an edge between nodes $v_i$
and $v_j$ if the link between them is on;
i.e., if the event $C_{ij}$ takes place.

Finally, the graph $\mathbb{G}_{on}(n, K_n, P_n, p_n)$ is
defined on the vertices $\mathcal{V}$ such that two distinct nodes
$v_i$ and $v_j$ have an edge in between, denoted $E_{i j}$, if
the events $K_{ij}$ and $C_{ij}$ take place at the same time. In other
words, we have
\begin{equation}
E_{ij} = K_{ij} \cap C_{ij}, \quad 1\leq i < j \leq n
\label{eq:E_is_K_cap_C_oy}
\end{equation}
so that
\begin{equation}
\mathbb{G}_{on}(n, K_n, P_n, p_n) = G(n, K_n, P_n) \cap G(n, p_n).
\label{eq:G_on_is_RKG_cap_ER_oy}
\end{equation}
Throughout, we simplify the notation by writing $\mathbb{G}_{on}$ instead
of $\mathbb{G}_{on}(n, K_n, P_n, p_n)$. Thus, our main model $\mathbb{G}_{on}$
is an {\em intersection} of a random key graph and an ER graph.

Throughout, we let $p_{s} (K_n, P_n)$ be the probability that the key rings of two
distinct nodes share at least one key and let ${p_e}(K_n, P_n, p_n)$
be the probability that there exists a link between two distinct
nodes in $\mathbb{G}_{on}$. For simplicity, we write $p_{s} (K_n, P_n)$ as ${p_s}$ and
write ${p_e}(K_n, P_n, p_n)$ as ${p_e}$. Then for any two distinct
nodes $v_i$ and $v_j$, we have
\begin{align}
p_{s} &:=  \mathbb{P} [{K}_{i j} ]. \label{eq_ps}
\end{align}
It is easy to derive $p_{s}$ in terms of $K_n$ and
$P_n$
as shown in previous work \cite{r1,ryb3,yagan}.
In fact, we have
\begin{align}
p_{s} & = \mathbb{P}[S_i \cap S_j \neq \emptyset]  = \begin{cases}
1- \frac{\binom{P_n- K_n}{K_n} } {\binom{P_n}{K_n}}, &\textrm{if
}P_n \geq 2 K_n , \\ 1 &\textrm{if }P_n < 2 K_n . \end{cases}
\label{hh2psps}
\end{align}
Given (\ref{eq:E_is_K_cap_C_oy}), the
independence of the events ${C}_{i j} $ and $ {K}_{i j} $
gives
\begin{align}
{p_e}  & :=  \mathbb{P} [E_{i j} ]  =  \mathbb{P} [{C}_{i j} ]
\cdot \mathbb{P} [{K}_{i j} ] =  p_n\cdot
p_{s} \label{eq_pre}
\end{align}
from (\ref{eq_prob_p}) and (\ref{eq_ps}).
Substituting (\ref{hh2psps}) into (\ref{eq_pre}), we obtain 
\begin{align}
{p_e}  & = p_n\cdot \left[1- \frac{\binom{P_n-
K_n}{K_n} } {\binom{P_n}{K_n}}\right] \quad \textrm{if }P_n \geq 2 K_n.
\label{eq:link_prob_oy}
\end{align}

%
%
%
%


\section{Main Results and Discussion}
\label{sec:MainResult}

\subsection{The Main Result}\label{seca}

The main result of this paper, given below, establishes
zero-one laws for $k$-connectivity and for the property that the
minimum node degree is no less than $k$ in graph ${\mathbb{G}}_{on} $.
Throughout this paper, $k$ is a positive integer and does
not scale with $n$. Also, we let $\mathbb{N}$ (resp., $\mathbb{N}_0$) stand for
the set of all non-negative (resp., positive) integers.

We refer to any pair of mappings $K,P: \mathbb{N}_0
\rightarrow \mathbb{N}_0$ as a {\em scaling}
as long as it satisfies the natural conditions
\begin{equation}
K_n \leq P_n, \quad n=1,2, \ldots . \label{eq:ScalingDefn}
\end{equation}
Similarly, any mapping $p: \mathbb{N}_0 \rightarrow (0,1)$ defines a
scaling.

\begin{thm}\label{thm:unreliableq1}
Consider scalings $K,P:
\mathbb{N}_0 \rightarrow \mathbb{N}_0$, $p:
\mathbb{N}_0 \rightarrow (0,1)$ such that
$K_n \geq 2$ for all $n$ sufficiently large.
 We define a sequence $\alpha:
\mathbb{N}_0 \rightarrow \mathbb{R}$ such that for any $n \in
\mathbb{N}_0$, we have
\begin{align}
p_e &  = \frac{\ln  n + {(k-1)} \ln \ln n + {\alpha_n}}{n}
.\label{eqn:unreliable}
\end{align}
The properties (a) and (b) below hold.

(a) If $\frac{K_n ^2}{P_n} = o(1)$ and {\em either}
there exists $\epsilon>0$ such that $p_e n > \epsilon$
holds for all $n$ sufficiently large {\em or} $\lim_{n \to \infty} p_e n=0$,
then
\begin{align}
\lim_{n \to \infty}\mathbb{P}\left[\mathbb{G}_{on} \textrm{ is
}k\textrm{-connected}\hspace{2pt}\right]
  &  = 0
  \quad \textrm{if}~\lim_{n \to \infty}{\alpha_n} = -\infty,
   \label{thm1kc}
 \end{align}
 and
\begin{align}
\lim_{n \to \infty} \mathbb{P}\left[ \begin{array}{c}
\textrm{Minimum~node~degree} \\  \mbox{of ${\mathbb{G}}_{on}$ is no less than
}k \end{array}
\right]  &  = 0 \:\:\: \textrm{if $\lim_{n \to \infty}{\alpha_n} =-\infty$} .
\label{mnd_zero}
 \end{align}
\indent (b) If $ P_n = \Omega(n)$ and
$\frac{K_n}{P_n} = o(1)$, then
\begin{align}
\lim_{n \to \infty}\mathbb{P}\left[\mathbb{G}_{on} \textrm{ is
}k\textrm{-connected}\hspace{2pt}\right]
  &  = 1
  \:\: \textrm{if $\lim_{n \to \infty}{\alpha_n} = \infty$},
  \label{eq:OneLaw+kConnectivity_olp}
 \end{align}
 and
  \begin{align}
\lim_{n \to \infty} \mathbb{P}\left[ \begin{array}{c}
\textrm{Minimum~node~degree} \\  \mbox{of ${\mathbb{G}}_{on}$ is no less than
}k \end{array}
\right] &  = 1\quad\textrm{if $\lim_{n \to \infty}{\alpha_n} =
\infty$} . \label{mnd_one}
 \end{align}
\end{thm}

Note that if we combine (\ref{thm1kc}) and (\ref{eq:OneLaw+kConnectivity_olp}),
we obtain
the zero-one law for $k$-connectivity in $ \mathbb{G}_{on} $, whereas
combining (\ref{mnd_zero}) and (\ref{mnd_one}) leads to the
zero-one law for the minimum node degree. 
Therefore, Theorem
\ref{thm:unreliableq1} presents the zero-one laws of
$k$-connectivity and the minimum node degree in graph $ \mathbb{G}_{on} $.
We also see from (\ref{eqn:unreliable}) that the critical scaling for both properties
is given by $ p_e = \frac{\ln  n + (k-1) \ln \ln n}{n}$. The
sequence $\alpha_{n}: \mathbb{N}_0 \rightarrow \mathbb{R}$ defined
through (\ref{eqn:unreliable}) therefore measures by how much the probability
$p_e$ deviates from the critical scaling.

In case (b) of Theorem \ref{thm:unreliableq1}, the conditions $ P_n
= \Omega(n)$ and $\frac{K_n}{P_n} = o(1)$ indicate that the size of
the key pool $P_n$ should grow at least linearly with the number of
sensor nodes in the network, and should grow unboundedly with the
size of each key ring. These conditions are enforced here
merely for technical reasons, but they hold trivially in practical
wireless sensor network applications
\cite{adrian,DiPietroTissec,virgil}.
Again, the condition $\frac{K_n^2}{P_n}=o(1)$ enforced for the zero-law
in Theorem \ref{thm:unreliableq1} is not a stringent one since the
$P_n$ is expected to be several orders of magnitude larger
than $K_n$. Finally, the condition that either
$p_e n > \epsilon> 0$ for all $n$ large or $\lim_{n \to \infty} p_e n = 0$ is imposed
to avoid degenerate situations. In most cases of interest
it holds that $ p_e n > \epsilon > 0$ as otherwise
the graph $\mathbb{G}_{on}$ becomes {\em trivially} disconnected. To see this,
notice that $p_e n$ is an upper-bound on the {\em expected} degree of a node
and that the {\em expected} number of edges in the graph is less than $p_e n^2$;
yet, a connected graph on $n$ nodes must have at least $n-1$ edges.

\subsection{Results with an approximation of probability $p_s$}
\label{sec:Cor1}

An analog of Theorem \ref{thm:unreliableq1} can be given with a
simpler form of the scaling (\ref{eqn:unreliable}); i.e., with
$p_s$ replaced by the more easily expressed quantity $K_n^2/P_n$,
and hence with {$p_e = p_n K_n^2/P_n$}. In fact, in the case of
random key graph $G(n,K_n,P_n)$ it is a common practice
\cite{r1,ryb3,yagan} to replace $p_s$ by $\frac{K_n^2}{P_n}$, owing
 to the fact \cite{yagan} that
\begin{equation}
p_s \sim \frac{K_n^2}{P_n}  \quad \textrm{if} \quad \frac{K_n^2}{P_n} = o(1).
\label{eq:oy_simple}
\end{equation}
However, when random key graph $G(n,K_n,P_n)$
is intersected with an ER graph $G(n,p_n)$ (as in the case of $\mathbb{G}_{on}$)
the simplification does not occur
naturally (even under (\ref{eq:oy_simple})), and
as seen below,
simpler forms of the zero-one laws are obtained at the expense of extra conditions enforced
on the parameters
$K_n$ and $P_n$.

\begin{cor}\label{cor:unreliableq1}
Consider a positive integer $k$, and scalings $K,P:
\mathbb{N}_0 \rightarrow \mathbb{N}_0$, $p:
\mathbb{N}_0 \rightarrow (0,1)$ such that
$K_n \geq 2$ for all $n$ sufficiently large.
 We define a sequence $\alpha:
\mathbb{N}_0 \rightarrow \mathbb{R}$ such that for any $n \in
\mathbb{N}_0$, we have
\begin{align}
p_n \cdot \frac{K_n^2}{P_n} &  = \frac{\ln  n + {(k-1)} \ln \ln n + {\alpha_n}}{n}
.\label{eqn:unreliable_cor}
\end{align}
The properties (a) and (b) below hold.

(a) If $\frac{K_n ^2}{P_n} = O(\frac{1}{\ln n})$ and
$\lim_{n \to \infty} (\ln n + (k-1) \ln \ln n + \alpha_n) = \infty$,
then
\begin{align}
\lim_{n \to \infty}\mathbb{P}\left[\mathbb{G}_{on} \textrm{ is
}k\textrm{-connected}\hspace{2pt}\right]
  &  = 0
  \quad \textrm{if}~\lim_{n \to \infty}{\alpha_n} = -\infty,
   \label{thm1kc_cor}
 \end{align}
 and
\begin{align}
\lim_{n \to \infty} \mathbb{P}\left[ \begin{array}{c}
\textrm{Minimum~node~degree} \\  \mbox{of ${\mathbb{G}}_{on}$ is no less than
}k \end{array}
\right]  &  = 0 \:\:\: \textrm{if $\lim_{n \to \infty}{\alpha_n} =-\infty$} .
\label{mnd_zero_cor}
 \end{align}
\indent (b) If $ P_n = \Omega(n)$ and
$\frac{K_n ^2}{P_n} = O(\frac{1}{\ln n})$, then
\begin{align}
\lim_{n \to \infty}\mathbb{P}\left[\mathbb{G}_{on} \textrm{ is
}k\textrm{-connected}\hspace{2pt}\right]
  &  = 1
  \:\: \textrm{if $\lim_{n \to \infty}{\alpha_n} = \infty$},
  \label{eq:OneLaw+kConnectivity_olp_cor}
 \end{align}
 and
  \begin{align}
\lim_{n \to \infty} \mathbb{P}\left[ \begin{array}{c}
\textrm{Minimum~node~degree} \\  \mbox{of ${\mathbb{G}}_{on}$ is no less than
}k \end{array}
\right] &  = 1\quad\textrm{if $\lim_{n \to \infty}{\alpha_n} =
\infty$} . \label{cor_mnd_one}
 \end{align}
\end{cor}
A proof of
Corollary \ref{cor:unreliableq1} can be found in Section \ref{subsec:proof_cor_1_yy}.
Note that the condition $\frac{K_n^2}{P_n}=O(\frac{1}{\ln n})$ enforced in
Corollary \ref{cor:unreliableq1} implies both
$\frac{K_n}{P_n}=o(1)$ and $\frac{K_n^2}{P_n}=o(1)$, and thus it is a stronger
condition than those enforced in Theorem \ref{thm:unreliableq1}.

\subsection{A Zero-One Law for $k$-Connectivity in Random Key Graphs}
\label{sec:Cor2}

We now provide a useful corollary of Theorem \ref{thm:unreliableq1}
that gives a zero-one law for $k$-connectivity in the random key graph
$G(n,K_n,P_n)$. As discussed in Section \ref{secref} below, this result
improves the one given {\em implicitly} by Rybarczyk \cite{zz}.

\begin{cor}\label{cor:k_con_rkg}
Consider a positive integer $k$, and scalings $K,P:
\mathbb{N}_0 \rightarrow \mathbb{N}_0$ such
that $K_n \geq 2$ for all $n$ sufficiently large.
With $\alpha:
\mathbb{N}_0 \rightarrow \mathbb{R}$ given by
\begin{align}
\frac{K_n^2}{P_n} &  = \frac{\ln  n + {(k-1)} \ln \ln n + {\alpha_n}}{n}, \quad n=1, 2, \ldots,
\label{eqn:scaling_rkg}
\end{align}
the following two properties hold.

(a) If either there exists an $\epsilon > 0$ such that $n \frac{K_n^2}{P_n} > \epsilon$
for all $n$ sufficiently large, or $\lim_{n \to \infty} n \frac{K_n^2}{P_n} = 0$, then we have
\begin{align}\nonumber
\lim_{n \to \infty}\mathbb{P}\left[{G}(n,K_n,P_n) \textrm{ is
}k\textrm{-connected}\hspace{2pt}\right] = 0~\textrm{if}~\lim_{n \to
\infty}{\alpha_n} =-\infty.
\end{align}

\indent (b) If $P_n = \Omega(n)$, then we have
\begin{align}\nonumber
\lim_{n \to \infty}\mathbb{P}\left[{G}(n,K_n,P_n) \textrm{ is
}k\textrm{-connected}\hspace{2pt}\right] = 1 \:\:
\textrm{if}~\lim_{n \to \infty}{\alpha_n} =\infty.
\end{align}
\end{cor}
A proof of Corollary \ref{cor:k_con_rkg} can be found in Section \ref{subsec:proof_cor_2_yy}.

\subsection{Discussion and Comparison with Related Results} \label{secref}

As already noted in the literature
\cite{r1,citeulike:4012374,erdos61conn,ryb3,zz,yagan},
Erd\H{o}s-R\'enyi graph $G(n,p_n)$ and random key graph $G(n, K_n, P_n)$ have similar
$k$-connectivity properties when they are {\em matched}
through their link probabilities; i.e. when $p_n=p_s$ with $p_s$ as
defined in (\ref{hh2psps}).
In particular, Erd\H{o}s and R\'{e}nyi
\cite{erdos61conn} showed that if $ p_n =\frac{\ln n + (k-1) \ln \ln n +
{\alpha_n}}{n}$, then $G(n, p_n)$ is asymptotically almost surely $k$-connected
(resp., not $k$-connected) if $\lim_{n\to \infty}{\alpha_n}=+\infty$
(resp., $\lim_{n\to \infty}{\alpha_n}=-\infty$). Similarly,
Rybarczyk \cite{zz} has shown under some extra
conditions (i.e., $P_n = \Theta(n^{\xi})$ with $\xi>1$)
that if $p_s =\frac{\ln n + (k-1) \ln \ln n +
{\alpha_n}}{n}$, then $G(n,K_n,P_n)$ is almost surely $k$-connected
(resp., not $k$-connected) if $\lim_{n\to \infty}{\alpha_n}=+\infty$
(resp., $\lim_{n\to \infty}{\alpha_n}=-\infty$).

The analogy between these two results could be exploited
to conjecture similar $k$-connectivity results for our system model $\mathbb{G}_{on}$.
To see this, recall from (\ref{eq:G_on_is_RKG_cap_ER_oy})
that
\begin{align}
{\mathbb{G}}_{on}  & =G(n,K_n,P_n) \cap G(n,{p_n\iffalse_{on}\fi}).
\label{eq_gon}
\end{align}
Since $G(n,K_n,P_n)$ and $G(n, p_s)$ have similar $k$-connectivity
properties, it would seem intuitive to replace $G(n,K_n,P_n)$ with $G(n, p_s)$
in the above equation (\ref{eq_gon}). Then, using
\[
{\mathbb{G}}_{on} \simeq G(n, p_s)\cap G(n,{p_n}) = G(n, p_n p_s) = G(n,p_e),
\]
we would automatically obtain Theorem \ref{thm:unreliableq1} via the aforementioned
results of Erd\H{o}s and R\'{e}nyi
\cite{erdos61conn}. Unfortunately, such heuristic approaches
can not be taken for granted as $G(n,K_n,P_n) \neq G(n, p_s)$ in general. For instance,
the two graphs are shown \cite{YaganTriangle,5383986} to exhibit quite different characteristics
in terms of properties including {\em clustering coefficient, number of triangles}, etc.
To this end, Theorem
\ref{thm:unreliableq1} formally validates the above intuition for the
$k$-connectivity property, and it is worth mentioning that we
establish Theorem
\ref{thm:unreliableq1} with a direct proof that does not rely on
coupling arguments between random key graph and ER graph.

We now compare our results with those of Rybarczyk \cite{zz}
for the $k$-connectivity of random key graph $G(n,K_n, P_n)$.
As already noted, Rybarczyk \cite[Remark 1, p. 5]{zz}
has established an analog of
Corollary \ref{cor:k_con_rkg}, but under assumptions much {\em stronger} than ours.
In particular,  her result requires that $P_n = \Theta(n^{\xi})$ where $\xi>1$.
In comparison, Corollary \ref{cor:k_con_rkg} established here
enforces only that $P_n \geq \Omega(n)$, which is clearly a much weaker condition
than $P_n = \Theta(n^{\xi})$ with $\xi>1$. More importantly, our condition
$P_n \geq \Omega(n)$ requires (from (\ref{eqn:scaling_rkg}))
only that $K_n = \Omega(\sqrt{\ln n})$ for the one-law to hold; i.e., for $\mathbb{G}_{on}$
to be $k$-connected. However,
the condition $P_n = \Theta(n^{\xi})$ with $\xi>1$ enforced in
\cite{zz} requires the key ring sizes to satisfy $K_n  = \Omega(\sqrt{n^{\xi-1} \ln n})$
with $\xi-1>0$. This condition not only constitutes a much stronger
requirement than $K_n = \Omega(\sqrt{\ln n})$, but it also renders the $k$-connectivity
result given in \cite{zz} {\em not applicable} in the context of WSNs. This is because
$K_n$ controls the number of keys kept in each
sensor's memory, and should be very small \cite{virgil}
due to limited memory and computational
capability of sensor nodes; in general $K_n=O(\ln n)$ is accepted \cite{DiPietroTissec}
as a reasonable bound on the key ring sizes.

%

\begin{figure}[!t]
\vspace{-5mm}
 \hspace{-0.2 cm}
\includegraphics[totalheight=0.29\textheight]{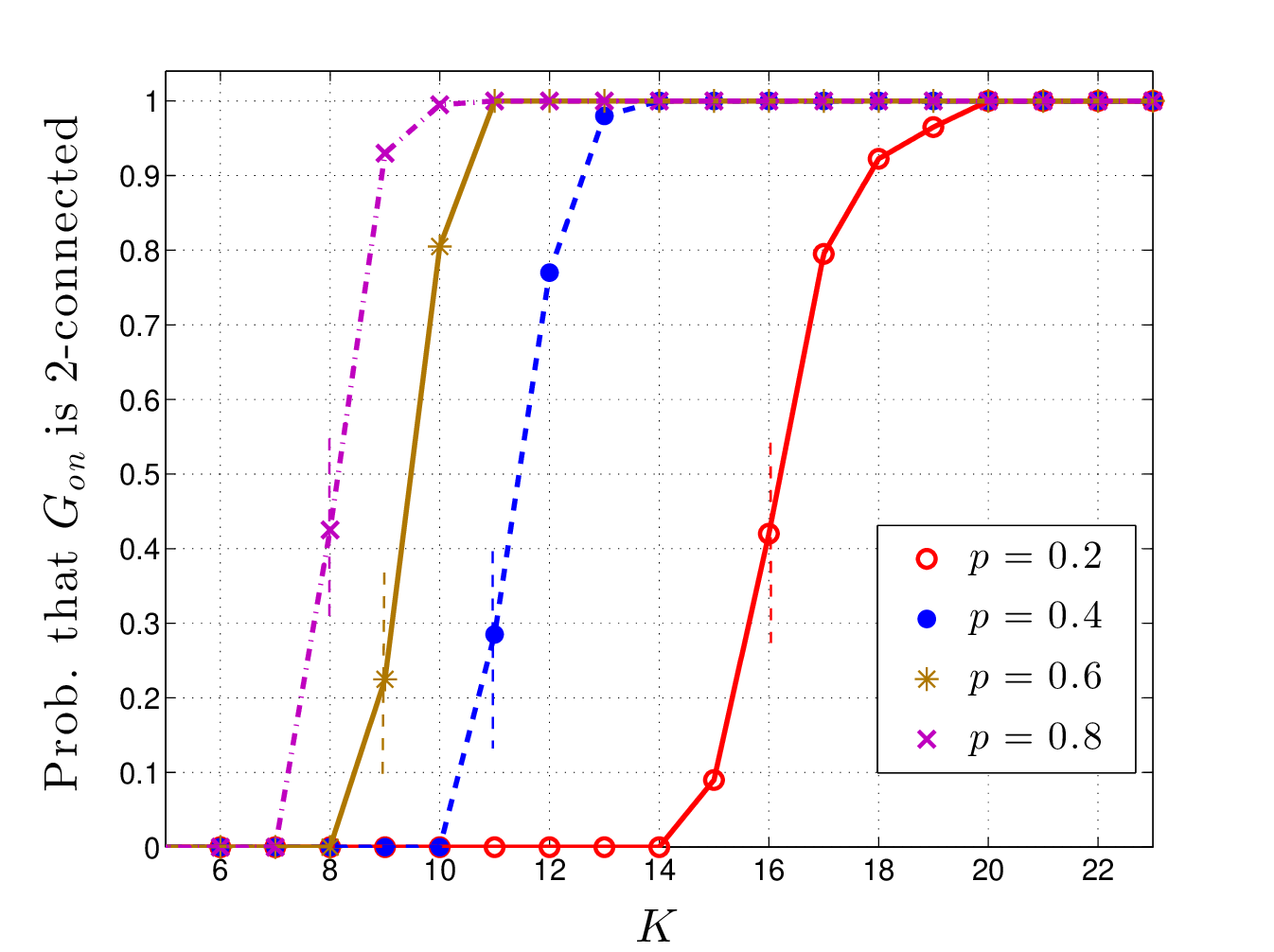} \vspace{-8.2mm} \caption{\sl Empirical
probability that $\mathbb{G}_{on}(n,K,P,p)$
         is $2$-connected for
         $p=0.2$, $p=0.4$, $p=0.6$, $p=0.8$
         with $n=2000$ and $P=10,000$. Vertical dashed lines
stand for the critical threshold of $2$-connectivity asserted by
Theorem \ref{thm:unreliableq1}.
\vspace{-4mm}} \label{figure:k_2_on_off}
\end{figure}

Finally, we compare Theorem
\ref{thm:unreliableq1} with the zero-one law
given by Ya\u{g}an \cite{yagan_onoff}
for the $1$-connectivity of $\mathbb{G}_{on}$.
As mentioned in Section \ref{related} above,
he shows that if
\begin{equation}
p_e \sim c \frac{\ln n }{n} = \frac{\ln n + (c-1) \ln n}{n} 
\label{eq:scaling_oy_comp}
\end{equation}
then $\mathbb{G}_{on}$ is a.a.s. connected if $c>1$, and
it is a.a.s. not connected if $c<1$. This was done under the
additional conditions that $P_n = \Omega(n)$ (required only for the one-law)
and that $\lim_{n \to \infty}p_n \ln n$ exists (required only for the
zero-law). On the other hand,
Theorem \ref{thm:unreliableq1} given here establishes
(by setting $k=1$) that, if
\begin{equation}
p_e = \frac{\ln n + \alpha_n}{n}
\label{eq:scaling_jz_comp}
\end{equation}
then $\mathbb{G}_{on}$ is a.a.s. connected if $\lim_{n \to \infty} \alpha_n= \infty$, and
it is a.a.s. not connected if $\lim_{n \to \infty} \alpha_n= - \infty $. This
result relies on the extra conditions $P_n =\Omega(n)$
and $\frac{K_n}{P_n}=o(1)$ for the one-law and on
$\frac{K_n^2}{P_n}=o(1)$ for the zero-law.

Comparing (\ref{eq:scaling_oy_comp}) and (\ref{eq:scaling_jz_comp}),
we see that our $1$-connectivity result for $\mathbb{G}_{on}$ is
somewhat more fine-grained than Ya\u{g}an's \cite{yagan_onoff}.
This is because,  a deviation of $\alpha_n= \pm \Omega(\ln n)$
is required to get the zero-one law in the form
(\ref{eq:scaling_oy_comp}), whereas in our formulation (\ref{eq:scaling_jz_comp}),
it suffices to have an unbounded deviation; e.g., even $\alpha_n = \pm \ln\ln \cdots \ln n$ will do.
Put differently, we cover the case of $c=1$ in (\ref{eq:scaling_oy_comp})
(i.e., the case when $p_e \sim \frac{\ln n}{n}$)
and show that $\mathbb{G}_{on}$ could be
almost surely connected or not connected, depending on the limit
of ${\alpha_n}$; in fact, if  (\ref{eq:scaling_oy_comp}) holds with
$c>1$, we see from Theorem \ref{thm:unreliableq1} that $\mathbb{G}_{on}$ is
not only $1$-connected but also $k$-connected for any $k=1,2, \ldots$.
However, it is worth noting that
the additional conditions assumed in \cite{yagan_onoff} are {\em weaker} than those
we enforce in Theorem \ref{thm:unreliableq1} for $k=1$.

\subsection{Numerical Results}
\label{subsec:Numerical}

\begin{figure}[!t]
\vspace{-5mm}
 \hspace{-0.2 cm}
\includegraphics[totalheight=0.29\textheight]{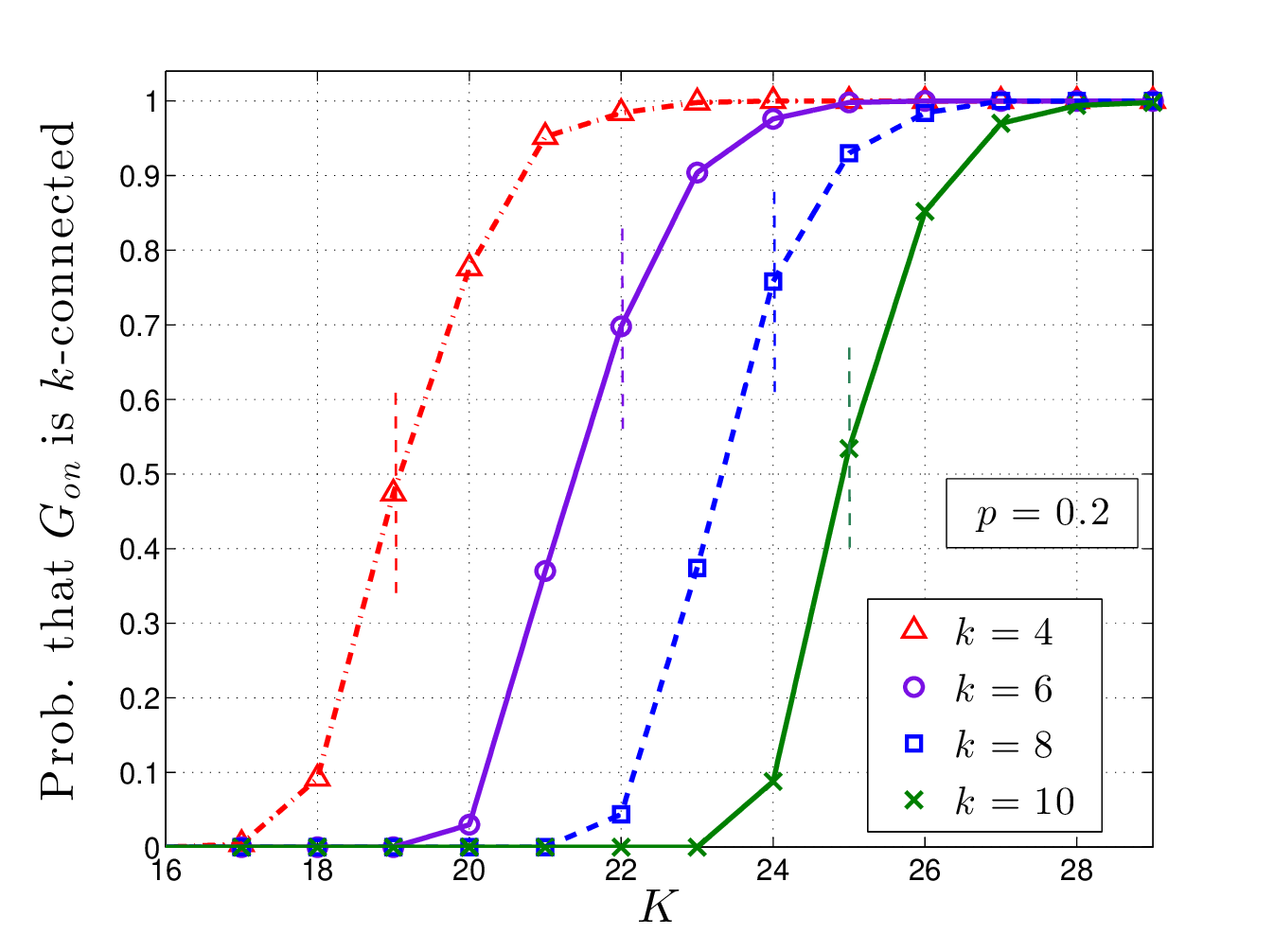}\vspace{-4mm} \caption{\sl Empirical probability
that $\mathbb{G}_{on}(n,K,P,p)$
         is $k$-connected for
         $k=4, 6, 8$, and $10$. We take $n=2000$, $P=10,000$ and
          $p=0.2$.  Vertical dashed
lines stand for the critical threshold of $k$-connectivity asserted by
Theorem \ref{thm:unreliableq1}.
\vspace{-4mm}} \label{figure:varying_k_on_off}
\end{figure}


We now present numerical results to check the
validity of Theorem \ref{thm:unreliableq1}, particularly in the non-asymptotic regime.
In all experiments, we fix the number of nodes at $n=2000$ and the
size of the key pool at $P=10,000$.  For Figure
\ref{figure:k_2_on_off}, we
consider several different probabilities of links being {\em on};
specifically, we have $p=0.2,0.4,0.6,0.8$, while varying the
parameter $K$ from $5$ to $23$; recall that $K$ stands for the
number of keys per node. For Figure \ref{figure:varying_k_on_off}, we fix
$p=0.2$ and vary $K$
from $16$ to $29$. For each parameter pair $(K, p)$, we generate $200$
independent samples of the graph
$\mathbb{G}_{on}(n,K,P,p)$ and count the number of times (out of a
possible 200) that the obtained graphs i) have minimum node degree
no less than $k$
and ii) are $k$-connected, for $k=1, 2, \ldots$. Dividing the counts by $200$, we obtain the
(empirical) probabilities for the events of interest. In all
cases, we observe that $ \mathbb{G}_{on}$
is $k$-connected whenever its minimum node degree is no less than $k$,
yielding the same empirical probability for both events. This confirms the
asymptotic equivalence of the properties of $k$-connectivity and the minimum node degree being
no less than $k$ in $\mathbb{G}_{on}$ as
stated in Theorem \ref{thm:unreliableq1}.

Figure \ref{figure:k_2_on_off} plots the empirical
probability of $2$-connectivity in $\mathbb{G}_{on}$ versus $K$ for
different $p$ values, while Figure \ref{figure:varying_k_on_off}
depicts the empirical probability of $k$-connectivity in
$\mathbb{G}_{on}$ versus $K$ for different $k$.
For each curve, we also show the critical
threshold of $k$-connectivity asserted by Theorem
\ref{thm:unreliableq1} (viz. (\ref{eqn:unreliable})) by a vertical dashed line. Namely,
the vertical dashed lines stand for the minimum integer value of
$K$ that satisfies
\begin{equation}
p_e= p \cdot \left( 1-{{{P-K} \choose K} \bigg/ {P \choose K}}\right) >
\frac{\ln n + \ln\ln n}{n}. \label{eq:threshold}
\end{equation}
Even with $n=2000$, the threshold
behavior in the probability of
$k$-connectivity is evident; it
transitions from {\em zero} to {\em one} with $K$ varying very slightly from a certain value that is close to the analytical
prediction obtained from (\ref{eq:threshold}).
Hence, we conclude that the
experimentally observed thresholds of
$k$-connectivity are in good
agreement with our theoretical results.

\subsection{A proof of Corollary \ref{cor:unreliableq1}}
\label{subsec:proof_cor_1_yy}
Consider $p_n$, $K_n$ and $P_n$
as in the statement of Corollary \ref{cor:unreliableq1} such that
(\ref{eqn:unreliable_cor}) holds.
As explained above, conditions
$\frac{K_n}{P_n}=o(1)$ and $\frac{K_n^2}{P_n}=o(1)$ both hold.
The proof is based on Theorem
\ref{thm:unreliableq1}. Namely, we will show that if the sequence $\alpha':
\mathbb{N}_0 \rightarrow \mathbb{R}$ is defined such that
\begin{equation}
p_e  = \frac{\ln  n + {(k-1)} \ln \ln n + {\alpha'_n}}{n}
\label{eq:oy_aux}
\end{equation}
for any $n \in
\mathbb{N}_0$, then it holds that
\begin{equation}
\alpha'_n=\alpha_n \pm O(1)
\label{eq:oy_aux2}
\end{equation}
under the enforced assumptions. In view of
$\lim_{n \to \infty} (\ln n + (k-1) \ln \ln n + \alpha_n) = \infty$
and (\ref{eq:oy_aux2}),
we get
$
\lim_{n \to \infty} p_e n =\infty
$
from (\ref{eq:oy_aux}).
Thus, for any $\epsilon > 0$, we have $p_e n > \epsilon$ for all $n$ sufficiently large.
Hence, all the conditions enforced by Theorem \ref{thm:unreliableq1}
are met, and
under (\ref{eq:oy_aux})
and (\ref{eq:oy_aux2}), Corollary \ref{cor:unreliableq1} follows from
Theorem \ref{thm:unreliableq1} since
$\lim_{n \to \infty}{\alpha'_n} =
\pm \infty$ if $\lim_{n \to \infty}{\alpha_n} = \pm \infty$.

We now establish (\ref{eq:oy_aux2}).
First, as seen by the analysis given in Section
\ref{commonideaforproof} below, we can introduce the extra condition
${\alpha_n} = o ( \ln n)$ in proving part (b) of Corollary \ref{cor:unreliableq1}; i.e.,
in proving the one-law under the condition $\lim_{n \to \infty} \alpha_n = \infty$.
This yields $p_n \frac{K_n ^2 }{P_n} = O(\frac{\ln n}{n})$ under (\ref{eqn:unreliable_cor}).
Also, in the case $\lim_{n \to \infty} \alpha_n = - \infty$, we have $\alpha_n < 0$
for all $n$ sufficiently large so that $p_n \frac{K_n ^2 }{P_n} = O(\frac{\ln n}{n})$.
Now, in order to establish (\ref{eq:oy_aux2}), we observe from part (a) of
Lemma \ref{jzremfb}\footnote{Except Fact 1 and Lemmas 1-6, the statements of other
facts and lemmas are all given in Appendix \ref{sec:addi:f:l}.} that
\begin{align}
p_{s} & =  \frac{K_n^2}{P_n} \pm O\left(
 \frac{K_n^4}{P_n^2} \right). \label{p_s_K_n_P_n}
\end{align}
Then, from (\ref{p_s_K_n_P_n}) and the fact that $p_e = p_s {p_n\iffalse_{on}\fi}$, we
get
\begin{align}
p_e &  =  {p_n\iffalse_{on}\fi} \cdot \frac{K_n^2}{P_n} \pm
{p_n\iffalse_{on}\fi} \cdot \frac{K_n^2}{P_n} \cdot O\left(
 \frac{K_n^2}{P_n} \right)  .\label{p_e_K_n_P_n}
\end{align}
Substituting (\ref{eqn:unreliable_cor}), $p_n \frac{K_n^2}{P_n} = O(\frac{\ln n}{n})$
and $\frac{K_n^2}{P_n} = O \left( \frac{1}{\ln
n}\right)$ into (\ref{p_e_K_n_P_n}), we find
\begin{align}
p_e &  = \frac{\ln  n + {(k-1)} \ln \ln n + {\alpha_n} \pm
O(1)}{n}.
\end{align}
Comparing the above relation with (\ref{eq:oy_aux}), the desired
conclusion (\ref{eq:oy_aux2}) follows.
\hfill $\blacksquare$

\subsection{A proof of Corollary \ref{cor:k_con_rkg}}
\label{subsec:proof_cor_2_yy}
We first establish the zero-law. Pick $K_n$, $P_n$
such that (\ref{eqn:scaling_rkg}) holds with $\lim_{n \to \infty} \alpha_n=-\infty$.
It is clear that we have $\alpha_n<0$ for all $n$ sufficiently large
so that
$\frac{K_n^2}{P_n} = O(\frac{\ln n}{n})=o(1)$. In view of
(\ref{p_s_K_n_P_n}) we thus get
\[
p_s =  \frac{\ln  n + {(k-1)} \ln \ln n + {\alpha_n} \pm o(1)}{n}, \quad n=1, 2, \ldots
\]
Let $p_n=1$ for all $n$. In this case, graph
$\mathbb{G}_{on}$ becomes equivalent to $G(n,K_n,P_n)$ with
\begin{equation}\label{eq:aux_scaling_rkg_oy}
p_e =  \frac{\ln  n + {(k-1)} \ln \ln n + {\alpha_n} \pm o(1)}{n}, \quad n=1, 2, \ldots
\end{equation}
From
(\ref{eq:aux_scaling_rkg_oy}) and (\ref{eqn:scaling_rkg}), we have
$p_e n = n \frac{K_n^2}{P_n} \pm o(1)$ so that i) if there exists an
$\epsilon>0$ such that $n \frac{K_n^2}{P_n} > \epsilon$, then there
exists an $\epsilon' > 0$ such that $p_e n > \epsilon'$ for all $n$
sufficiently large and ii) if $\lim_{n \to \infty} n
\frac{K_n^2}{P_n} = 0$, then $\lim_{n \to \infty} p_e n = 0$. Thus,
all the conditions enforced by part (a) of Theorem
\ref{thm:unreliableq1} are satisfied for the given $K_n$, $P_n$ and
$p_n$. Comparing (\ref{eq:aux_scaling_rkg_oy}) with
(\ref{eqn:unreliable}), we get $\lim_{n \to \infty} \alpha_n \pm
o(1) = -\infty$ and the zero law $\lim_{n \to \infty}
\bP{G(n,K_n,P_n) \textrm{ is }k\textrm{-connected}\hspace{2pt}} = 0$
follows from (\ref{thm1kc}) of Theorem \ref{thm:unreliableq1}.

We now establish the one-law. Pick $K_n$, $P_n$
such that (\ref{eqn:scaling_rkg}) holds with $\lim_{n \to \infty} \alpha_n= + \infty$,
$P_n = \Omega(n)$ and $K_n \geq 2$ for all $n$ sufficiently large.
In view of \cite[Lemma 6.1]{yagan}, there exists $\tilde{K}_n$, $\tilde{P}_n$
such that $\tilde{K}_n \geq 2$ for all $n$ sufficiently large,
\[
\tilde{K}_n \leq K_n \quad \textrm{and} \quad \tilde{P}_n=P_n, \quad n=1,2,\ldots,
\]
and
\begin{equation}\label{eq:scaling_with_tilde}
\frac{\tilde{K}_n^2}{\tilde{P}_n}   = \frac{\ln  n + {(k-1)} \ln \ln n + {\tilde{\alpha}_n}}{n}, \quad n=1, 2, \ldots,
\end{equation}
with
\[
\tilde{\alpha}_n = O(\ln n) \quad \textrm{and} \quad \lim_{n \to \infty} \tilde{\alpha}_n =\infty.
\]
By an easy coupling argument, it is easy to
check that
\begin{align}\nonumber
\lefteqn{\mathbb{P}\left[{G}(n, \tilde{K}_n, \tilde{P}_n) \textrm{
is }k\textrm{-connected}\hspace{2pt}\right]}   \\ \nonumber
 &\leq \mathbb{P}\left[{G}(n,K_n,P_n) \textrm{ is
}k\textrm{-connected}\hspace{2pt}\right].
\end{align}
Therefore, the one-law proof will be completed upon showing
\begin{align}\nonumber
\lim_{n \to \infty} \bP{G(n, \tilde{K}_n, \tilde{P}_n) \textrm{ is
}k\textrm{-connected}\hspace{2pt}} & = 1.
\end{align}
Under
(\ref{eq:scaling_with_tilde}) we have
$\frac{\tilde{K}_n^2}{\tilde{P}_n} = O(\frac{\ln n}{n})=o(1)$ since
$\tilde{\alpha}_n=O(\ln n)$. It also follows that
$\frac{\tilde{K}_n}{\tilde{P}_n}=o(1)$.
 In view of
(\ref{p_s_K_n_P_n}), we get
\[
\tilde{p_s} =  \frac{\ln  n + {(k-1)} \ln \ln n + {\tilde{\alpha}_n} \pm o(1)}{n}, \quad n=1, 2, \ldots,
\]
and with $p_n=1$ for all $n$ sufficiently large, we obtain
\[
\tilde{p_e} =  \frac{\ln  n + {(k-1)} \ln \ln n + {\tilde{\alpha}_n} \pm o(1)}{n}, \quad n=1, 2, \ldots,
\]
It is clear that $\lim_{n \to \infty} \tilde{\alpha}_n \pm o(1) = \infty$. Thus, we
get the desired one-law by applying (\ref{eq:OneLaw+kConnectivity_olp})
of Theorem \ref{thm:unreliableq1}.
\hfill $\blacksquare$

\section{Basic Ideas for Proving Theorem \ref{thm:unreliableq1}}
\label{sec:BasicIdeas}


\subsection{$k$-Connectivity vs. Minimum Node Degree} \label{sec:any_graph_k_connected}

It is easy to see that if a graph $G$ is $k$-connected, then
the minimum node degree of $G$ is at least $k$ \cite{penrose}.
Therefore, we have
\[
\left[{G} \textrm{ is
}k\textrm{-connected}\hspace{2pt}\right] \subseteq \left[ \begin{array}{c}
\textrm{Minimum~node~degree} \\  \mbox{of ${G}$ is no less than
}k \end{array}
\right]
\]
and the inequality
\[
\mathbb{P}\left[{G} \textrm{ is
}k\textrm{-connected}\hspace{2pt}\right] \leq \mathbb{P}\left[ \begin{array}{c}
\textrm{Minimum~node~degree} \\  \mbox{of ${G}$ is no less than
}k \end{array}
\right]
\]
follows immediately.

It is now clear that (\ref{mnd_zero}) implies
(\ref{thm1kc}) and (\ref{eq:OneLaw+kConnectivity_olp}) implies
(\ref{mnd_one}). Thus, in order to prove Theorem \ref{thm:unreliableq1}, we only
need to show (\ref{mnd_zero}) under the conditions of case (a), and
(\ref{eq:OneLaw+kConnectivity_olp}) under the conditions of case
(b).

\subsection{Confining $\alpha_n$} \label{commonideaforproof}

As seen in Section
\ref{sec:any_graph_k_connected}, Theorem
\ref{thm:unreliableq1} will follow
if we show (\ref{mnd_zero}) and
(\ref{eq:OneLaw+kConnectivity_olp})
under the appropriate conditions.
In this subsection, we show that
the extra condition ${\alpha_n} =  o ( \ln n)$
can be introduced in the proof of
(\ref{eq:OneLaw+kConnectivity_olp}). Namely, we will show that
\begin{align}
\lefteqn{\textrm{part (b) of Theorem \ref{thm:unreliableq1} under $\alpha_n = o(\ln n)$}}
\nonumber \\
&\Rightarrow
\textrm{part (b) of Theorem \ref{thm:unreliableq1}} \hspace{2cm}
\label{eq:partb_with_extra_implies_partb}
\end{align}

We write $\mathbb{G}_{{on}}$ as
$\mathbb{G}_{{on}}( n,K_n, P_n, p_n)$ and remember that given
$K_n$, $P_n$ and $p_n$, one can determine
$\alpha_n$ from (\ref{eqn:unreliable}); just use (\ref{eq:link_prob_oy}).

Assume that  part (b) of Theorem \ref{thm:unreliableq1} holds under the extra condition
$\alpha_n = o(\ln n)$. The desired result (\ref{eq:partb_with_extra_implies_partb})
will follow if we establish
\begin{equation}
\lim_{n \to \infty}\mathbb{P}\left[{G} (n, \tilde{K}_n, \tilde{P}_n, \tilde{p}_n)\textrm{ is
}k\textrm{-connected}\hspace{2pt}\right] =1
\label{eq:scalings_with_tilde_k_con}
\end{equation}
for any $\tilde{K}_n$, $\tilde{P}_n$ and $\tilde{p}_n$ such that
$\frac{\tilde{K}_n}{\tilde{P}_n}=o(1)$, $\tilde{P}_n= \Omega(n)$, and
\begin{equation}
\tilde{p}_e = \frac{\ln n + (k-1) \ln \ln n + \tilde{\alpha}_n }{n}
\label{eq:scaling_p_e_tilde}
\end{equation}
holds with $\lim_{n \to \infty}{\tilde{\alpha}_n} = + \infty$. We will
prove (\ref{eq:scalings_with_tilde_k_con}) by a coupling argument.
Namely, we will show that there exist
scalings $\hat{K}_n$, $\hat{P}_n$ and $\hat{p}_n$ such that
\begin{align}
\frac{\hat{K}_n}{\hat{P}_n}=o(1) \quad \textrm{and} \quad \hat{P}_n= \Omega(n)
\label{eq:refine_alpha_cond1}
\end{align}
and
\begin{align}
\hat{p}_e = \frac{\ln n + (k-1) \ln \ln n + \hat{\alpha}_n}{n}
\label{eq:refine_alpha_scaling}
\end{align}
with
\begin{align}
\hat{\alpha}_n=o(\ln n) \quad \textrm{and} \quad \lim_{n \to \infty} \hat{\alpha}_n=\infty,
\label{eq:refine_alpha_cond2}
\end{align}
and that we have
\begin{align}
 \lefteqn{\mathbb{P}[\mathbb{G}_{{on}}( n, \tilde{K}_n,\tilde{P}_n,{\tilde{p}_n} ) \textrm{ is $k$-connected }]} &
 \nonumber \\
  & \geq  \mathbb{P}[\mathbb{G}_{{on}}( n,\hat{K}_n, \hat{P}_n, \hat{p}_n )\textrm{ is $k$-connected }] .
  \label{eq:to_show_refine_alpha}
\end{align}
Notice that $\hat{K}_n$, $\hat{P}_n$ and $\hat{p}_n$ satisfy all the conditions enforced by
part (b) of Theorem \ref{thm:unreliableq1} together with the extra condition
$\hat{\alpha}_n = o(\ln n)$. Thus, we get
\begin{align}\label{eq:scaling_with_hat_k_con}
\lim_{n \to \infty} \mathbb{P}[\mathbb{G}_{{on}}( n,\hat{K}_n, \hat{P}_n, \hat{p}_n )\textrm{ is $k$-connected }]  =1
\end{align}
by the initial assumption, and (\ref{eq:scalings_with_tilde_k_con})
follows immediately from (\ref{eq:to_show_refine_alpha}) and (\ref{eq:scaling_with_hat_k_con}).
Therefore, given any $\tilde{K}_n$, $\tilde{P}_n$ and $\tilde{p}_n$ as stated above,
if we can show the existence
of $\hat{K}_n$, $\hat{P}_n$ and $\hat{p}_n$ that satisfy (\ref{eq:refine_alpha_cond1})-(\ref{eq:to_show_refine_alpha}),
then the desired conclusion (\ref{eq:partb_with_extra_implies_partb}) will follow.

We now establish the existence of $\hat{K}_n$, $\hat{P}_n$ and $\hat{p}_n$
that satisfy (\ref{eq:refine_alpha_cond1})-(\ref{eq:to_show_refine_alpha}).
Let $\hat{P}_n=\tilde{P}_n$ and $\hat{K}_n =\tilde{K}_n$ so that  (\ref{eq:refine_alpha_cond1})
is satisfied automatically. Let $\hat{\alpha}_n = \min\left\{ \tilde{\alpha}_n, \ln \ln n\right\}$. Hence,
we have $\hat{\alpha}_n \leq  \tilde{\alpha}_n$, $\hat{\alpha}_n = o ( \ln
n)$ and $\lim_{n \to \infty}{\hat{\alpha}_n} =  + \infty$ so that
(\ref{eq:refine_alpha_cond2}) is also satisfied.
The remaining parameter $\hat{p}_n$ will be defined through
\begin{align}
\hat{p}_n \cdot  \left[1- \frac{\binom{\hat{P}_n-
\hat{K}_n}{\hat{K}_n} } {\binom{\hat{P}_n}{\hat{K}_n}}\right] =  \frac{\ln n + (k-1) \ln \ln n + \hat{\alpha}_n}{n}
\label{eq:defn_of_p_n_refining}
\end{align}
so that $\hat{p}_e = \hat{p}_n \cdot \left[1- \frac{\binom{\hat{P}_n-
\hat{K}_n}{\hat{K}_n} } {\binom{\hat{P}_n}{\hat{K}_n}}\right] $ satisfies (\ref{eq:refine_alpha_scaling}).
Thus, it remains to establish (\ref{eq:to_show_refine_alpha}).

Comparing (\ref{eq:defn_of_p_n_refining})
with (\ref{eq:scaling_p_e_tilde}), it follows that $\hat{p}_n \leq \tilde{p}_n$
since $\hat{K}_n=\tilde{K}_n$, $\hat{P}_n = \tilde{P}_n$ and
$\hat{\alpha}_n \leq \tilde{\alpha}_n$.
Consider graphs $\mathbb{G}_{{on}}(n, \tilde{K}_n, \tilde{P}_n, \tilde{p}_n )$,
$\mathbb{G}_{{on}}(n, \tilde{K}_n, \tilde{P}_n,{\hat{p}_n} )$
that have the same number of nodes $n$, the same key ring size
$\tilde{K}_n$ and the same key pool size $\tilde{P}_n$, but have different
probabilities $\tilde{p}_n$ and $\hat{p}_n$ for a link to be {\em on}.
We will show that there exists a
coupling such that $\mathbb{G}_{{on}}( n, \tilde{K}_n, \tilde{P}_n,{\hat{p}_n} )$ is a spanning
{\em subgraph} of $\mathbb{G}_{{on}}( n, \tilde{K}_n, \tilde{P}_n,{\tilde{p}_n} )$ so that,
as shown by Rybarczyk \cite[pp. 7]{zz}, we have
\begin{align}
 \lefteqn{\mathbb{P}[\mathbb{G}_{{on}}( n, \tilde{K}_n, \tilde{P}_n,{\hat{p}_n} ) \textrm{ has property }\mathscr{P}]} &
 \nonumber \\
  & \leq  \mathbb{P}[\mathbb{G}_{{on}}( n, \tilde{K}_n, \tilde{P}_n, \tilde{p}_n )\textrm{ has property }\mathscr{P}] .
  \label{graph_g_a_g_b}
\end{align}
for any monotone increasing\footnote{A graph property
is called monotone increasing if it holds under the addition of
edges in a graph. \label{fnote:gr_prop}}  graph property $\mathscr{P}$.
The properties of being $k$-connected and having a minimum
node degree of at least $k$ can easily be seen to be
 monotone increasing graph properties. Therefore, (\ref{eq:to_show_refine_alpha})
will follow immediately (with $\hat{K}_n=\tilde{K}_n$ and $\hat{P}_n=\tilde{P}_n$) if
(\ref{graph_g_a_g_b}) holds.

We now give the coupling argument that leads to (\ref{graph_g_a_g_b}).
As seen from (\ref{eq:G_on_is_RKG_cap_ER_oy}), $\mathbb{G}_{on}$ is
the intersection of a random key graph $G(n,K_n,P_n)$ and an
Erd\H{o}s-R\'enyi graph
$G(n,{p_n\iffalse_{on}\fi})$. Using graph coupling, we use the same
random key graph $G(n, \tilde{K}_n, \tilde{P}_n)$ to help construct both $\mathbb{G}_{on}(
n, \tilde{K}_n, \tilde{P}_n, {\tilde{p}_n})$ and $\mathbb{G}_{on}(n, \tilde{K}_n, \tilde{P}_n,{\hat{p}_n})$. Then
we have
\begin{align}
\mathbb{G}_{on}(
n, \tilde{K}_n, \tilde{P}_n, {\tilde{p}_n}) & = G(n, \tilde{K}_n, \tilde{P}_n) \cap
G(n,{\tilde{p}_n}) \label{on1}
\\ \mathbb{G}_{on}( n, \tilde{K}_n, \tilde{P}_n,{\hat{p}_n}) & =  G(n, \tilde{K}_n, \tilde{P}_n)
\cap G(n, \hat{p}_n). \label{on2}
\end{align}
Since ${\hat{p}_n} \leq {\tilde{p}_n}$, we
couple $G(n, \hat{p}_n)$ and
$G(n, \tilde{p}_n)$ in the following manner. Pick independent
Erd\H{o}s-R\'enyi graphs $G(n, {\hat{p}_n/ \tilde{p}_n})$ and
$G(n, \tilde{p}_n)$ on the same vertex set.
It is clear that the intersection $G(n, {\hat{p}_n/ \tilde{p}_n}) \cap G(n, \tilde{p}_n)$
will still be an Erd\H{o}s-R\'enyi graph (due to independence) with an edge probability
given by $\tilde{p}_n \cdot \frac{\hat{p}_n}{\tilde{p}_n} = \hat{p}_n$. In other words,
we have $G(n, {\hat{p}_n/ \tilde{p}_n}) \cap G(n, \tilde{p}_n) = G(n, \hat{p}_n)$.
Consequently, under this coupling, $G(n,{\hat{p}_n\iffalse_{on}\fi})$ is a
spanning subgraph of $G(n, \tilde{p}_n)$. Then from
(\ref{on1}) and (\ref{on2}), $\mathbb{G}_{on}( n, \tilde{K}_n,\tilde{P}_n, \hat{p}_n)$ is a
spanning subgraph of $\mathbb{G}_{on}( n, \tilde{K}_n, \tilde{P}_n, \tilde{p}_n)$ and  (\ref{graph_g_a_g_b})
follows.

%

\subsection{The Method of First and Second Moments}
\label{sec:Method_of_Moments_oy}

The following fact is based on the method of
the first and second moments and will be useful
in deriving zero-one laws for the minimum node degree of a
graph. We use $\mathbb{E}[\cdot]$ to denote the expectation operator.
{\begin{fact}\label{lem:graph_degreegeneral}

For any graph $G$ with $n$ nodes, let $X_{\ell}$ be the number of nodes having
degree $\ell$ in $G$, where $\ell = 0, 1, \ldots, n-1$; and
let $\delta$ be the minimum node degree of $G$. Then the following
three properties hold for any positive integer $k$.

(a) For any non-negative 
 integer $\ell$, if $\mathbb{E}[X_{\ell}] =
o(1)$, then
\begin{align}
\lim_{n \rightarrow  \infty} \bP{ \delta = \ell} =0 .
\label{eq_delta_s}
\end{align}

(b) If (\ref{eq_delta_s}) holds for $\ell=0,1,\ldots, {k-1}$, then
\begin{align}
\lim_{n\to \infty}\mathbb{P}[\delta \geq k] & =   1. \nonumber
\end{align}

(c) If $\mathbb{E}\left[{\big(X_{\ell}\big)}^2\right] \sim
\big\{\mathbb{E}\big[X_{\ell}\big]\big\}^2$ and $
\mathbb{E}\big[X_{\ell}\big] \to +\infty$ as $n\to \infty$
hold for some $\ell = 0, 1, \ldots, k-1$, then
\begin{align}
\lim_{n\to \infty}\mathbb{P}[\delta \geq k] & = 0.\nonumber
\end{align}
\end{fact}}
\noindent A proof of Fact \ref{lem:graph_degreegeneral} is given in Appendix
\ref{prf:lem:graph_degreegeneral}.

\subsection{Useful Notation for Graph $\mathbb{G}_{on} $}
\label{sec:othernotation}

We collect in this section some notation that will be used throughout.
For any event $A$, we let $\overline{A}$ be the complement of $A$. Also,
for sets $S_a$ and $S_b$, the
relative complement of $S_a$ in $S_b$ is given by $S_a \setminus S_b$.

In graph $\mathbb{G}_{on} $, for each node $v_i \in \mathcal {V}$, we define
$N_{i}$ as the set of neighbors of node $v_i$. For any two distinct
nodes $v_x$ and $v_y$, there are $(n-2)$ nodes other than $v_x$ and
$v_y$ in graph $\mathbb{G}_{on}$. These $(n-2)$ nodes can be split into the four
sets $ N_{x y} $, $N_{x \overline{y}}$, $ N_{\overline{x}y } $ and $
N_{\overline{x} \hspace{1.5pt} \overline{y}} $ as follows.
Let $N_{x y} $ be the set of nodes that are neighbors of both $v_{x}$
and $v_{y}$; i.e., $N_{x y} = N_{x} \cap N_y$. Let $N_{x \overline{y}} $
denote the set of nodes in $\mathcal{V}\setminus \{v_x,v_y\}$
that are neighbors of $v_x$, but are
not neighbors of $v_y$. Similarly, $N_{\overline{x}
y}$ is defined as the set of nodes in $\mathcal{V}\setminus \{v_x,v_y\}$
that are not
neighbors of $v_x$, but are neighbors of $v_y$. Finally, $N_{\overline{x}
\hspace{1.5pt} \overline{y}}$ is the set of nodes in $\mathcal{V}\setminus \{v_x,v_y\}$
that are not connected to either $v_x$ or $v_y$.

For any three distinct nodes $v_x, v_y$ and $v_j$, recalling that
$E_{x j}$ (resp., $E_{y j}$) is the event that there exists a link
between nodes $v_x$ (resp., $v_y$) and $v_j$, we define
 \begin{align}
E_{{x} j \cap {y} j} &: = E_{{x} j} \cap E_{{y} j}, \textrm{~~~~~~}
E_{{x} j\cap \overline{ {{y} j}}} : = E_{{x} j} \cap
\overline{E_{{y} j}}, \nonumber
\\ E_{\overline{{x} j}\cap {y} j} & : = \overline{E_{{x} j}} \cap E_{{y} j},
\textrm{ and } E_{\overline{{x} j}\cap \overline{{{y} j}}} : =
\overline{E_{{x} j}}\cap \overline{E_{{y} j}}.  \nonumber
\end{align}
In graph $\mathbb{G}_{on} $, for any non-negative integer $\ell$, let
$X_{\ell}$ be the number of nodes having degree $\ell$; let
$D_{x,\ell}$ be the event that node $v_x$ has degree $\ell$. We
define $\delta$ as the minimum node degree of graph $\mathbb{G}_{on} $, and
define $\kappa$ as the {\em connectivity} of graph $\mathbb{G}_{on} $.
The connectivity of a graph is defined as the minimum number of
nodes whose deletion renders the graph disconnected; thus, a
graph is $k$-connected if and only if its connectivity is at least
$k$. Finally, a graph is said to be simply {\em connected} if its connectivity
is at least $1$, i.e., if  it is $1$-connected.

\section{Establishing (\ref{mnd_zero}) (The Zero-Law for the Minimum Node Degree in $\mathbb{G}_{on}  $)}
\label{sec:proof_of_zero_law_oy}

Our main goal in this section is to establish (\ref{mnd_zero}) under the following
conditions:
 \begin{align}
&(\ref{eqn:unreliable}), K_n  \geq 2\textrm{ for all $n$ sufficiently large }, \frac{K_n^2}{P_n} = o(1) \label{eq_alpha1} \\
 &\lim_{n\to +\infty}
{\alpha_n} = -\infty \textrm{ and } p_e n > \epsilon>0  ~\textrm{or}~ {\lim_{n \to \infty} p_e n} =  0.
\label{eq_alpha2}
\end{align}
%
From property (c) of
Fact \ref{lem:graph_degreegeneral}, we see that the proof will be completed
if we demonstrate the following two results under the conditions (\ref{eq_alpha1}) and
(\ref{eq_alpha2}):
\begin{align}
 \lim_{n\to \infty}\mathbb{E}\big[X_{\ell}\big]  & = +\infty ,
 \label{gl1chry}
\end{align}
and
\begin{align}
\mathbb{E}\left[{\big(X_{\ell}\big)}^2\right]  &\sim
\big\{\mathbb{E}\big[X_{\ell}\big]\big\}^2. \label{gl1}
\end{align}
for some $\ell =0 ,1, \ldots, k-1$.

The first step in establishing (\ref{gl1chry}) and (\ref{gl1}) is to
compute the moments $\mathbb{E}\left[X_{\ell}\right]$ and
$\mathbb{E}\left[{\left(X_{\ell}\right)}^2\right]$. This step is taken
in the next Lemma. Recall
that in graph $\mathbb{G}_{on} $,
$X_{\ell}$ stands for the number of nodes with degree $\ell$ for each
$\ell = 0, 1, \ldots$. Also,
$D_{x,\ell}$ is the event that node $v_x$ has degree $\ell$ for each
$x=1,2,\ldots,n$.

{\begin{lem} \label{lemrst} In $\mathbb{G}_{on}  $, for any non-negative
integer $\ell$ and any two distinct nodes $v_x$ and $v_y$, we have
\begin{align}
\mathbb{E}\big[X_{\ell}\big] & = n
\mathbb{P}\left[D_{x,\ell}\right], \label{gl2lem} \\
\mathbb{E}\left[{\big(X_{\ell}\big)}^2\right]  & = n
\mathbb{P}\left[D_{x,\ell}\right] +
n(n-1)\mathbb{P}\left[D_{x,\ell}\textstyle{\bigcap} D_{y,\ell
}\right]. \label{gl3lem}
\end{align}
\end{lem}}
\noindent Lemma \ref{lemrst} follows from the exchangeability of the indicator random variables $\{\mathbf{1}[D_{i,\ell}]; i=1,\ldots, n\}$
upon writing $X_{\ell} = \sum_{i=1}^{n}\mathbf{1}[D_{i,\ell}]$.
Interested reader is referred to the full version \cite{ZhaoYaganGligorArxiv} for details.

In view of (\ref{gl2lem}), we will obtain (\ref{gl1chry}) once we show that
\begin{align}
\lim_{n\to +\infty} \left( n \mathbb{P}\left[D_{x, \ell}\right]
\right) & = +\infty .\label{gl5}
\end{align}
under (\ref{eq_alpha1}) and (\ref{eq_alpha2}).
Also, from (\ref{gl2lem}) and (\ref{gl3lem}), we get
\begin{align}
 \frac{\mathbb{E}\left[{\big(X_{\ell}\big)}^2\right]}
 {\big\{\mathbb{E}\big[X_{\ell}\big]\big\}^2} & = \frac{1}{n
\mathbb{P}\left[D_{x,\ell}\right]} +  \frac{n-1}{n} \cdot
\frac{\mathbb{P}\left[D_{x, \ell}\textstyle{\bigcap}
D_{y, \ell}\right]}{\big\{\mathbb{P}\left[D_{x, \ell}\right]\big\}^2}.
\label{gl4}
\end{align}
Thus, (\ref{gl1}) will follow upon showing (\ref{gl5})
and
\begin{align}
 \mathbb{P}\left[D_{x, \ell}
 \textstyle{\bigcap} D_{y, \ell}\right]
& \sim \big\{\mathbb{P}\left[D_{x, \ell }\right]\big\}^2 \label{gl6}
\end{align}
for some $\ell =0,1, \ldots, k-1$
under (\ref{eq_alpha1}) and
(\ref{eq_alpha2}).

We establish (\ref{gl5}) and
(\ref{gl6}) from of the following two results.
\begin{lem} \label{lsn1}
If ${p_e} = o\left(\frac{1}{\sqrt{ n}}\right)$, then for any
non-negative integer constant $\ell$ and any node $v_x$,
\begin{align}
\mathbb{P}\left[D_{x,\ell}\right] & \sim \left( \ell! \right)^{-1}
\left({p_e} n\right)^{\ell}e^{- {p_e}n}. \label{gl7ott}
\end{align}
\end{lem}
\noindent A proof of Lemma \ref{lsn1} is given in Appendix \ref{prf:lsn1}.

\begin{lem} \label{p1sim}
Let $p_s = o(1)$, $K_n \geq 2$ for all $n$ sufficiently large,
$p_e = \frac{\ln n + (k-1) \ln \ln n + \alpha_n}{n}$
with $\lim_{n \to \infty} \alpha_n= - \infty$. Then,
properties (a) and (b) below hold.

(a) If there exist an $\epsilon>0$ such that $p_e n > \epsilon$
for all $n$ sufficiently large,
then for any non-negative integer constant $\ell$ and any two distinct nodes
$v_x$ and $v_y$, we have
\begin{align}
\mathbb{P}\left[D_{x,{\ell}}\cap D_{y,{\ell}}\right] & \sim ({\ell}
!)^{-2} \left({p_e} n\right)^{2\ell}e^{-2{p_e}n} .\label{gl7ot2}
\end{align}

\indent (b)  For any two distinct nodes
$v_x$ and $v_y$, we have
\begin{align}
\mathbb{P}\left[D_{x,{0}}\cap D_{y,{0}}\right] & \sim e^{-2{p_e}n} .\label{gl7ot2_oy_new}
\end{align}
\end{lem}
\myproof Recalling that $E_{{x} {y}}$ is the event that nodes $v_{x}$ and $v_{y}$ are adjacent, we have
\begin{align}
\lefteqn{\mathbb{P}\left[D_{x,{\ell}}\cap D_{y,{\ell}}\right] }
\nonumber
\\ & = \mathbb{P}[D_{x,{\ell}}\cap D_{y,{\ell}}\cap \overline{E_{{x} {y}}}
] + \mathbb{P}[D_{x,{\ell}}\cap D_{y,{\ell}} \cap E_{{x}
  {y}} ].
  \label{eq:key_bound_for_zero_law_osy}
\end{align}
Thus, Lemma \ref{p1sim} will follow from the following two results.

\begin{proposition} \label{clpra}
Let $p_s = o(1)$, $K_n \geq 2$ for all $n$ sufficiently large and
$p_e = \frac{\ln n + (k-1) \ln \ln n + \alpha_n}{n}$
with $\lim_{n \to \infty} \alpha_n= - \infty$. Then, the following two
properties hold.

(a) If there exist an $\epsilon>0$ such that $p_e n > \epsilon$
for all $n$ sufficiently large, then for any
non-negative integer constant $\ell$, we have
\begin{align}
\mathbb{P}[D_{x,{\ell}}\cap D_{y,{\ell}}\cap \overline{E_{{x} {y}}}
] &\sim ({\ell} !)^{-2} \left({p_e} n\right)^{2\ell}e^{-2{p_e}n}.
\label{praaa}
\end{align}

\indent (b) We have
\begin{align}
\mathbb{P}[D_{x,{0}}\cap D_{y,{0}}\cap \overline{E_{{x} {y}}} ]
&\sim  e^{-2{p_e}n}. \label{praaa_l_zero_oy}
\end{align}
\end{proposition}

\begin{proposition}  \label{clpra:pr11.8aaabb}
Let $p_s = o(1)$, $K_n \geq 2$ for all $n$ sufficiently large and
$p_e = \frac{\ln n + (k-1) \ln \ln n + \alpha_n}{n}$
with $\lim_{n \to \infty} \alpha_n= - \infty$.
If there exists an $\epsilon > 0$ such that $ p_e n > \epsilon$ for
all $n$ sufficiently large, then for any
positive integer constant $\ell$, we have
\begin{align}
  \mathbb{P}[D_{x,{\ell}}\cap
D_{y,{\ell}} \cap E_{{x}
  {y}} ] &=   o\left(\mathbb{P}[D_{x,{\ell}}\cap D_{y,{\ell}}\cap \overline{E_{{x}
{y}}} ]\right).
   \label{pr11.8aaabb}
\end{align}
\end{proposition}
Propositions \ref{clpra} and
\ref{clpra:pr11.8aaabb} are established in Section \ref{sec:proof_Prop1} and
Section \ref{sec:proof_Prop2}, respectively. Now, we complete the proof of
Lemma \ref{p1sim}. Under the condition
$ p_e n > \epsilon >0$, (\ref{gl7ot2}) follows from
(\ref{praaa}) and (\ref{pr11.8aaabb}) in view of (\ref{eq:key_bound_for_zero_law_osy}).
For $\ell =0$, we obtain (\ref{gl7ot2_oy_new}) by using (\ref{praaa_l_zero_oy})
in (\ref{eq:key_bound_for_zero_law_osy}) and noting that
$\mathbb{P}[D_{x,{0}}\cap D_{y,{0}}\cap {E_{{x} {y}}}] =0$ always holds;
it is not possible for nodes $v_x$ and $v_y$
to have degree zero and yet to have an edge in between.
\myendpf

We now complete the proof of (\ref{gl5}) and
(\ref{gl6}) under (\ref{eq_alpha1}) and (\ref{eq_alpha2}).
First, in view of (\ref{eqn:unreliable}) and
the condition
$\lim_{n \to \infty}{\alpha_n} = - \infty$, we
obtain
${p_e} \leq \frac{\ln n + (k-1) \ln \ln n}{n}$
for all $n$ sufficiently large.
Thus, ${p_e} = o\big(\frac{1}{\sqrt{ n}}\big)$, and we use Lemma
\ref{lsn1} to get
\begin{align}
 n \mathbb{P}\left[D_{x, \ell}\right] & \sim n \cdot  \left( \ell ! \right)^{-1} \left({p_e} n\right)^{\ell}e^{- {p_e}n}
\label{gl7ottn2}
\end{align}
for each $\ell = 0,1, \ldots$. The proof will be given in
two steps. First, in the case where there exists
an $\epsilon >0$ such that
$p_e n > \epsilon$ for all $n$ sufficiently large, we will establish
 (\ref{gl5}) and (\ref{gl6}) for $\ell =k-1$. Next,
 for the case where $\lim_{n \to \infty} p_e n = 0$,
 we will show that  (\ref{gl5}) and (\ref{gl6}) hold for $\ell = 0$.

Assume now that $ p_e n > \epsilon > 0$ for all
$n$ sufficiently large. Substituting (\ref{eqn:unreliable})
into (\ref{gl7ottn2}) with $\ell=k-1$, we get
\begin{align} \label{eq:second_moment_calc_oy}
 \lefteqn{ n
\mathbb{P}\left[D_{x,k-1}\right] } \\  & \sim n \cdot
\left[ \left( k-1 \right)! \right]^{-1}  \left(p_e n \right)^{k-1}e^{- \ln n - (k-1) \ln \ln n - {\alpha_n}} \nonumber
\\ &= \left[ \left( k-1 \right)! \right]^{-1}
\nonumber \\
& ~~ \times \left(\ln
n + (k-1) \ln \ln n+ \alpha_n \right)^{k-1} e^{- (k-1)\ln \ln n -{\alpha_n}}
\nonumber .
\end{align}
Let
\begin{eqnarray}
\lefteqn{f_n(k; \alpha_n)} &
\nonumber \\
& := \left(\ln
n + (k-1) \ln \ln  n+ \alpha_n \right)^{k-1} e^{-(k-1) \ln \ln n- {\alpha_n}},
\nonumber
\end{eqnarray}
and observe that we have
$\ln
n + (k-1) \ln \ln n + \alpha_n \geq  \epsilon$
for all
$n$ sufficiently large since $p_e n  > \epsilon$.
On that range, fix $n$, pick $0<\gamma<1$
and consider the cases $\alpha_n \leq -(1-\gamma) \ln n$ and
$\alpha_n > -(1-\gamma) \ln n$. In the former case, we have
\begin{eqnarray}\nonumber
f_n(k; \alpha_n)  \geq  \epsilon \cdot
 e^{-(k-1) \ln \ln n +(1-\gamma) \ln n },
\end{eqnarray}
whereas in the latter we obtain
\begin{eqnarray}\nonumber
f_n(k; \alpha_n)  \geq
 \left(\gamma \ln n  \right)^{k-1} e^{-(k-1) \ln \ln n - \alpha_n} = \gamma^{k-1} e^{-\alpha_n}.
\end{eqnarray}
Thus, for all $n$ sufficiently large, we have
\begin{align}\nonumber
f_n(k; \alpha_n)
 \geq \min\left\{ \epsilon \cdot
 e^{-(k-1) \ln \ln n +(1-\gamma) \ln n }, \gamma^{k-1} e^{-\alpha_n} \right\}.
 \nonumber
\end{align}
It is now easy to see that
$
\lim_{n \to \infty} f_n(k; \alpha_n) = \infty
$
since $0<\gamma<1$ and $\lim_{n \to \infty} \alpha_n=-\infty$.
Substituting this into (\ref{eq:second_moment_calc_oy}),
we obtain  (\ref{gl5}) with $\ell =k-1$.
In addition, from (\ref{gl7ott}) of Lemma
\ref{lsn1}, and (\ref{gl7ot2}) of Lemma \ref{p1sim}, it is clear
that (\ref{gl6}) follows with $\ell =k -1$. As mentioned already, (\ref{gl5}) and
(\ref{gl6}) imply (\ref{gl1chry}) and (\ref{gl1}) in view of Lemma \ref{lemrst},
and the zero-law (\ref{mnd_zero}) is now established
for the case when $p_e n > \epsilon > 0$.

We now turn to the case where $\lim_{n \to \infty} p_e n = p_e^{\star} = 0$.
This time, we let $\ell = 0$ in (\ref{gl7ottn2}) and obtain
\[
n \mathbb{P}\left[D_{x, 0}\right]  \sim n e^{-p_e n} \sim n.
\]
We clearly have (\ref{gl5}) for $\ell = 0$. Also, from (\ref{gl7ott}) of Lemma
\ref{lsn1} with $\ell=0$, and (\ref{gl7ot2_oy_new}) of Lemma \ref{p1sim},
we obtain (\ref{gl6}) for $\ell = 0$. Having obtained (\ref{gl5}) and
(\ref{gl6}) for $\ell = 0$, we get (\ref{gl1chry}) and (\ref{gl1}) and
the zero-law (\ref{mnd_zero}) is now established from
Fact \ref{lem:graph_degreegeneral} (c). \hfill $\blacksquare$

\section{A Proof of Proposition \ref{clpra}}
\label{sec:proof_Prop1}

We start by noting that $D_{x,{\ell}}\cap
D_{y,{\ell}}\cap \overline{E_{{x} {y}}}$ stands for the event that
nodes $v_{x}$ and $v_{y}$ both have $\ell$ neighbors but are not
neighbors with each other. To compute its probability,
we specify all the possible cardinalities of sets $
N_{x y} $, $N_{x \overline{y}}$ and $ N_{\overline{x}y } $, defined
in Section \ref{sec:othernotation}. 
To this end,
we define the series of events $A_h$ in the following manner
\begin{align}
  A_h  & =   \left[|N_{x y}|=h\right] \textstyle\bigcap
 \left[|N_{x \overline{y}}|=\ell-h\right] \textstyle\bigcap
   \left[|N_{\overline{x} y }|=\ell-h\right]
  \label{defah}
\end{align}
for each $h = 0, 1, \ldots, \ell$; here, $|S|$ denotes the
cardinality of the discrete set $S$.

It is now a simple matter to check that
\begin{align}
D_{x,\ell}\cap D_{y,\ell } \cap \overline{E_{{x} {y}}}
= \bigcup_{h=0}^{{\ell}}\left(A_h\cap \overline{E_{{x}
{y}}}\right) . \label{eq:old_prop_oy}
\end{align}
for each $\ell= 0,1, \ldots$.
Using (\ref{eq:old_prop_oy}) and the fact that the events $A_h$ ($h = 0, 1,  \ldots, \ell$)
are mutually exclusive,
we obtain
\begin{align}
  \mathbb{P}\left[D_{x,\ell}\cap D_{y,\ell } \cap
\overline{E_{{x} {y}}}\right] &  =
\sum_{h=0}^{{\ell}}\mathbb{P}\left[A_h\cap \overline{E_{{x}
{y}}}\hspace{2pt}\right] . \label{di1di2pr}
\end{align}
We begin computing the right hand side (R.H.S.) of (\ref{di1di2pr}) by evaluating
$\overline{E_{{x} {y}}}$. From (\ref{eq:E_is_K_cap_C_oy}),
we have ${E_{{x} {y}}} = {K_{{x} {y}}
\cap C_{{x} {y}}}$. Hence
\begin{align}
\overline{E_{{x} {y}}} & =
 \overline{K_{{x} {y}}} \cup
\overline{C_{{x} {y}}}=
 \overline{K_{{x} {y}}} \cup
(K_{{x} {y}} \cap \overline{C_{{x} {y}}}).
\label{ense}
\end{align}
Also, by definition we have
\begin{align}
K_{{x} {y}}  & = \bigcup_{u=1}^{K_n} (|S_{xy}|=u).
\label{txtse}
\end{align}

For each $u=1,2,\ldots, K_n$, we define event $\mathcal{X}_u$ as follows:
\begin{align}
\mathcal{X}_u & = (|S_{xy}|=u)\cap \overline{C_{{x} {y}}}
\label{defxu}
\end{align}
 Applying (\ref{txtse}) to (\ref{ense}) and using (\ref{defxu}), we obtain
\begin{align}
\overline{E_{{x} {y}}} & =
 \overline{K_{{x} {y}}} \cup
\left\{\left[\bigcup_{u=1}^{K_n} (|S_{xy}|=u)\right] \cap
\overline{C_{{x} {y}}}\right\} \nonumber \\ & =
 \overline{K_{{x} {y}}} \cup
\left(\bigcup_{u=1}^{K_n} \mathcal{X}_u \right). \label{eq811}
\end{align}
From (\ref{eq811}) and the fact that the events
$\overline{K_{{x} {y}}}, \mathcal{X}_1, \mathcal{X}_2, \ldots, \mathcal{X}_{K_n}$ are mutually disjoint, we obtain
\begin{align}
 \mathbb{P}\left[A_h\cap \overline{E_{{x} {y}}}\right]
 & =  \mathbb{P}\left[A_h\cap \overline{K_{{x}
{y}}}\right]  + \sum_{u=1}^{K_n}\mathbb{P}\left[A_h\cap \mathcal{X}_u \right].
\label{prahle}
\end{align}
Substituting (\ref{prahle}) into (\ref{di1di2pr}), we get
\begin{align}
\lefteqn{\mathbb{P}\left[D_{x,\ell}\cap D_{y,\ell } \cap
\overline{E_{{x} {y}}}\right]} \nonumber \\ & =
\sum_{h=0}^{\ell}\mathbb{P}\left[A_h\cap \overline{K_{{x}
{y}}}\right] +
\sum_{h=0}^{\ell}\sum_{u=1}^{K_n}\mathbb{P}\left[A_h\cap \mathcal{X}_u
\right]. \label{prahle2222}
\end{align}
Proposition \ref{clpra} will follow from the
next two results.
%


\begin{proposition1.1}
Let $\ell$ be a non-negative integer constant. If
$p_s = o(1)$, $p_e = \frac{\ln n + (k-1) \ln\ln n + \alpha_n}{n}$ with
$\lim_{n \to \infty} \alpha_n = -\infty$, then
\begin{align}
\sum_{h=0}^{\ell}\mathbb{P}\left[A_h\cap \overline{K_{{x}
{y}}}\right] &\sim ({\ell} !)^{-2} \left({p_e}
n\right)^{2\ell}e^{-2{p_e}n}.  \label{pra162219}
\end{align}
\end{proposition1.1}

\begin{proposition1.2}
Let $\ell$ be a non-negative integer constant. Consider
$p_s = o(1)$, $K_n \geq 2$
for all $n$ sufficiently large and
$p_e = \frac{\ln n + (k-1) \ln\ln n + \alpha_n}{n}$ with
$\lim_{n \to \infty} \alpha_n = -\infty$. Then, the following two
properties hold.

(a) If there exists an $\epsilon>0$ such that
$p_e n  > \epsilon$ for all $n$ sufficiently large,
then  we have
\begin{align}
\sum_{h=0}^{\ell}\sum_{u=1}^{K_n}\mathbb{P}\left[A_h\cap \mathcal{X}_u \right]
&= o\left(\sum_{h=0}^{\ell}\mathbb{P}\left[A_h\cap
\overline{K_{{x} {y}}}\right]\right). \label{pra262219}
\end{align}

\indent (b) We have
\begin{align}
 \sum_{u=1}^{K_n}\mathbb{P}\left[A_0\cap \mathcal{X}_u \right]
&= o\left( \mathbb{P}\left[A_0\cap
\overline{K_{{x} {y}}}\right]\right). \label{pra262219_l_zero_oy}
\end{align}

\end{proposition1.2}

In order to see why Proposition \ref{clpra} follows from
Propositions 1.1 and 1.2, consider $p_s$ and $p_e$ as
stated in Proposition
\ref{clpra}. Then from Propositions 1.1 and
1.2, (\ref{pra162219}) and (\ref{pra262219}) hold. Substituting
(\ref{pra162219}) and (\ref{pra262219}) into (\ref{prahle2222}), we get
(\ref{praaa}). Also, using  (\ref{pra162219}) with $\ell = 0$
we get $\mathbb{P}\left[A_0\cap \overline{K_{{x} {y}}}\right] \sim e^{-2 p_e n}$.
Using this and (\ref{pra262219_l_zero_oy}) in  (\ref{prahle2222})
with $\ell =0$, we obtain (\ref{praaa_l_zero_oy})
and Proposition \ref{clpra} is then established.
\myendpf

The rest of this section is devoted to establishing Propositions 1.1 and 1.2.
We will establish Proposition \ref{clpra:pr11.8aaabb}
in the next Section \ref{sec:proof_Prop2}, and this will complete the proof
of Lemma \ref{p1sim} and thus the zero-law (\ref{mnd_zero}).

\subsection{A Proof of Proposition 1.1}\label{proof_proposition1.2}

Given $\mathbb{P}[\overline{K_{{x} {y}}}] = 1-p_{s} \to 1
$ as $n \to  \infty$, it is clear that
\begin{align}
\sum_{h=0}^{\ell}\mathbb{P}\left[A_h\cap \overline{K_{{x}
{y}}}\right] &  \sim
\sum_{h=0}^{\ell}\mathbb{P}\left[A_h \boldsymbol{\mid}
\overline{K_{{x} {y}}}\right] \label{cl1}
\end{align}
The next result evaluates a generalization of
$\mathbb{P}\left[A_h \boldsymbol{\mid} \overline{K_{{x}
{y}}}\right]$. In addition to the proof of Proposition 1.1 here, the
proofs of Propositions 1.2 and 2.1 also use Lemma
\ref{lem:event_f_result}.

\begin{lem} \label{lem:event_f_result}
Let $m_1, m_2$ and $m_3$ be non-negative integer constants. We
define event $\mathcal {F}$ as follows.
\begin{align}
 \mathcal {F}  & : =   \left[|N_{x y}|=m_1\right] \textstyle\bigcap
 \left[|N_{x \overline{y}}|=m_2 \right] \textstyle\bigcap
   \left[|N_{\overline{x} y }|=m_3\right]. \label{def_event_f}
\end{align}
Then given $u$ in $\{0, 1, \ldots, K_n \}$ and $p_e = \frac{\ln n + (k-1) \ln \ln n +\alpha_n}{n}$
with $\lim_{n \to \infty} \alpha_n = -\infty$, we have
\begin{align}
 \mathbb{P}\left[\mathcal {F} \boldsymbol{\mid}
(|S_{xy}| = u) \right]   &\sim
\frac{n^{m_1+m_2+m_3}}{m_1! m_2! m_3! }  \cdot
e^{-2p_en + \frac{ {p_e p_n \iffalse_{on}\fi}u }{ K_n}n  } \nonumber\\
&\quad ~ \times \left\{\mathbb{P}[E_{{x} j \cap {y} j}
\boldsymbol{\mid} (|S_{xy}| = u)]\right\}^{m_1}  \nonumber\\
&\quad ~ \times  \{\mathbb{P}[E_{{x} j \cap {\overline{y j}}}
\boldsymbol{\mid} (|S_{xy}| = u)]\}^{m_2}\nonumber\\  & \quad ~ \times
\{\mathbb{P}[E_{\overline{{x} j}\cap {y} j}\boldsymbol{\mid}
(|S_{xy}| = u)]\}^{m_3} \label{event_f_result}
\end{align}
with $j$ distinct from $x$ and $y$.
\end{lem}
\noindent A proof of Lemma \ref{lem:event_f_result} is given in
Appendix \ref{proof:lem:event_f_result}.

Given the definition of $A_h$ in (\ref{defah}) and
$\overline{K_{{x} {y}}} \Leftrightarrow (|S_{xy}| = 0)$, we
let $m_1 = h, m_2 = m_3 = \ell-h $ and $u = 0$ in Lemma
\ref{lem:event_f_result} in order to compute $\mathbb{P}\left[A_h
\boldsymbol{\mid} \overline{K_{{x} {y}}}\right]$. We get
\begin{align}
 \lefteqn{\mathbb{P}\left[A_h \boldsymbol{\mid}
\overline{K_{{x} {y}}}\right] }\nonumber \\ &\sim
\frac{n^{2\ell-h}}{h![(\ell-h)!]^2} \cdot e^{-2 {p_e} n} \cdot
\left\{\mathbb{P}[E_{{x} j \cap {y} j} \boldsymbol{\mid}
\overline{K_{{x} {y}}}]\right\}^{h} \nonumber\\  & \quad
\times \{\mathbb{P}[E_{\overline{{x} j}\cap {y} j}\boldsymbol{\mid}
\overline{K_{{x} {y}}}]\}^{{\ell}-h} \{\mathbb{P}[E_{{x} j
\cap {\overline{y j}}} \boldsymbol{\mid} \overline{K_{{x}
{y}}}]\}^{{\ell}-h} . \label{pr_ah}
\end{align}
In order to compute the R.H.S. of (\ref{pr_ah}), we evaluate the
following three terms in turn:
\begin{align}
 \mathbb{P}[E_{{x} j \cap {y} j} \boldsymbol{\mid}
\overline{K_{ {x} {y}}}] ,  \mathbb{P}[E_{{x} j \cap
{\overline{y j}}} \boldsymbol{\mid} \overline{K_{ {x}
{y}}}] ,  &\textrm{ and }\mathbb{P}[E_{\overline{{x} j}\cap {y}
j}\boldsymbol{\mid} \overline{K_{ {x} {y}}}]. \nonumber
\end{align}
For the first term $\mathbb{P}[E_{{x} j \cap {y} j}
\boldsymbol{\mid} \overline{K_{{x} {y}}}]$, we use $E_{{x}
j} = K_{{x} j} \cap C_{{x} j}$ and $ E_{{y} j} =
K_{{y} j} \cap C_{{y} j}$ to obtain
\begin{align}
\lefteqn{\mathbb{P}[E_{{x} j \cap {y} j} \boldsymbol{\mid}
\overline{K_{{x} {y}}}]} \nonumber \\ & =
\mathbb{P}[(C_{{x} j} \cap C_{{y} j}) \cap
(K_{xj} \cap K_{{y} j}) \boldsymbol{\mid}
\overline{K_{{x} {y}}}]. \nonumber
 \\
 & =  {p_n\iffalse_{on}\fi} ^2
\cdot \mathbb{P}[ K_{{x} j} \cap K_{{y} j}
 \boldsymbol{\mid} \overline{K_{{x} {y}}}]
\label{eqn:oversec}
\end{align}
Applying Lemma \ref{lem:secprob} (Appendix \ref{sec:additional_lemmas_oy}) to (\ref{eqn:oversec}) and
using the definition $p_e =p_n p_s$, we get
\begin{align}
\mathbb{P}[E_{{x} j \cap {y} j} \boldsymbol{\mid}
\overline{K_{{x} {y}}}]  & \leq  {p_e} ^2.
\label{eqn:acapb2xyz}
\end{align}

We now evaluate the second term $\mathbb{P}[E_{{x} j \cap
{\overline{y j}}} \boldsymbol{\mid} \overline{K_{{x} {y}}}
]$. It is clear that $E_{{x}
j}$ is independent of $\overline{K_{{x} {y}}}$. Hence,
\begin{align}
\mathbb{P}[E_{{x} j} \boldsymbol{\mid} \overline{K_{{x}
{y}}}] & = p_e. \label{pr_exj}
\end{align}
Since $p_e = \frac{\ln n + (k-1) \ln \ln n +\alpha_n}{n}$
with $\lim_{n \to \infty} \alpha_n =- \infty$, we have
$p_e=o\left(\frac{1}{\sqrt{ n}}\right)$.
Together with (\ref{eqn:acapb2xyz}), (\ref{pr_exj}) this yields
\begin{align}
 \mathbb{P}[E_{{x} j\cap \overline{ {{y}
j}}}\boldsymbol{\mid} \overline{K_{{x} {y}}}] & =
\mathbb{P}[E_{{x} j}  \boldsymbol{\mid} \overline{K_{{x}
{y}}}]  - \mathbb{P}[E_{{x} j \cap {y} j} \boldsymbol{\mid}
\overline{K_{{x} {y}}}] \nonumber \\ & = {p_e}  -
O\left({p_e} ^2\right) \sim {p_e}.\label{acapoverb}
\end{align}
Similarly, for the third term $\mathbb{P}[E_{\overline{{x} j}\cap {y}
j}\boldsymbol{\mid} \overline{K_{ {x} {y}}}]$, we have
\begin{align}
 \mathbb{P}[E_{\overline{{x} j}\cap {y} j}\boldsymbol{\mid}
\overline{K_{{x} {y}}}]& \sim  {p_e}.\label{overacapb}
\end{align}

Now we compute the R.H.S. of (\ref{pr_ah}). Substituting
(\ref{acapoverb}) and (\ref{overacapb}) into R.H.S. of (\ref{pr_ah}),
given constant $\ell$, we obtain
\begin{align}
 \lefteqn{\mathbb{P}\left[A_h \boldsymbol{\mid}
\overline{K_{{x} {y}}}\right] }\nonumber \\ &\sim
\frac{n^{2\ell-h}}{h![(\ell-h)!]^2}\cdot e^{-2 {p_e} n} \cdot
\left\{\mathbb{P}[E_{{x} j \cap {y} j} \boldsymbol{\mid}
\overline{K_{{x} {y}}}]\right\}^{h}\cdot
  p_e ^{2(\ell-h)} . \label{eqn:eal}
\end{align}
for each $h=0,1,\ldots,\ell$.
Thus, for $h=0$, we have
\begin{align}
 \mathbb{P}\left[A_0 \boldsymbol{\mid}
\overline{K_{{x} {y}}}\right] &\sim (\ell!)^{-2}
(p_en)^{2\ell} e^{-2 p_e  n}  \label{papapa} .
\end{align}
For $h=1,2,\ldots,\ell$, we use (\ref{eqn:acapb2xyz}) and
(\ref{eqn:eal}) to get
\begin{align}
 \frac{\mathbb{P}\left[A_h \boldsymbol{\mid}
\overline{K_{{x} {y}}}\right]}{\mathbb{P}\left[A_0
\boldsymbol{\mid} \overline{K_{{x} {y}}}\right]}   &\sim
\frac{n^{ -h}(\ell!)^2}{h![(\ell-h)!]^2} \left\{\mathbb{P}[E_{{x} j
\cap {y} j} \boldsymbol{\mid} \overline{K_{{x}
{y}}}]\right\}^{h}
  p_e ^{ -2h } \nonumber \\ &\leq \frac{n^{
-h}(\ell!)^2}{h![(\ell-h)!]^2} = o(1). \nonumber
\end{align}
Thus, we have
\begin{align}
\mathbb{P}\left[A_h \boldsymbol{\mid} \overline{K_{{x}
{y}}}\right] & = o\left(\mathbb{P}\left[A_0 \boldsymbol{\mid}
\overline{K_{{x} {y}}}\right]\right), \quad h=1,2,\ldots,\ell.
\label{pbbp}
\end{align}
Applying (\ref{papapa}) and (\ref{pbbp}) to (\ref{cl1}),
we obtain the desired conclusion (\ref{pra162219}) (for Propostion 1.1)
since $\ell$ is constant.\hfill $\blacksquare$

\subsection{A Proof of Proposition 1.2}

Notice that (\ref{pra262219_l_zero_oy}) can be obtained from
(\ref{pra262219}) by setting $\ell = 0$.
Thus, in the discussion given below,
we will establish (\ref{pra262219})
for each $\ell = 0,1, \ldots$
under $ p_e n  = \Omega(1)$, and show that
this extra condition is {\em not} needed if $\ell = 0$.

We start by finding an upper bound on the
left hand side (L.H.S.) of (\ref{pra262219}). Given the definition of
$\mathcal{X}_u$ in (\ref{defxu}), 
 we obtain
\begin{align}
\mathbb{P}\left[A_h \cap \mathcal{X}_u\right] & \leq \mathbb{P}\left[A_h \cap
(|S_{xy}|=u ) \right] .
\nonumber
\end{align}
Then, we have
\begin{align}
\lefteqn{\sum_{h=0}^{\ell}\sum_{u=1}^{K_n}\mathbb{P}\left[A_h\cap
\mathcal{X}_u \right] }\nonumber\\ &  \leq
\sum_{h=0}^{\ell}\sum_{u=1}^{K_n}\mathbb{P}\left[A_h\cap (|S_{xy}|=u
) \right] \nonumber\\ &  = \sum_{u=1}^{K_n} \bigg\{
\mathbb{P}[|S_{xy}|=u] \cdot \sum_{h=0}^{\ell} \mathbb{P}\left[A_h
\boldsymbol{\mid} (|S_{xy}|=u ) \right] \bigg\}. \label{eval}
\end{align}
To compute the R.H.S. of (\ref{eval}), we
first use Lemma \ref{lemb} to get
\begin{align}
\mathbb{P}[|S_{xy}|=u ] & \leq \frac{1}{u!}
\left(\frac{K_n^2}{P_n-K_n}\right)^u. \label{sxyu}
\end{align}
Next, we compute $ \mathbb{P}\left[A_h
\boldsymbol{\mid}(|S_{xy}|=u )\right]$. Given (\ref{defah}), we let $m_1 = h$ and $m_2 = m_3 = \ell-h$
in Lemma \ref{lem:event_f_result} and obtain
\begin{align}
 \mathbb{P}\left[A_h \boldsymbol{\mid}
(|S_{xy}|=u ) \right]  &\sim
\frac{n^{2\ell-h}}{h![(\ell-h)!]^2} \cdot e^{-2p_e n + \frac{ {p_e
p_n\iffalse_{on}\fi }u}{ K_n}n }  \nonumber\\
& \quad \times \left\{\mathbb{P}[E_{{x} j
\cap {y} j} \boldsymbol{\mid} (|S_{xy}|=u )]\right\}^{h} \nonumber\\
& \quad \times \{\mathbb{P}[E_{\overline{{x} j}\cap {y}
j}\boldsymbol{\mid} (|S_{xy}|=u )]\}^{{\ell}-h} \nonumber\\  & \quad
\times \{\mathbb{P}[E_{{x} j \cap {\overline{y j}}}
\boldsymbol{\mid} (|S_{xy}|=u )]\}^{{\ell}-h} . \label{pr_ah_sxy_u}
\end{align}

 From $ {E_{{x}
j}} =  {C_{{x} j}} \cap {K_{{x} j}}$ and $ {E_{{y}
j}} = {C_{{y} j}} \cap
 {K_{{y} j}}$, it is clear that $E_{{x} j}$ and $E_{{y}
j}$ are independent of $(|S_{xy}|=u)$. 
This leads
\begin{align}
 \mathbb{P}[E_{{x} j \cap {y} j} \boldsymbol{\mid}
(|S_{xy}|=u)]  & \leq  \mathbb{P}[E_{{x} j} \boldsymbol{\mid}
(|S_{xy}|=u)]
   =
 {p_e} \label{acapbzzzj}
\\
 \mathbb{P}[E_{{x} j \cap {\overline{y j}}}
\boldsymbol{\mid} (|S_{xy}|=u)]  & \leq\mathbb{P}[E_{{x} j}
\boldsymbol{\mid} (|S_{xy}|=u)]
     =
  {p_e} \label{acapoverbzzzj} \\
 \mathbb{P}[E_{\overline{{x} j}\cap {y} j}
\boldsymbol{\mid} (|S_{xy}|=u)]  & \leq \mathbb{P}[E_{{y} j}
\boldsymbol{\mid} (|S_{xy}|=u)]
   =
 {p_e}. \label{overacapbzzzj}
\end{align}
Applying (\ref{acapbzzzj}), (\ref{acapoverbzzzj})
 and (\ref{overacapbzzzj}) to (\ref{pr_ah_sxy_u}), we obtain
\begin{align}
 \mathbb{P}\left[A_h \boldsymbol{\mid}(|S_{xy}|=u
)\right] & \leq  2 n^{{2\ell}-h } \cdot e^{-2p_e n + \frac{ {p_e
p_n\iffalse_{on}\fi n}u}{ K_n} } \cdot (p_e)^{{2\ell}-h }
 \nonumber\\
&  = 2
  e^{-2p_e n + \frac{ {p_e p_n\iffalse_{on}\fi n}u}{ K_n} }
   \left(p_e  n\right)^{{2\ell}-h } \label{eqn937}
\end{align}
for all $n$ sufficiently large.

Applying
 (\ref{eqn937}) to (\ref{eval}), we derive for all $n$ sufficiently large
\begin{align}
\lefteqn{\sum_{h=0}^{\ell}\sum_{u=1}^{K_n}\mathbb{P}\left[A_h\cap
\mathcal{X}_u \right]} \nonumber\\  &\leq \sum_{u=1}^{K_n} \bigg\{
\mathbb{P}[|S_{xy}|=u] \cdot  2 e^{-2p_e n + \frac{
{p_n\iffalse_{on}\fi}u}{ K_n} \cdot {p_e n }} \cdot
\sum_{h=0}^{\ell} \left(p_e  n\right)^{{2\ell}-h }
  \bigg\}. \label{zjprf}
\end{align}
Given (\ref{zjprf}), it is clear that (\ref{pra262219}) follows once
we prove
\begin{align}
\textrm{R.H.S. of (\ref{zjprf})}& = o
\left({\sum_{h=0}^{\ell}\mathbb{P}\left[A_h\cap
\overline{K_{{x} {y}}}\right]} \right).
\label{parti_to_prove}
\end{align}
Using $ p_e n = \Omega(1)$, \fo
\begin{align}
 \sum_{h=0}^{\ell} \left(p_e  n\right)^{{2\ell}-h }  
= O \left(p_e n\right)^{2\ell}.
  \label{eqn937sla}
\end{align}
 Notice that (\ref{eqn937sla}) follows trivially for $\ell =0$ without
 requiring $p_e n = \Omega(1)$.
  Applying (\ref{sxyu}) and (\ref{eqn937sla}) to R.H.S. of
(\ref{zjprf}), we get
\begin{eqnarray}
 \lefteqn{\textrm{R.H.S. of (\ref{zjprf})}}  \nonumber \\
& = 
O(1) \cdot \left(p_e  n\right)^{2\ell}e^{-2p_e n} \cdot \sum_{u=1}^{K_n}
 \bigg(\frac{K_n^2}{P_n-K_n}\cdot
e^{ \frac{ {p_n\iffalse_{on}\fi} }{ K_n} \cdot {p_e  n }}\bigg)^u
\label{eq1132}
\end{eqnarray}
 From (\ref{pra162219}) and (\ref{eq1132}), we have
\begin{align}
\lefteqn{ \textrm{R.H.S. of (\ref{zjprf})}} \nonumber \\
  & =  {\sum_{h=0}^{\ell}\mathbb{P}\left[A_h\cap
\overline{K_{{x} {y}}}\right]} \cdot  O( ({\ell} !)^{2})
\cdot \sum_{u=1}^{K_n}
 \bigg(\frac{K_n^2}{P_n-K_n}\cdot
e^{ \frac{ {p_n\iffalse_{on}\fi}{p_e  n } }{ K_n} }\bigg)^u.
  \label{ponkn}
\end{align}

 If we show that
\begin{align}
\frac{K_n^2}{P_n-K_n}\cdot e^{ \frac{ {p_n\iffalse_{on}\fi} }{ K_n}
\cdot {p_e  n }} & = o(1), \label{ponkn2}
\end{align}
then we obtain
\begin{align}
\sum_{u=1}^{K_n}
 \bigg(\frac{K_n^2}{P_n-K_n}\cdot
e^{ \frac{ {p_n\iffalse_{on}\fi}{p_e  n } }{ K_n} }\bigg)^u & \leq
\frac{\frac{K_n^2}{P_n-K_n}\cdot e^{ \frac{ {p_n\iffalse_{on}\fi} }{
K_n} \cdot {p_e n }}}{1-\frac{K_n^2}{P_n-K_n}\cdot e^{ \frac{
{p_n\iffalse_{on}\fi} }{ K_n} \cdot {p_e  n }}} = o(1),
\label{ponkn2sfwer}
\end{align}
leading to (\ref{pra262219}) given (\ref{ponkn}) and the fact that
$\ell$ is constant. Now we prove (\ref{ponkn2}). Given $ p_e = \frac{\ln
n + (k-1) \ln \ln n + \alpha_n}{n}$ with $\lim_{n \to \infty} \alpha_n = - \infty$
we have $p_e \leq \frac{3}{2}  \cdot \frac{\ln n}{n} $ for all sufficiently large $n$.
Recalling also that $K_n \geq 2$,
we get
\begin{align}
e^{ \frac{ {p_n\iffalse_{on}\fi}{p_e  n } }{ K_n} } & \leq
e^{\frac{3}{4} {p_n\iffalse_{on}\fi} \ln n} .\label{ja1}
\end{align}
on the same range.
From Lemma \ref{jzremfb}, property (c) (Appendix \ref{sec:additional_lemmas_oy}),
it holds under
$p_s = o(1)$ that $p_s \sim \frac{K_n^2}{P_n}$
so that $\frac{K_n^2}{P_n}=o(1)$ and
$\frac{K_n}{P_n}=o(1)$. We now obtain
\begin{align}
\frac{K_n^2}{P_n-K_n} & \sim \frac{K_n^2}{P_n} \sim p_s . \nonumber
\end{align}
Then, $\frac{K_n^2}{P_n-K_n}  \leq
2 p_s$ for all $n$ sufficiently large.
Hence, on the same range, we see from (\ref{ja1}) that
\begin{align}
 \frac{K_n^2}{P_n-K_n}\cdot e^{ \frac{
{p_n\iffalse_{on}\fi} }{ K_n} \cdot {p_e  n }} &  \leq 2 p_s
\cdot e^{\frac{3}{4}
{p_n\iffalse_{on}\fi} \ln  n}. \label{gc}
\end{align}

In order to evaluate the R.H.S. of (\ref{gc}), we define
\begin{equation}
F(n)=
2 p_s
\cdot e^{\frac{3}{4}
{p_n\iffalse_{on}\fi} \ln  n}.
\label{eq:defn_F_n_oy}
\end{equation}
With $p_n p_s = p_e  \leq  \frac{3}{2}  \cdot \frac{\ln n}{n} $
for all $n$ sufficiently large, we note that
\begin{equation}
p_s \leq \frac{3}{2} \frac{\ln n}{n p_n}.
\label{eq:new_bound_on_ps_oy}
\end{equation}
Now, fix $n$ large enough such that (\ref{gc}) and (\ref{eq:new_bound_on_ps_oy}) hold.
We consider the
cases $p_n \leq \frac{1}{\ln n}$ and $p_n > \frac{1}{\ln n}$, separately.
In the former case, we have
$
F(n) \leq 2 p_s e^{3/4}
$
immediately from (\ref{eq:defn_F_n_oy}).
In the latter case we use the bound (\ref{eq:new_bound_on_ps_oy})
to get
\[
F(n) \leq  3 \frac{\ln n}{n p_n}  e^{\frac{3}{4}
{p_n\iffalse_{on}\fi} \ln  n} < 3 \frac{(\ln n)^2}{n}  \cdot n^{3/4}
\]
upon noting that $p_n \leq 1$. Combining the two bounds, we have
\begin{equation}\label{eq:new_bound_F_oy}
F(n) \leq \max\left\{2 p_s e^{3/4}  \:,\:  3 n^{-1/4} (\ln n)^2 \right\}
\end{equation}
for all $n$ sufficiently large. Letting $n$ grow large and recalling that
$p_s = o(1)$ we obtain $\lim_{n \to \infty} F(n) = 0$. This establishes
(\ref{ponkn2}) in view of (\ref{gc}), and  (\ref{parti_to_prove}) follows from
(\ref{ponkn}) and (\ref{ponkn2sfwer}) for constant $\ell$. From
(\ref{zjprf}) and (\ref{parti_to_prove}), we finally establish the desired
conclusion (\ref{pra262219}).  Note that (\ref{pra262219_l_zero_oy})
also follows since the extra condition
 $ p_e n = \Omega(1)$ is used only once
 in obtaining (\ref{eqn937sla}) which holds trivially for $\ell =0$.
The proof of Proposition 1.2 is
thus completed.
\hfill $\blacksquare$

\section{A Proof of Proposition \ref{clpra:pr11.8aaabb}}
\label{sec:proof_Prop2}

Given (\ref{prahle2222}) and Proposition 1.2 (property (a)), it is clear that
Proposition \ref{clpra:pr11.8aaabb} will follow if we show
for each $\ell =1, 2 \ldots$ that
\begin{equation}
\mathbb{P}[D_{x,{\ell}}\cap D_{y,{\ell}} \cap E_{{x}
  {y}} ]  =
o\left(\sum_{h=0}^{\ell}\mathbb{P}\left[A_h\cap
\overline{K_{{x} {y}}}\right]\right).
\label{proposition2goal}
\end{equation}

In order to establish (\ref{proposition2goal}), we evaluate
$\mathbb{P}[D_{x,{\ell}}\cap D_{y,{\ell}} \cap E_{{x}
  {y}} ]$ proceeding similarly as in the proof of Proposition \ref{clpra}.
To this end, we define the series of events $B_h$ in the
following manner
\begin{align}
B_h  =& \left(|N_{x y}|=h\right) \textstyle\bigcap \left(|N_{x
\overline{y}}|=\ell-h-1\right) \nonumber \\ &  \textstyle\bigcap
\left(|N_{\overline{x}y }|=\ell-h-1\right).  \label{defahbh}
\end{align}
for each $h = 0, 1, \ldots, \ell- 1$. An analog of (\ref{eq:old_prop_oy})
follows immediately for any positive integer $\ell$.
\begin{align}
D_{x,\ell}\cap D_{y,\ell }\cap {E_{{x} {y}}} =
\bigcup_{h=0}^{\ell-1}\left(B_h\cap {E_{{x} {y}}}\right).
\label{nonzjbcs}
\end{align}
The minus one term on $\ell$ is due to the fact that $x$ and $y$
are adjacent on event $E_{{x} {y}}$;
there can be at most $\ell -1$ nodes that are neighbors of both
$x$ and $y$ on $D_{x,\ell}\cap D_{y,\ell} \cap E_{{x} {y}}$.

 Given (\ref{nonzjbcs}) and mutually exclusive events $B_h$ ($h = 0, 1, \ldots, \ell -
1$), we obtain
\begin{align}
  \mathbb{P}[D_{x,\ell}\cap D_{y,\ell
} \cap E_{{x}
  {y}} ] & = \sum_{h=0}^{\ell-1}\mathbb{P}\left[B_h\cap
{E_{{x} {y}}}\right] .\label{di1di2prbh}
\end{align}
We will establish Proposition \ref{clpra:pr11.8aaabb} by obtaining
the following result which evaluates the R.H.S. of
(\ref{di1di2prbh}).

\begin{proposition2.1}
Let $\ell$ be a positive integer constant. If
$p_s = o(1)$,
$p_e = \frac{\ln n + \ln\ln n + \alpha_n}{n}$ with
$\lim_{n \to \infty} \alpha_n = -\infty$ and
$ p_e n = \Omega(1)$,
then
\begin{align}
\sum_{h=0}^{\ell-1}\mathbb{P}\left[B_h\cap {E_{{x} {y}}}\right] &=
o\left(\sum_{h=0}^{\ell}\mathbb{P}\left[A_h\cap
\overline{K_{{x} {y}}}\right]\right).
\label{proposition2.1_result}
\end{align}
\end{proposition2.1}

In order to see why Proposition \ref{clpra:pr11.8aaabb} follows
from Proposition 2.1, observe that
(\ref{proposition2.1_result}) establishes
(\ref{proposition2goal}) with the help of (\ref{di1di2prbh}).
As noted before,
this establishes Proposition \ref{clpra:pr11.8aaabb}.

\myproof
As given in (\ref{txtse}), $K_{{x} {y}} =
\bigcup_{u=1}^{K_n} [|S_{xy}|=u]$. Using this and the fact
that $ E_{{x} {y}} =
{K_{{x} {y}} \cap C_{{x} {y}}}$, we get
\[
 {E_{{x} {y}}} =
 \bigcup_{u=1}^{K_n} \big[ (|S_{xy}|=u) \bigcap
{C_{{x} {y}}}\big].
\]
We use $\mathcal{Y}_u$ to denote the event $ (|S_{xy}|=u)\cap {C_{{x}
{y}}}$, where $u = 1, 2, \ldots, K_n$. Thus, $E_{{x} {y}}
= \bigcup_{u=1}^{K_n} \mathcal{Y}_u$. Then considering that
the events $Y_1, Y_2, \ldots, Y_{K_n}$ are disjoint, we get
\begin{align}
 \mathbb{P}\left[B_h\cap {E_{{x} {y}}}\right] & =
 \mathbb{P}\left[B_h\cap \left(\bigcup_{u=1}^{K_n} \mathcal{Y}_u\right)\right]
=\sum_{u=1}^{K_n}\mathbb{P}\left[B_h\cap \mathcal{Y}_u \right].
\label{di1di2prbhzzbh}
\end{align}
Given $\mathcal{Y}_u = [(|S_{xy}|=u)\cap {C_{{x} {y}}}]$, we obtain
\begin{align}
\mathbb{P}\left[B_h\cap \mathcal{Y}_u\right] & \leq \mathbb{P}\left[ B_h \cap
(|S_{xy}|=u ) \right] . \label{sjtu}
\end{align}
Applying (\ref{sjtu}) to (\ref{di1di2prbhzzbh}), \fo
\begin{align}
\lefteqn{\sum_{h=0}^{\ell-1}\mathbb{P}\left[B_h\cap {E_{{x}
{y}}}\right]}\nonumber\\ &  \leq \sum_{h=0}^{\ell-1}
\sum_{u=1}^{K_n}
 \mathbb{P}\left[B_h\cap (|S_{xy}|=u) \right]
 \nonumber\\ &  = \sum_{u=1}^{K_n} \left\{\mathbb{P}[|S_{xy}|=u]
 \cdot \sum_{h=0}^{\ell-1}
 \mathbb{P}\left[B_h
\boldsymbol{\mid} (|S_{xy}|=u ) \right] \right\}. \label{prdi1di2bh}
\end{align}

R.H.S. of (\ref{prdi1di2bh}) is similar to the
R.H.S. of (\ref{eval}), whence it will be computed in a
similar manner. We first calculate $  \mathbb{P}\left[B_h \boldsymbol{\mid}
(|S_{xy}|=u ) \right]$. Given the definition of $B_h$ in
(\ref{defahbh}), we let $m_1 = h$ and $m_2 = m_3 = \ell-h - 1 $ in
Lemma \ref{lem:event_f_result} to obtain
\begin{align}
\mathbb{P}\left[B_h \boldsymbol{\mid}
(|S_{xy}|=u ) \right]  &\sim
\frac{n^{2\ell-h-2}}{h![(\ell-h-1)!]^2} \cdot e^{-2p_e n + \frac{
{p_e
p_n\iffalse_{on}\fi }u}{ K_n} n}  \nonumber\\
& \quad \times \left\{\mathbb{P}[E_{{x} j
\cap {y} j} \boldsymbol{\mid} (|S_{xy}|=u )]\right\}^{h} \nonumber\\
& \quad \times \{\mathbb{P}[E_{\overline{{x} j}\cap {y}
j}\boldsymbol{\mid} (|S_{xy}|=u )]\}^{{\ell}-h-1} \nonumber\\  &
\quad \times \{\mathbb{P}[E_{{x} j \cap {\overline{y j}}}
\boldsymbol{\mid} (|S_{xy}|=u )]\}^{{\ell}-h-1} .
 \label{eqn:f2hbhaf}
\end{align}
Substituting (\ref{acapbzzzj}), (\ref{acapoverbzzzj}) and
(\ref{overacapbzzzj}) into (\ref{eqn:f2hbhaf}), we obtain
\begin{equation}\label{eqn937sbwse}
 \mathbb{P}\left[B_h \boldsymbol{\mid} (|S_{xy}|=u )
\right] \leq 2
  e^{-2p_e n + \frac{ {p_e p_n\iffalse_{on}\fi} n u}{ K_n} } \left(p_e
  n\right)^{{2\ell}-h-2
  }.
\end{equation}
for all $n$ sufficiently large.

Returning to the evaluation of the R.H.S. of (\ref{prdi1di2bh}), we
apply
 (\ref{eqn937sbwse}) to (\ref{prdi1di2bh}) and obtain for all $n$
 sufficiently large,
\begin{align}
\lefteqn{\sum_{h=0}^{\ell-1}\mathbb{P}\left[B_h\cap {E_{{x}
{y}}}\right]} \nonumber\\  &\leq \sum_{u=1}^{K_n} \bigg\{
\mathbb{P}[|S_{xy}|=u] \cdot  2 e^{-2p_e n + \frac{
{p_n\iffalse_{on}\fi}u}{ K_n} \cdot {p_e  n }}  \cdot
\sum_{h=0}^{\ell} \left(p_e n\right)^{ {2\ell}-h-2 }
  \bigg\} \nonumber\\  & =\left(p_e  n\right)^{-2 } \times \textrm{R.H.S. of (\ref{zjprf})}.
\end{align}
From $ {p_e n} = \Omega(1)$, \fo
\begin{align}
 \sum_{h=0}^{\ell-1}\mathbb{P}\left[B_h\cap {E_{{x}
{y}}}\right] & =   O \left(\textrm{R.H.S. of
(\ref{zjprf})}\right).\label{eqn937sbwse_oy}
\end{align}
Given (\ref{parti_to_prove}) and (\ref{eqn937sbwse_oy}), we obtain
(\ref{proposition2.1_result}) and this completes the proof of
Proposition 2. \hfill $\blacksquare$

Having established Propositions \ref{clpra} and
\ref{clpra:pr11.8aaabb}, we prove Lemma \ref{p1sim}, and the zero-law (\ref{mnd_zero}) follows as explained in
Section \ref{sec:proof_of_zero_law_oy}.

\section{Establishing (\ref{eq:OneLaw+kConnectivity_olp}) (The One-Law for
$k$-Connectivity in $\mathbb{G}_{on}$)}
\label{sec:Proof_One_Law}

As shown in Section \ref{commonideaforproof}, we can enforce the
extra condition ${\alpha_n} = o ( \ln n)$ in establishing
(\ref{eq:OneLaw+kConnectivity_olp}) (i.e., the one-law for
$k$-connectivity in $\mathbb{G}_{on}$). Therefore,
we will establish (\ref{eq:OneLaw+kConnectivity_olp}) under the following
conditions:
\begin{eqnarray}\label{eq:new_conditions_oy}
(\ref{eqn:unreliable}), K_n  \geq 2\textrm{ for all $n$ sufficiently
large }, P_n = \Omega(n),  \\ \label{eq:new_conditions2_oy}
\frac{K_n}{P_n} =
o(1), \lim_{n\to \infty} {\alpha_n} = + \infty\textrm{ and
}{\alpha_n} = o ( \ln n). \qquad
\end{eqnarray}

In graph $\mathbb{G}_{on}$, consider scalings $K,P: \mathbb{N}_0
\rightarrow \mathbb{N}_0$ and $p: \mathbb{N}_0 \rightarrow
(0,1)$ as in Theorem \ref{thm:unreliableq1}.
We find it useful to define a sequence $\beta_{\ell,n} :
\mathbb{N} \times \mathbb{N}_0 \rightarrow \mathbb{R}$ through the
relation
\begin{align}
{p_e} = \frac{ \ln  n + \ell \ln  \ln  n + \beta_{\ell,n} }{n}
\label{eq:DeviationCondition}
\end{align}
for each $n \in \mathbb{N}_0$ and each $\ell \in \mathbb{N}$.
(\ref{eq:DeviationCondition}) follows by just setting
\begin{equation}
\beta_{\ell,n} := n {p_e} - \ln  n - \ell \ln  \ln n .
\label{eq:beta_l_oy}
\end{equation}
%
%

The one-law (\ref{eq:OneLaw+kConnectivity_olp}) will
follow from the next key result. Recall that, as defined in Section
\ref{sec:othernotation}, $\kappa$ is the connectivity of the graph
$\mathbb{G}_{on} $, namely the minimum number nodes whose
deletion makes it disconnected.

\begin{lem} \label{lem:OneLaw+c_v}

Let $\ell$ be a non-negative constant integer. If $K_n \geq 2$ for
any sufficiently large $n$, $ P_n = \Omega(n)$, $\frac{K_n}{P_n} =
o(1)$, and (\ref{eq:DeviationCondition}) holds with $\beta_{{\ell},n} = o(
\ln n)$ and $\lim_{n \to \infty} {\beta_{{\ell},n}} = + \infty$,
then 
%
%
\begin{align}
\lim_{n \to  \infty} \bP{\kappa= \ell} = 0 . \label{eq:OneLaw+c_v}
\end{align}
\end{lem}
\noindent We now explain why the one-law
(\ref{eq:OneLaw+kConnectivity_olp}) follows from
Lemma \ref{lem:OneLaw+c_v}.  Consider $p_n$, $K_n$ and
$P_n$ such that (\ref{eq:new_conditions_oy}) and (\ref{eq:new_conditions2_oy})
hold. Comparing (\ref{eqn:unreliable}) and
(\ref{eq:DeviationCondition}), we get
\begin{align}
 \beta_{\ell, n}& = ( k  -1 - \ell ) \ln  \ln  n + \alpha_{ n} .\label{eq:DeviationConditionsb}
\end{align}
Since $\alpha_n = o(\ln n)$ and $\lim_{n \to  \infty} \alpha_n=+ \infty$, we have for each
$\ell =0, 1, \ldots, k-1$ that
\begin{align}
\lim_{n\to \infty}
{\beta_{{\ell},n}} = + \infty  \quad  \textrm{and} \quad   \beta_{{\ell},n} = o( \ln n).
\label{eq:DeviationConditionsbq}
\end{align}
Given (\ref{eq:DeviationConditionsbq}), we use Lemma
\ref{lem:OneLaw+c_v} and obtain
\begin{align}
\lim_{n \to  \infty} \bP{\kappa= \ell} = 0, \quad \ell =0, 1,
\ldots, k-1. \nonumber
\end{align}
For any constant $k$, this implies
$\lim_{n \to \infty} \bP{\kappa \geq k} = 1$,
or equivalently
\begin{align}
\lim_{n \to \infty}\mathbb{P}\left[\mathbb{G}_{on} \textrm{ is
}k\textrm{-connected}\hspace{2pt}\right]
  &  = 1.
\nonumber
 \end{align}
This completes the proof of the one-law (\ref{eq:OneLaw+kConnectivity_olp}).
\hfill $\blacksquare$

The remaining part of this section is devoted to the proof of
Lemma \ref{lem:OneLaw+c_v}.
\myproof
We present the steps of proving Lemma \ref{lem:OneLaw+c_v} below.
First, by
a crude bounding argument, we get
\begin{eqnarray}\nonumber
\bP{\kappa = {\ell} } &\leq& \bP{ (\kappa = {\ell}) ~\cap~ (\delta >
{\ell}) } + \bP{ \delta \leq {\ell}},
\label{eq:FromConnectivityToMinimumDegree}
\end{eqnarray}
where $\delta$ is the minimum node degree of graph $\mathbb{G}_{on}$, as
defined in Section \ref{sec:othernotation}. We will prove
Lemma \ref{lem:OneLaw+c_v} by establishing the
following two results under the enforced assumptions:
\begin{align}
\lim_{n \rightarrow \infty} \bP{ \delta \leq {\ell}}
=0\quad\textrm{if }\lim_{n \to \infty} \beta_{{\ell},n}=+ \infty,
\label{eq:OneLawForMinDegree}
\end{align}
and
\begin{equation}
\lim_{n \rightarrow  \infty} \bP{  \kappa = {\ell} ~\cap~ \delta >
{\ell} } = 0 \quad\textrm{if }\lim_{n \to \infty}
\beta_{{\ell},n}=+ \infty. \label{eq:OneLawAfterReduction}
\end{equation}

We first establish (\ref{eq:OneLawForMinDegree}).
First, from $ \ell \ln \ln n =  o(\ln n) $,
$\beta_{{\ell},n} = o( \ln n)$ and ${p_e} = \frac{ \ln  n + \ell \ln
\ln  n + \beta_{\ell,n} }{n} $, it is clear that ${p_e} \sim \frac{
\ln n}{n}$. Then ${p_e} = o\big(\frac{1}{\sqrt{ n}}\big)$. Thus,
from Lemmas \ref{lemrst} and \ref{lsn1}, we get
\begin{align}
\mathbb{E}\big[X_{\ell}\big] = n \mathbb{P}\left[D_{x,\ell}\right] &
\sim n \cdot  \left( \ell ! \right)^{-1} \left({p_e}
n\right)^{\ell}e^{- {p_e}n}.\label{olp:gl7ottn2}
\end{align}
 Substituting ${p_e} \sim \frac{
\ln n}{n}$
  and (\ref{eq:DeviationCondition}) into (\ref{olp:gl7ottn2}), we get
\begin{align}\nonumber
 \mathbb{E}\big[X_{\ell}\big]  & \sim n
 \left( \ell ! \right)^{-1}  \left(\ln
n\right)^{\ell}e^{- \ln n - \ell \ln \ln n - \beta_{\ell,n}}
 = \left( \ell ! \right)^{-1}  e^{  - \beta_{\ell,n} }.
\end{align}
In view of the fact that $\lim_{n\to \infty} {\beta_{{\ell},n}} = + \infty$, we
thus obtain $\mathbb{E}\big[X_{\ell}\big] = o(1)$.
 Then from property (a) of Fact \ref{lem:graph_degreegeneral} (Section
\ref {sec:Method_of_Moments_oy}),
we get
\begin{equation}
\lim_{n \to \infty} \mathbb{P}[\delta = \ell] =0.
\label{eq:delta_is_not_ell_oy}
\end{equation}
As seen from (\ref{eq:beta_l_oy}), $\beta_{\ell, n}$ is decreasing in $\ell$. Thus,
we have $\lim_{n \to \infty }\beta_{\ell ^{\star}, n} = +\infty$ for each $\ell^{\star} = 0, 1, \ldots, \ell$.
It is also immediate from (\ref{eq:beta_l_oy}) that $\beta_{\ell ^{\star}, n} =o(\ln n)$ since
$\beta_{\ell, n} =o(\ln n)$. Therefore, using the same arguments that lead to
(\ref{eq:delta_is_not_ell_oy}), we obtain
\[
\lim_{n \to \infty} \mathbb{P}[\delta = \ell^{\star}] =0, \quad \ell^{\star}=0,1,\ldots, \ell,
\]
and (\ref{eq:OneLawForMinDegree}) follows immediately.

As (\ref{eq:OneLawForMinDegree}) is established,
it remains to prove (\ref{eq:OneLawAfterReduction}) in order to complete
the proof of Lemma \ref{lem:OneLaw+c_v}.
The basic idea in establishing (\ref{eq:OneLawAfterReduction}) is to
find a sufficiently tight upper bound on the probability $\bP{
\kappa = {\ell} ~\cap~ \delta > {\ell}  }$ and then to show that this bound
tends to zero as $n$ goes to $ +\infty$. This approach is similar to
the one used for proving the one-law for $k$-connectivity in Erd\H{o}s-R\'enyi
graphs \cite{erdos61conn}, as well as to the approach used by
Ya\u{g}an \cite{yagan_onoff} to establish the one-law for
connectivity in the graph $\mathbb{G}_{on}$.

We start by obtaining the needed upper bound. Let $\mathcal{N}$ denote
the collection of all non-empty subsets of $\{ v_1, \ldots , v_n \}$. We
define $\mathcal{N}_{*} = \{T \boldsymbol{\mid} T \in
\mathcal{N}, |T| \geq 2\}$ and $\mathcal{K}_T = \cup_{v_i \in T} S_i $.
For the reasons that will later become apparent we find it useful to
introduce the event ${\mathcal{E} (\boldsymbol{J})}$ in the following manner:
\begin{align}
{\mathcal{E} (\boldsymbol{J})}  &= \bigcup_{T \in \mathcal{N}^*} ~
\left[\left|\mathcal{K}_T\right|~\leq~{J}_{ |T|} \right],
\label{eq:E_n_defnex}
\end{align}
where $\boldsymbol{J} =[{J}_{2} , {J}_{3}, \ldots, {J}_{n } ]$ is an
$(n-1)$-dimensional integer valued array. Let
\begin{align}
r_n  & := \min \left ( {\left \lfloor \frac{P_n}{K_n} \right
\rfloor}, \left \lfloor \frac{n}{2} \right \rfloor \right ).
\label{olp_defrn}
\end{align}
We define $J_{i}$ as follows:
\begin{align}
 J_{i} &=
\begin{cases}
\max\{ \left \lfloor (1+\varepsilon) K_n \right \rfloor ,
\left \lfloor \lambda K_n i \right \rfloor \} & i=2,\ldots, r_n,\\
 \left \lfloor\mu P_n \right \rfloor & i=r_n+1, \ldots, n.
\end{cases} \label{olp_xjdef}
\end{align}
for some arbitrary constant $0<\varepsilon<1$ and
constants $\lambda, \mu$ in $(0,\frac{1}{2})$ that
will be specified later; see
(\ref{eq:ConditionOnLambda})-(\ref{eq:ConditionOnMU+1}) below.

By a crude bounding argument we now get
\begin{align}\nonumber
\lefteqn{\bP{ (\kappa = {\ell}) ~\cap~ (\delta > {\ell} )}}  \\
\label{eq:to_show_one_law_oy}
 &\leq
\bP{{\mathcal{E} (\boldsymbol{J})}} + \bP{ (\kappa = {\ell}) ~ \cap~ (\delta >
{\ell} )~ \cap~ \overline{\mathcal{E} (\boldsymbol{J})} }.
\end{align}
Hence, a proof of (\ref{eq:OneLawAfterReduction}) consists of
establishing the following two propositions.

\begin{proposition}
{Let $\ell$ be a non-negative constant integer. If
(\ref{eq:DeviationCondition}) holds with $\beta_{\ell, n}>0$, $K_n \geq 2$ and $ P_n \geq
\sigma n$ for some $\sigma>0$ for all $n$ sufficiently large
and $\frac{K_n}{P_n} = o(1)$, then
\begin{align}
  \lim_{n \to \infty} \bP{\mathcal{E} (\boldsymbol{J})} = 0,
\label{eq:OneLawAfterReductionPart1}
\end{align}
where $\boldsymbol{J} =[{J}_{2} , {J}_{3}, \ldots, {J}_{n } ]$ is as
specified in (\ref{olp_xjdef}) with arbitrary $\varepsilon$ in $(0,1)$,
constant $\lambda $ in $(0,
\frac{1}{2})$ is selected small enough to
ensure
\begin{align}
\max \left ( 2 \lambda \sigma , \lambda \left( \frac{e^2}{\sigma}
\right) ^{\frac{ \lambda }{ 1 - 2 \lambda } } \right ) < 1,
\label{eq:ConditionOnLambda}
\end{align}
and constant $\mu$ in $(0, \frac{1}{2})$ is selected so that
\begin{align}
\max \left ( 2 \left ( \sqrt{\mu} \left ( \frac{e}{ \mu } \right
)^{\mu} \right )^\sigma, \sqrt{\mu} \left ( \frac{e}{ \mu }
\right)^{\mu} \right ) < 1 . \label{eq:ConditionOnMU+1}
\end{align}
 }
\label{prop:OneLawAfterReductionPart1}
\end{proposition}
\noindent A proof of Proposition \ref{prop:OneLawAfterReductionPart1} is given
in Section \ref{sec:OneLawAfterReductionPart2aa} below. Note that for any $\sigma
>0$, $\lim_{\lambda \downarrow 0} \lambda \left( \frac{e^2}{\sigma}
\right) ^{\frac{ \lambda }{ 1 - 2 \lambda } } =0 $ so that the
condition (\ref{eq:ConditionOnLambda}) can always be met by suitably
selecting constant $\lambda > 0$ small enough. Also, we have
$\lim_{\mu \downarrow 0} \left ( \frac{e}{ \mu } \right)^{\mu} =1$,
whence $\lim_{\mu \downarrow 0} \sqrt{\mu} \left ( \frac{e}{ \mu }
\right)^{\mu} = 0$, and (\ref{eq:ConditionOnMU+1}) can be made to
hold for any constant $\sigma>0$ by taking $\mu > 0$ sufficiently
small. Finally, we remark that
the condition $P_n \geq \sigma n$ for some $\sigma>0$ is equivalent to
having $P_n = \Omega(n)$.

\begin{proposition}
{
Let $\ell$ be a non-negative constant integer. If $K_n \geq 2$ and
$ P_n \geq \sigma n$ for some $\sigma>0$ for all $n$ sufficiently large,
$\frac{K_n}{P_n} =
o(1)$, and (\ref{eq:DeviationCondition}) holds with $\beta_{{\ell},n} = o(
\ln n)$ and $\lim_{n \to \infty} {\beta_{{\ell},n}} = + \infty$,
then
\begin{align}\nonumber
\lim_{n \to \infty} \bP{ (\kappa ={\ell}) ~\cap~ (\delta >
{\ell}) ~\cap~ \overline{\mathcal{E} (\boldsymbol{J})} } = 0,
\label{eq:OneLawAfterReductionPart2}
\end{align}
where $\boldsymbol{J} =[{J}_{2} , {J}_{3}, \ldots, {J}_{n } ]$ is as
specified in (\ref{olp_xjdef}) with arbitrary
$\varepsilon$ in $(0,1)$,
constant $\mu$ in $(0,
\frac{1}{2})$ selected small enough to ensure
(\ref{eq:ConditionOnMU+1}) and constant $\lambda \in
(0,\frac{1}{2})$ selected such that it satisfies
(\ref{eq:ConditionOnLambda}). \label{prop:OneLawAfterReductionPart2}
}
\end{proposition}
\noindent A proof of Proposition \ref{prop:OneLawAfterReductionPart2} is given
in Section \ref{sec:OneLawAfterReductionPart2} below.

Using Proposition \ref{prop:OneLawAfterReductionPart1} and
Proposition \ref{prop:OneLawAfterReductionPart2} (with the same constants $\varepsilon$, $\lambda$, $\mu$)
in (\ref{eq:to_show_one_law_oy}),
we obtain the desired conclusion
(\ref{eq:OneLawAfterReduction}). The
proof of Lemma \ref{lem:OneLaw+c_v}
is now completed.
\hfill $\blacksquare$

\section{A Proof of Proposition \ref{prop:OneLawAfterReductionPart1}}\label{sec:OneLawAfterReductionPart2aa}

We begin by finding an upper bound on the probability  $\bP{\mathcal{E} (\boldsymbol{J})}$.
To this end, we define
\begin{equation}
 Y_{i}
 =\label{eq:X_S_thetaa}
\begin{cases}
 \left \lfloor \lambda K_n i \right \rfloor & i=2,\ldots, r_n,\\
\left \lfloor \mu P_n \right \rfloor & i=r_n+1, \ldots, n.
\end{cases}
\end{equation}
From (\ref{olp_xjdef}) and (\ref{eq:X_S_thetaa}), we get
\begin{align}
J_{i}&= \label{olpxja} \begin{cases}
\max\{ \left \lfloor (1+\varepsilon) K_n \right \rfloor , Y_{i} \} & i=2,\ldots, r_n,\\
 Y_{i} & i=r_n+1, \ldots, n.
\end{cases}
\end{align}
We also define
\begin{align}
\mathcal{N}_{-}& : = \{T \boldsymbol{\mid} T \in \mathcal{N},
2\leq |T| \leq r_n\}, \nonumber
\end{align}
and
\begin{align}
\mathcal{N}_{+} & : = \{T \boldsymbol{\mid}  T \in
\mathcal{N}, |T| > r_n\} .\nonumber
\end{align}
Using the definition (\ref{eq:E_n_defnex}) and the fact that
$J_{i} =Y_{i}  $ for $i=r_n+1, r_n+2, \ldots, n$, we get
\begin{align}
{\mathcal{E} (\boldsymbol{J})} & = \left (  \bigcup_{T \in \mathcal{N}_{-}}
\left[\left|\mathcal{K}_T\right| \leq {J}_{ |T|} \right] \right ) \cup \left (
\bigcup_{T \in \mathcal{N}_{+} } \left[\left|\mathcal{K}_T\right| \leq {Y}_{
|T|} \right] \right ). \label{olpeq2}
\end{align}
Given $J_{i} = \max\{ \lfloor (1+\varepsilon) K_n \rfloor , Y_{i} \}$ for
$i=2,3,\ldots, r_n$, we have
\begin{align} \label{olpeq1}
\lefteqn{\left (  \bigcup_{T \in \mathcal{N}_{-}}
\left[\left|\mathcal{K}_T\right| \leq {J}_{ |T|} \right] \right )} \\
&= \left (  \bigcup_{T \in \mathcal{N}_{-}} \left[\left|\mathcal{K}_T\right|
\leq (1+\varepsilon) K_n \right] \right ) \cup \left (  \bigcup_{T \in
\mathcal{N}_{-} } \left[\left|\mathcal{K}_T\right| \leq {Y}_{ |T|} \right]
\right ).\nonumber
\end{align}

From (\ref{olpeq2}), (\ref{olpeq1}) and the fact that $\mathcal{N}^* =
\mathcal{N}_{-} \cup \mathcal{N}_{+}$, we obtain
\begin{align}\label{olpxj2ar}
\lefteqn{{\mathcal{E} (\boldsymbol{J})}} \\ & = \left (  \bigcup_{T
\in \mathcal{N}_{-}} \left[\left|\mathcal{K}_T\right| \leq (1+\varepsilon) K_n
\right] \right ) \cup \left ( \bigcup_{T \in \mathcal{N}^* }
\left[\left|\mathcal{K}_T\right| \leq {Y}_{ |T|} \right] \right ).
\nonumber
\end{align}
It is easy to check by direct inspection that
\begin{equation}
  \bigcup_{T \in \mathcal{N}_{-}}
\left[\left|\mathcal{K}_T\right| \leq (1+\varepsilon) K_n \right]  =
   \bigcup_{T \in \mathcal{N}_{n,2}}
\left[\left|\mathcal{K}_T\right| \leq (1+\varepsilon) K_n \right]\label{olpeq2b}
\end{equation}
where $\mathcal{N}_{n,2} $ denotes the collection
of all subsets of $\{v_1, \ldots , v_n \}$ with exactly two elements.
With $\boldsymbol{Y} =[{Y}_{2} , {Y}_{3}, \ldots, {Y}_{n } ]$
and
\begin{align}
{\mathcal{E}(\boldsymbol{Y})}= \bigcup_{T \in \mathcal{N}^*} ~
\left[\left|\mathcal{K}_T\right|~\leq~{Y}_{ |T|} \right]
\label{eq:E_n_defn22}
\end{align}
it is also easy to see that
\begin{align}
  {\mathcal{E} (\boldsymbol{J})} & =  \left ( \bigcup_{T \in \mathcal{N}_{n,2}}
\left[\left|\mathcal{K}_T\right| \leq (1+\varepsilon) K_n \right] \right )  \cup
\mathcal{E}(\boldsymbol{Y}). \nonumber
\end{align}
upon using (\ref{olpeq2b}) and
(\ref{eq:E_n_defn22}) in (\ref{olpxj2ar}).

Using a standard union bound, we now get
\begin{align}
 \bP{ {\mathcal{E} (\boldsymbol{J})} } & \leq
\bP{ {\mathcal{E}(\boldsymbol{Y})} } + \sum_{T \in \mathcal{N}_{n,2}} \bP{
\left|\mathcal{K}_T\right|
\leq (1+\varepsilon) K_n  } .\nonumber
\end{align}
It was shown in \cite[Proposition 7.2]{yagan_onoff} that
given $P_n = \Omega(n)$ and $\lim_{n \to
\infty} K_n = \infty$, we have
\begin{align}
 \bP{\mathcal{E}(\boldsymbol{Y})} & = o(1).
\label{eq:first_part_oy}
\end{align}
Noting that $\lim_{n \to
\infty} K_n = \infty$ holds
in view of Lemma \ref{lem:usefulcons2} and $P_n = \Omega(n)$
by assumption, we conclude that (\ref{eq:first_part_oy}) holds under
the assumptions enforced in
Proposition \ref{prop:OneLawAfterReductionPart1}.

In order to compute $\sum_{T \in \mathcal{N}_{n,2}} \left[\left|\mathcal{K}_T\right| \leq
(1+\varepsilon) K_n \right]$, we use exchangeability and the fact that
$|\mathcal{N}_{n,2} | = {n \choose 2}$. With $\mathcal{K}_{1, 2} = S_1 \cup
S_2$, we find
\begin{align}
 \bP{ {\mathcal{E} (\boldsymbol{J})} } & \leq
o(1) + {n \choose 2}  \bP{ \mathcal{K}_{1, 2} \leq \lfloor (1+\varepsilon) K_n
\rfloor }. \label{olpexde}
\end{align}
Then, from
(\ref{olpexde}), the desired conclusion
(\ref{eq:OneLawAfterReductionPart1}) (for Proposition \ref{prop:OneLawAfterReductionPart1})
will follow if we show that
\begin{align}
  n^2 \bP{ \mathcal{K}_{1, 2} \leq \lfloor (1+\varepsilon) K_n \rfloor} =
  o(1).
\label{eq:E_n_part2}
\end{align}
This will also be established by means of the bounds
given in \cite{YaganThesis}. To this end, it was shown
\cite[Proposition 7.4.11, pp. 137--139] {YaganThesis} under the
condition $\frac{K_n}{P_n} = o(1)$ that
\[
\bP{\mathcal{K}_{1, 2}  \leq \lfloor (1+\varepsilon) K_n \rfloor} \leq
\left(\Gamma(\varepsilon)\frac{K_n}{P_n}\right)^{K_n (1-\varepsilon)},
\]
with
$\Gamma(\varepsilon):=(1+\varepsilon)e^{\frac{1+\varepsilon}{1-\varepsilon}}$.
Using this bound, we now obtain
\begin{align}\label{eq:this_bound_oy}
n^2 \bP{\mathcal{K}_{1, 2}  \leq \lfloor (1+\varepsilon) K_n \rfloor} \leq
\left(\Gamma(\varepsilon) n^{\frac{2}{
(1-\varepsilon)K_n}}\frac{K_n}{P_n}\right)^{K_n  (1-\varepsilon)}.
\end{align}
Given $ P_n \geq \sigma n$ and $\frac{K_n}{P_n} = o(1)$, there exist
a sequence $w_n$ satisfying $\lim_{n \to
+\infty} w_n = \infty$ such that for all $n$ sufficiently large, we have
\[
P_n \geq \max\{\sigma n,  K_n w_n \}.
\]
As noted before, it also holds that $\lim_{n \to
\infty} K_n = \infty$ in view of Lemma \ref{lem:usefulcons2}. It is
now easy to see that
\begin{align}
n^{\frac{2}{K_n (1-\varepsilon)}}\frac{K_n}{P_n} &\leq \min\left\{
n^{-1+\frac{2}{K_n (1-\varepsilon)}} \frac{K_n}{\sigma},
\frac{e^{\frac{2 \ln n}{K_n(1-\varepsilon)}}}{w_n}\right\} \nonumber
\\ \nonumber
&\leq \max \left\{  \frac{n^{-\frac{1}{2}} \ln n}{\sigma},
\frac{e^{\frac{2}{(1-\varepsilon)}}}{w_n}\right\}
\end{align}
for all $n$ sufficiently large to ensure that $K_n \geq
4/(1-\varepsilon) $. The last inequality follows by considering the
cases $K_n \geq \ln n$ and $K_n < \ln n$ separately for each $n$
on the given range. 
 It follows that
\[
\lim_{n \to \infty} \Gamma(\varepsilon) n^{\frac{2}{K_n
(1-\varepsilon)}}\frac{K_n}{P_n} =0,
\]
and the desired conclusion (\ref{eq:E_n_part2}) follows from (\ref{eq:this_bound_oy}).
Proposition
\ref{prop:OneLawAfterReductionPart1} is now established.
\hfill $\blacksquare$

\section{A Proof of Proposition \ref{prop:OneLawAfterReductionPart2}}
\label{sec:OneLawAfterReductionPart2}

We start by introducing some notation. For any non-empty subset $U$
of nodes, i.e., $U \subseteq \{v_1, \ldots , v_n \}$, we define the graph ${\mathbb{G}_{on}} (U)$ (with
vertex set $U$) as the subgraph of ${\mathbb{G}_{on}} $ restricted
to the nodes in $U$. If all nodes in $U$ are deleted from
${\mathbb{G}_{on}} $, the remaining graph is given by
${\mathbb{G}_{on}} ({U}^{c})$ on the vertices ${U}^{c} = \{ v_1,
\ldots , v_n \} \setminus U$. Let $\mathcal{N}_{{U}^{c}}$ denote the
collection of all non-empty subsets of $\{ v_1, \ldots , v_n \}
\setminus U$. We say that a subset $T$ in $\mathcal{N}_{{U}^{c}}$ is
{\em isolated} in ${\mathbb{G}_{on}} ({U}^{c})$ if there are no
edges (in ${\mathbb{G}_{on}} $) between the nodes in $T$ and the
nodes in $ {U}^{c} \setminus T$. This is characterized by
\[
 \overline{E_{ij}} ,
\quad v_i \in T , \ v_j \in {U}^{c} \setminus T.
\]

With each non-empty subset $T \subseteq {U}^{c}$ of nodes, we
associate several events of interest: Let $\mathcal{C}_{T}$ denote the event
that the subgraph ${\mathbb{G}_{on}}  (T)$ is itself connected. The event
$\mathcal{C}_{T}$ is completely determined by the random
variables (rvs) $\{ S_i, \ v_i \in T \}$
and $\{ C_{ij} , \ v_i , v_j \in T \}$. We also introduce
the event $\mathcal{D}_{U,T}$ to capture the fact that $T$ is isolated in
${\mathbb{G}_{on}} ({U}^{c})$, i.e.,
\begin{align}
  \mathcal{D}_{U,T}  &:= \bigcap_{ \begin{subarray} ~~~v_i \in T \\ v_j \in
{U}^{c} \setminus T
\end{subarray}} \overline{E_{ij}}. \nonumber
\end{align}
Finally, we let $\mathcal{B}_{U,T}$ denote the event that each node in $U$
has an edge with at least one node in $T$, i.e.,
\begin{eqnarray}
\mathcal{B}_{U,T} :=  \bigcap_{v_i \in U} \bigcup_{v_j \in T } E_{ij}.
\nonumber
\end{eqnarray}
We also set
\[
\mathcal{A}_{U,T}:= \mathcal{B}_{U,T} \cap \mathcal{C}_{T} \cap \mathcal{D}_{U,T} .
\]

The proof starts with the following
observations: In graph $\mathbb{G}_{on}$, if the connectivity is ${\ell} $
(i.e., $\kappa = {\ell} $) and yet each node has degree at least
${\ell}+1$ (i.e., $\delta > {\ell} $), then there must exist subsets
$U$, $T$ of nodes with $U \in \mathcal{N}$, $|U| = {\ell}$ and $T \in
\mathcal{N}_{{U}^{c}}$, $|T| \geq 2$, such that ${\mathbb{G}_{on}}  (T)$ is
connected while $T$ is isolated in ${\mathbb{G}_{on}} ({U}^{c})$. This
ensures that ${\mathbb{G}_{on}} $ can be disconnected by deleting an
appropriately selected set of ${\ell}$ nodes; i..e, nodes in $U$. Notice that, this would not be
possible for sets $T$ in $\mathcal{N}_{{U}^{c}}$ with $|T|=1$, since the degree
of a node in $T$ is at least $\ell +1$ by virtue of the event $\delta > {\ell} $;
this ensures that a single node in $T$ is connected to at least one node
in $U^c \setminus T$.
Moreover, the event $\kappa =
{\ell} $ also enforces ${\mathbb{G}_{on}} $ to remain connected after the
deletion of any ${\ell}-1$ nodes. Therefore, if there exists a
subset $U$ (with $|U|={\ell}$) such that some $T$ in
$\mathcal{N}_{{U}^{c}}$ is isolated in ${\mathbb{G}_{on}} ({U}^{c})$, then
each of the ${\ell}$ nodes in $U$ should be connected to
at least one node in $T$ {\em and} to at least one node in
${U}^{c} \setminus T$. This can easily be seen by contradiction: Consider
subsets $U \in \mathcal{N}$ with $|U|={\ell}$, and $T \in
\mathcal{N}_{{U}^{c}}$ with $|T| \geq 2$, such that there exists no
edge between the nodes in $T$ and the nodes in ${U}^{c} \setminus T$. Suppose
there exists a node $v_i$ in $U$ such that $v_i$ is connected to at
least one node in ${U}^{c} \setminus T$ but is not connected to any node in
$T$. Then, ${\mathbb{G}_{on}} $ can be disconnected by deleting the
nodes in $U \setminus \{v_i\}$ since there will be no edge between the nodes in $T$
and the nodes in $\{v_i\} \cup {U}^{c} \setminus T$. But, $|U \setminus \{v_i\}| = {\ell}-1$, and this
contradicts the fact that $\kappa = {\ell}$.

The inclusion
\begin{align}
 [ (\kappa ={\ell}) ~\cap~ (\delta > {\ell}) ] &\subseteq \bigcup_{U \in \mathcal{N}_{n,{\ell}},\: T \in
 \mathcal{N}_{{U}^{c}}: ~ |T| \geq 2} ~ \mathcal{A}_{U,T} \nonumber
\end{align}
is now immediate with $\mathcal{N}_{n,r} $ denoting the collection
of all subsets of $\{ v_1, \ldots , v_n \}$ with exactly $r$ elements. It is also easy
to check that this union need
only be taken over all subsets $T$ of $\{v_1, \ldots , v_n \}$ with $2
\leq |T| \leq \lfloor \frac{n-{\ell}}{2} \rfloor $.

We now use a standard union bound argument to obtain
\begin{eqnarray}\nonumber
\lefteqn{\bP{(\kappa ={\ell}) ~\cap~ (\delta > {\ell})
~\cap~\overline{\mathcal{E} (\boldsymbol{J})} }}
\\ \nonumber
 &\leq & \hspace{-1mm} \sum_{ U \in \mathcal{N}_{n,{\ell}}, T \in
\mathcal{N}_{{U}^{c}}:~2 \leq |T| \leq \lfloor \frac{n-{\ell}}{2} \rfloor }
\hspace{-5mm} \bP{ \mathcal{A}_{U,T}~\cap~ \overline{\mathcal{E} (\boldsymbol{J})} }
\nonumber \\
&=& \hspace{-4mm} \sum_{r=2}^{ \lfloor \frac{n-{\ell}} {2} \rfloor }
\sum_{U \in \mathcal{N}_{n,{\ell}}, T \in \mathcal{N}_{{U}^{c},r} }
\hspace{-.3cm} \bP{ \mathcal{A}_{U,T}~\cap~ \overline{\mathcal{E} (\boldsymbol{J})}}
\label{eq:BasicIdea+UnionBound}
\end{eqnarray}
with $\mathcal{N}_{{U}^{c},r}$ denoting the collection of all
subsets of ${U}^{c}$ with exactly $r$ elements.

For each $r=1, \ldots , n-\ell-1$, we simplify the notation by writing
$\mathcal{A}_{{\ell},r} := \mathcal{A}_{ \{v_1, \ldots, v_{\ell}\}, \{ v_{{\ell}+1}, \ldots ,
v_{{\ell}+ r} \} }$, $\mathcal{D}_{{\ell},r} := \mathcal{D}_{\{v_1, \ldots, v_{\ell}\}, \{
v_{{\ell}+1}, \ldots , v_{{\ell}+ r} \}}$, $\mathcal{B}_{{\ell},r} := \mathcal{B}_{ \{v_1, \ldots,
v_{\ell}\}, \{ v_{{\ell}+1}, \ldots , v_{{\ell}+ r} \} }$ and
$\mathcal{C}_{r}  := \mathcal{C}_{\{ v_{{\ell}+1}, \ldots , v_{{\ell}+ r} \}}$. 
Under the enforced assumptions on the system model (viz. Section
\ref{sec:SystemModel}), exchangeability yields
\[
\bP{ \mathcal{A}_{U,T}} = \bP{ \mathcal{A}_{{\ell},r} }, \quad U \in
\mathcal{N}_{n,{\ell}}, \:\: T \in \mathcal{N}_{{U}^{c},r}
\]
and the expression
\begin{eqnarray} \nonumber
\lefteqn{\sum_{U \in \mathcal{N}_{n,{\ell}}, T \in
\mathcal{N}_{{U}^{c},r} } \bP{ \mathcal{A}_{U,T}~\cap~ \overline{\mathcal{E}
(\boldsymbol{J})} }} &&
\\
&=& {n \choose {\ell} }{ {n-{\ell}} \choose r} ~ \bP{ \mathcal{A}_{{\ell},r}
~\cap~ \overline{\mathcal{E} (\boldsymbol{J})} } \nonumber
\end{eqnarray}
follows since $|\mathcal{N}_{n,{\ell}} | = {n \choose {\ell}}$ and
$|\mathcal{N}_{{U}^{c},r} | = {n-{\ell} \choose r}$. Substituting
into (\ref{eq:BasicIdea+UnionBound}) we obtain the key bound
\begin{eqnarray}\nonumber
\lefteqn{\bP{(\kappa ={\ell}) ~\cap~ (\delta > {\ell}) ~\cap~
\overline{\mathcal{E} (\boldsymbol{J})} }}  &&
\\
 &\leq& \hspace{-4mm} \sum_{r=2}^{ \lfloor
\frac{n-{\ell}}{2} \rfloor } {n \choose {\ell} }{ {n-{\ell}} \choose
r} ~ \bP{ \mathcal{A}_{{\ell},r} ~\cap~ \overline{\mathcal{E} (\boldsymbol{J})}} .
\hspace{.7cm} \label{eq:BasicIdea+UnionBound2}
\end{eqnarray}
 The proof of Proposition \ref{prop:OneLawAfterReductionPart2} will
be completed once we show
\begin{align}
\lim_{n \to \infty} \sum_{r=2}^{ \lfloor
\frac{n-{\ell}}{2} \rfloor } {n \choose {\ell}} {n-{\ell} \choose r}
~ \bP{ \mathcal{A}_{{\ell},r} ~\cap~ \overline{\mathcal{E} (\boldsymbol{J})}} = 0.
\label{eq:OneLawToShow}
\end{align}
The means to do so are provided in the next section.

\section{Bounding Probabilities $\bP{ \mathcal{A}_{{\ell},r} ~\cap~ \overline{\mathcal{E} (\boldsymbol{J})}}$}

First, for $r=2, 3, \ldots , n-{\ell}-1$, observe the equivalence
\begin{align}
 \mathcal{D}_{{\ell},r} &= \bigcap_{j=r+{\ell}+1}^{n}\left [ \left ( \cup_{i \in
\nu_{r,j}} S_i \right ) \cap S_j = \emptyset \right ]
\end{align}
where $\nu_{r,j}$ is defined via
\begin{eqnarray}
\nu_{r,j} := \{ i= {\ell}+1, \ell + 2,\ldots, {\ell}+r :
C_{ij} \} \label{eq:v}
\end{eqnarray}
for each $j=1,2, \ldots, {\ell}$ and
$j=r+{\ell}+1,r+{\ell}+2,\ldots,n$. In words, $\nu_{r,j}$ is the set of
indices in $i={\ell}+1, \ell + 2,\ldots, {\ell}+r$ for which $v_i$ is connected to
the node $v_j$ in the communication graph ${G}(n; p_n)$. Thus,
the event $\left [ \left ( \cup_{i \in
\nu_{r,j}} S_i \right ) \cap S_j = \emptyset \right ]$ ensures that
node $v_j$ is not connected (in $\mathbb{G}_{on}$) to any of the
nodes $\{v_{\ell+1}, \ldots, v_{\ell +r}\}$.
Under the enforced assumptions on the rvs $S_1, S_2, \ldots ,
S_n $, we readily obtain the expression

\begin{eqnarray}\nonumber
\lefteqn{\bP{ \mathcal{D}_{{\ell},r} ~~\Bigg | ~~\begin{array}{r}
  S_i , \ i={\ell}+1, \ldots , {\ell}+r \\ C_{ij} , \ i={\ell}+1,\ldots, {\ell}+r, \\ j={\ell}+r+1,\ldots,n
  \\
\end{array}  }
} &&
\\ \nonumber
&=& \prod_{j=r+{\ell}+1}^n \left ( {{P_n- |\cup_{i \in \nu_{r,j}} S_i
|} \choose {K_n}} \over {{P_n \choose K_n}} \right ). \hspace{2.5cm}
\end{eqnarray}
In a similar manner, we find
\begin{eqnarray}\nonumber
\lefteqn{\bP{ \mathcal{B}_{\ell,r} ~~\Bigg | ~~\begin{array}{l}
  S_i , \ i={\ell}+1, \ldots , {\ell}+r \\ C_{ij} , \ i=1,\ldots, {\ell}, \\ \hspace{.7cm} j={\ell}+1,\ldots, {\ell}+r
  \\
\end{array}  }
} &&
\\ \nonumber
&=& \prod_{j=1}^{\ell} \left( 1- { {{P_n- |\cup_{i \in \nu_{r,j}}
S_i|} \choose {K_n}} \over {{P_n \choose K_n}}} \right ).
\hspace{2.5cm}
\end{eqnarray}

It is clear that the distributional properties of the term $|\cup_{i
\in \nu_{r,j}} S_i|$ will play an important role in efficiently
bounding $\bP{\mathcal{D}_{\ell,r}}$ and $\bP{\mathcal{B}_{\ell,r}}$. Note that
it is always the case that
\begin{align}
|\cup_{i \in \nu_{r,j}} S_i| \geq K_n \1{|\nu_{r,j}|>0}.
\label{eq:U_r_Trivial_Bound}
\end{align}
Also, on the event $\overline{\mathcal{E} (\boldsymbol{J})}$, we have
\begin{align}
|\cup_{i \in \nu_{r,j}} S_i| \geq \left( J_{|\nu_{r,j}|}+1\right)
\cdot \1{|\nu_{r,j}|>1} \label{eq:U_r_Difficult_Bound}
\end{align}
for each $j=r+{\ell}+1, \ldots, n$. Finally, we note the crude bound
\begin{align}
|\cup_{i \in \nu_{r,j}} S_i| \leq |\nu_{r,j}| K_n
\label{eq:U_r_Trivial_Bound2}
\end{align}
for each $j=1,\ldots, {\ell}$.

Conditioning on the rvs $S_{{\ell}+1},
\ldots , S_{r+{\ell}} $ and $\{ C_{ij} , \
i,j={\ell}+1,\ldots, {\ell}+r \}$ (which determine the event $\mathcal{C}_{r}
$), we conclude via
(\ref{eq:U_r_Trivial_Bound})-(\ref{eq:U_r_Trivial_Bound2}) that
\begin{eqnarray}\nonumber
\lefteqn{\bP{ \mathcal{A}_{\ell, r} \cap \overline{\mathcal{E} (\boldsymbol{J})}}
 }  &&
\\ \nonumber
 &=& \bP{ \mathcal{C}_{r} \cap \mathcal{B}_{\ell,r} \cap \mathcal{D}_{\ell,r} \cap
\overline{\mathcal{E} (\boldsymbol{J})} }
 \\ \nonumber
&\leq& \hspace{-3mm} \bE{ \begin{array}{l}
 \1{ \mathcal{C}_{r}  } \times  {\prod_{j=1}^{\ell} \left(1-{
 { {P_n- K_n |\nu_{r,j}| }  \choose {K_n} } \over{{P_n \choose K_n}}}\right)} \times \\
 \times \prod_{j=r+{\ell}+1}^n {
{{P_n- L(\nu_{r,j} )
 } \choose {K_n} }\over{{P_n \choose K_n}}}
\end{array} }, \label{eq:ComputePA_{n,k,r}1}
\end{eqnarray}
where for notational convenience we have set
\begin{eqnarray}\label{eq:DefinitionL}
L(\nu_{r,j}) &=& \max \left\{ K_n \cdot \1{|\nu_{r,j}|>0} , \right.
\\ \nonumber
& & ~ \left. (J_{|\nu_{r,j}|} +1) \cdot \1{|\nu_{r,j}|>1}\right\}.
\end{eqnarray}
It is immediate that the rvs $\{|\nu_{r,j}|\}_{j=r+1+{\ell}}^{n}$ (as well as
$\{|\nu_{r,j}|\}_{j=1}^{{\ell}}$) are independent and identically
distributed. Let $\nu_{r}$ denote a generic random variable
identically distributed with $\nu_{r,j}, \; j=1,\ldots, {\ell},
r+{\ell}+1, \ldots, n$. Then, we have
\begin{align}
|\nu_{r}| =_{\mbox{st}} \mbox{Bin}(r, p_n). \label{eq:v_r_alpha}
\end{align}
where we use the notation $=_{st}$ to indicate
distributional equality.
Then, we define $L(|\nu_r|)$ as follows:
\begin{align}
L(\nu_r)= \max \left\{ K_n \cdot \1{|\nu_r|>0} ,
(J_{|\nu_r|} +1) \cdot \1{|\nu_r|>1}\right\}. \label{olp:deflvr}
\end{align}

 Observe that the event $\mathcal{C}_{r} $ is independent from the set-valued
random variables $\nu_{r,j}$ for each $j=1,\ldots, {\ell}$ and for
each $j=r+\ell+1, \ldots, n$. Also, as noted before
$\{|\nu_{r,j}|\}_{j=r+1+{\ell}}^{n}$ (as well as
$\{|\nu_{r,j}|\}_{j=1}^{{\ell}}$) are independent and identically
distributed. Using these we obtain
\begin{align}\nonumber
  \lefteqn{\bP{ \mathcal{A}_{\ell, r} \cap \overline{\mathcal{E} (\boldsymbol{J})} } } \nonumber
\\  &\leq
 \bP{\mathcal{C}_{r} } \times \bE {
1-{{ {P_n-  K_n|\nu_{r} |}  \choose K_n }\over{{P_n \choose
K_n}}}}^{{\ell}} \times \bE { { {P_n- L(\nu_r)} \choose K_n
}\over{{P_n \choose K_n}}}^{n-r-{\ell}}. \label{olp:alr}
\end{align}

We will give sufficiently tight bounds for each term appearing in
the R.H.S. of (\ref{olp:alr}). First, note from Lemma
\ref{lem:ProbabilityOfC} (Appendix \ref{sec:additional_lemmas_oy})
that
\begin{equation}
\bP{\mathcal{C}_{r}} \leq r^{r-2} p_e^{r-1} , \qquad r=2, 3, \ldots , n.
\label{eq:new_bound_on_C_r_oy}
\end{equation}
 Next, we give an easy bound on the second term appearing in
the R.H.S. of (\ref{olp:alr}).
With
\begin{align}
r   & \leq \frac{P_n-K_n}{2K_n} \label{eq:cond_for_bound_on_B_r}
\end{align}
 it follows that $|\nu_{r}| \leq r \leq \frac{P_n-K_n}{2K_n}$. Then we
 use Fact \ref{fact:lemyagan2} and Fact \ref{fact1xynew}
successively to obtain
\[
1-{{{P_n-  K_n|\nu_{r}|}  \choose K_n }\over{{P_n \choose K_n}}} \leq
1 - {(1-p_s)}^{2 |\nu_{r}|} \leq 2|\nu_{r}| p_s.
\]
Taking the expectation in the above relation and noting that
$\bE{|\nu_{r}|}=r p_n$ via (\ref{eq:v_r_alpha}), we get
\begin{align}
\bE { 1-{{ {P_n-  K_n|\nu_{r}|}  \choose K_n }\over{{P_n \choose
K_n}}}} \leq
 2 r p_s p_n =   2 r {p_e}  \label{eq:BoundOnB_r}
\end{align}
under the condition
(\ref{eq:cond_for_bound_on_B_r}).
Finally, for the last term in the R.H.S. of (\ref{olp:alr}), we
establish in Lemma \ref{lem:bounding_expectation}
(Appendix \ref{sec:additional_lemmas_oy})
that if $\frac{K_n}{P_n} = o(1)$ and $p_e = o(1)$, then
\begin{align}\nonumber
 \lefteqn{\bE {
{ {P_n- L( \nu_{r} )} \choose K_n }\over{{P_n \choose K_n}}}}
\\
 &\leq
\min \left\{ e^{-p_e(1+\varepsilon/2)},  e^{-p_e\lambda r}  +
e^{-K_n\mu} \1{r>r_n}
 \right\}
 \label{eq:crucial_bound_expectation_text_oy}
\end{align}
for all $n$ sufficiently large and for each $r=2, 3, \ldots, n$.

Substituting the bounds (\ref{eq:new_bound_on_C_r_oy}), (\ref{eq:BoundOnB_r})
and (\ref{eq:crucial_bound_expectation_text_oy}) into
(\ref{olp:alr}), and noting that each of the terms
in the RHS of (\ref{olp:alr}) are trivially upper bounded by $1$,
we obtain the key bounds
on the probabilities $\bP{ \mathcal{A}_{{\ell},r} ~\cap~ \overline{\mathcal{E} (\boldsymbol{J})}}$
that are summarized in the following Lemma.
\begin{lem} \label{olp_lem1}
With $\boldsymbol{J} $ defined in (\ref{olp_xjdef}) for some
$\varepsilon$, $\lambda$ and $\mu$ in $(0, \frac{1}{2})$, if
$\frac{K_n}{P_n} = o(1)$ and $p_e = o(1)$, then the following two
properties hold.

(a) For all $n$ sufficiently large and for each $r = 2, 3, \ldots,
\left \lfloor \frac{P_n-K_n}{2K_n} \right \rfloor $, we have
\begin{align}
 \lefteqn{ \bP{ \mathcal{A}_{{\ell},r} ~\cap~ \overline{\mathcal{E} (\boldsymbol{J})} }}\nonumber \\  &\leq
 r^{r-2} \left ( p_e \right)^{r-1} \cdot (2 r {p_e})^{{\ell} }\nonumber \\ &
 \times \left[ \min \left\{e^{-p_e(1+\varepsilon/2)}, e^{-p_e\lambda r} + e^{-K_n\mu} \1{r>r_n} \right\} \right]^{n-r-{\ell}}
 . \nonumber
\end{align}

(b) For all $n$ sufficiently large and for each $r = 2, 3, \ldots,
n$, we have
\begin{align}
 \lefteqn{ \bP{ \mathcal{A}_{{\ell},r} ~\cap~ \overline{\mathcal{E} (\boldsymbol{J})} }}\nonumber \\  &\leq
 \min \left\{ r^{r-2} \left ( p_e \right)^{r-1}, 1 \right\} \nonumber \\ &
 \times \left[ \min \left\{e^{-p_e(1+\varepsilon/2)}, e^{-p_e\lambda r} + e^{-K_n\mu} \1{r>r_n} \right\} \right]^{n-r-{\ell}}
 . \nonumber
\end{align}
\end{lem}

\section{Establishing (\ref{eq:OneLawToShow})}
\label{sec:Last_step_One_law}

We now proceed as follows: Given $\frac{K_n}{P_n} = o(1)$ and
the definition of $r_n$ in (\ref{olp_defrn}), we necessarily have
$\lim_{n \to \infty} r_n=+\infty$, and for an given integer $R \geq
2$, we have
\begin{align}
r_n > R\textrm{ for any }n\geq n^{\star}(R) \label{eq:n_star_defn}
\end{align}
for some finite integer $n^{\star}(R)$. We define $ f_{n,{\ell},r}$
as follows.
\begin{align}
 f_{n,{\ell},r} & =  {n \choose
{\ell}}{n-{\ell} \choose r} \bP{ \mathcal{A}_{{\ell},r}  \cap \overline{\mathcal{E}
(\boldsymbol{J})} }. \nonumber
 \end{align}
 Then, we have
 \begin{align}
\textrm{L.H.S. of (\ref{eq:OneLawToShow})} & = \sum_{r=2}^{\lfloor
\frac{n-{\ell}}{2} \rfloor}  f_{n,{\ell},r}.
\end{align}
 For the time being, pick an \textit{arbitrarily large} integer $R \geq 2$ (to be specified in
Section \ref{sec:Last_Parts_2}), and on the range $n \geq
n^{\star}(R)$ consider the decomposition
\begin{align}
 \sum_{r=2}^{\lfloor \frac{n-{\ell}}{2} \rfloor}  f_{n,{\ell},r} & = \sum_{r=2}^{ R } f_{n,{\ell},r}
 +\sum_{r=R+1}^{ r_n}  f_{n,{\ell},r} +\sum_{r=r_n +1}^{\lfloor \frac{n-{\ell}}{2}
\rfloor} f_{n,{\ell},r} . \nonumber
\end{align}
Let $n$ go to infinity: The desired convergence
(\ref{eq:OneLawToShow}) (for Proposition \ref{prop:OneLawAfterReductionPart2})
will be established if we show
\begin{align}
 \sum_{r=2}^{ R } f_{n,{\ell},r} & = o(1),
\label{eq:StillToShow0}
\\
 \sum_{r=R+1}^{ r_n } f_{n,{\ell},r}  &=
 o(1), \label{eq:StillToShow1}
\end{align}
and
\begin{align}
  \sum_{ r=r_n +1}^{\lfloor
\frac{n-{\ell}}{2} \rfloor} f_{n,{\ell},r} &  =  o(1)  .
\label{eq:StillToShow2}
\end{align}

The next subsections are devoted to proving the validity of
(\ref{eq:StillToShow0}), (\ref{eq:StillToShow1}) and
(\ref{eq:StillToShow2}) by repeated applications of Lemma
\ref{olp_lem1}. Throughout, we also make repeated use of the
standard bounds
\begin{align}
{n \choose r} \leq \left ( \frac{e n}{r} \right )^r
\label{eq:CombinatorialBound1}
\end{align}
valid for all $r,n=1,2, \ldots $ with $r\leq n$. 

\subsection{Establishing (\ref{eq:StillToShow0})}
\label{sec:Last_Parts_1}

Positive scalar $\varepsilon$ in $(0,1)$ is picked arbitrarily as
stated in
Proposition \ref{prop:OneLawAfterReductionPart2}.
Consider $K_n$, $P_n$ and $p_e$ as in the statement of
Proposition \ref{prop:OneLawAfterReductionPart2}.
For any arbitrary integer $R\geq 2$, it is clear that
(\ref{eq:StillToShow0}) will follow upon showing
\begin{align}
\lim_{n \rightarrow \infty} f_{n,{\ell},r} = 0 \quad \textrm{ if } \quad \lim_{n
\to \infty} \beta_{{\ell},n}=+ \infty \label{eq:StillToShow0_a}
\end{align}
for each $r=2,3, \ldots, R$. On that range,
property (a) of Lemma \ref{olp_lem1} is valid since
$r \leq  \lfloor \frac{P_n-K_n}{2K_n} \rfloor $
for all $n$ sufficiently large by virtue of the fact that
$\frac{K_n}{P_n}=o(1)$.

From the easily obtained bounds
${n \choose {\ell}} \leq n^{{\ell} }$ and ${ n-{\ell} \choose r}
\leq n^{ r}$, we now get
\begin{align}
\lefteqn{f_{n,{\ell},r}} \nonumber
\\
&\leq n^{{\ell} } \cdot n^{ r} \cdot r^{r-2} p_e^{r-1}  (2 r
p_e)^{\ell} \cdot e^{-p_e(1+\varepsilon/2) (n-r-{\ell})} \nonumber \\ &
= (2 r )^{\ell} r^{r-2} \cdot n^{{\ell}+r} p_e^{ {\ell}+r -1} \cdot
e^{-p_e n (1+\varepsilon/2) } \cdot e^{ p_e(1+\varepsilon/2) ( r+{\ell})}.
\label{olp_zja}
\end{align}
for each
$r=2,3, \ldots, R$.
Given $p_e  = \frac{\ln  n + {\ell} \ln  \ln  n +
\beta_{{\ell},n}}{n} \sim \frac{\ln n}{n} = o(1)$ (since $\beta_{\ell, n} = o(\ln n)$),
we find
\begin{align}
\lefteqn{\frac{\textrm{R. H. S. of (\ref{olp_zja})}}{(2 r )^{\ell}
r^{r-2}}} \nonumber
\\  & = n^{{\ell}+r} p_e^{ {\ell}+r -1} \cdot e^{-p_e n
(1+\varepsilon/2) } \cdot e^{ p_e(1+\varepsilon/2) ( r+{\ell})} \nonumber
\\  & \sim n^{{\ell}+r} \left( \frac{\ln n}{n} \right)^{ {\ell}+r -1}
\cdot e^{-\left( \ln  n + {\ell} \ln  \ln  n + \beta_{{\ell},n}
\right) (1+\varepsilon/2) } \cdot e^{o(1)} \nonumber \\  &  = n \cdot
\left( \ln n \right)^{ {\ell}+r -1} \cdot \left[ n^{-1}(\ln
n)^{-{\ell}} e^{- \beta_{{\ell},n}} \right]^ {1+\varepsilon/2}
\nonumber
\\ & = n^{-\varepsilon/2} \left( \ln n \right)^{ r  - {\ell}\varepsilon/2
-1} e^{- \beta_{{\ell},n}(1+\varepsilon/2)} \nonumber \\ \nonumber & = o(1)
\end{align}
by virtue of the facts that $r$ is bounded and
$\lim_{n \to \infty} \beta_{{\ell},n}=+ \infty $.
We get (\ref{eq:StillToShow0_a}) and the desired result
(\ref{eq:StillToShow0}) is obtained. \hfill $\blacksquare$

\subsection{Establishing (\ref{eq:StillToShow1})}
\label{sec:Last_Parts_2}

Positive scalars $\lambda, \mu$ are given in the statement of
Proposition \ref{prop:OneLawAfterReductionPart2}. Note that $R$ can
be taken to be arbitrarily large by virtue of the previous section.
From ${n \choose {\ell}} \leq n^{{\ell} }$, ${ n-{\ell} \choose r}
\leq \left( \frac{e (n-{\ell}) }{r}\right)^r$ and property (b) of
Lemma \ref{olp_lem1}, for $n \geq n^{\star}(R)$ (with $n^{\star}(R)$
as specified in (\ref{eq:n_star_defn})) and for each $r=R+1, \ldots,
r_n$, we obtain
\begin{align}\nonumber
 f_{n,{\ell},r}
 &\leq n^{\ell} \cdot \left( \frac{e (n-{\ell}) }{r}\right)^r \cdot
r^{r-2}\left(p_e\right)^{r-1}  e^{-{p_e} r \lambda (n-r-{\ell})}\nonumber \\
 &\leq  n^{{\ell}+r} e^{r} \left(p_e\right)^{r-1}  e^{-{p_e} r
\lambda (n-r-{\ell})}. \label{olpewrwfr2tae}
\end{align}
 Now, observe that on the range $r=R+1, R+2, \ldots, \lfloor
\frac{n-{\ell}}{2}\rfloor$, from $r\leq \frac{n-{\ell}}{2}$, we have
for all $n$ sufficiently large, $ n-r-{\ell} \geq \frac{1}{2}(n-{\ell}) \geq \frac{n}{3}$.
This yields
\begin{align}
e^{-{p_e} r \lambda (n-r-{\ell})} & \leq e^{-{p_e} r \lambda n /3}.
\label{olp2}
\end{align}
Substituting $p_e = \frac{\ln  n + {\ell} \ln \ln  n +
\beta_{{\ell},n}}{n}$ into (\ref{olp2}), we also get
\begin{align}
 e^{-{p_e} r \lambda n /3} &  = e^{-r \lambda (\ln  n + {\ell}
\ln \ln  n + \beta_{{\ell},n} ) / 3} \nonumber \\  & = n^{-r \lambda
/3 } (\ln n)^{-r \lambda {\ell} /3} e^{-r \lambda \beta_{{\ell},n} /
3}.\label{olpewrwfr2}
\end{align}
Applying (\ref{olp2}), (\ref{olpewrwfr2}) and $p_e \leq \frac{2\ln
n}{n}$ to (\ref{olpewrwfr2tae}), we get
\begin{align}
\lefteqn{f_{n,{\ell},r}}\nonumber \\ \nonumber & \leq n^{{\ell}+r}
e^{r}
 \cdot \left(\frac{2\ln n}{n}\right)^{r-1} \cdot n^{-r \lambda /3 } (\ln
n)^{-r \lambda {\ell} /3} e^{-r \lambda \beta_{{\ell},n} / 3}\\
\nonumber & \leq n^{{\ell}+1-r \lambda /3} \cdot (2e\ln n)^r \\  & =
n^{{\ell}+1} \cdot (2en^{-\lambda /3} \ln n)^r.
 \label{olp_an3}
\end{align}
Given $2en^{-\lambda /3} \ln n = o(1)$ and (\ref{olp_an3}), we obtain
\begin{align}
 \sum_{r=R+1}^{  r_n } f_{n,{\ell},r} & \leq
 \sum_{r=R+1}^{ + \infty } n^{{\ell}+1} \cdot (2e n^{-\lambda/3} \ln n)^r
  \nonumber \\   &  = n^{{\ell}+1} \cdot
\frac{(2en^{-\lambda /3} \ln n)^{R+1}}{1-
 2en^{-\lambda /3} \ln n }
 \nonumber \\
  &  \sim n^{{\ell}+1 -\lambda (R+1) /3 }
  (2e \ln n)^{R+1} .\label{olp_an3ar}
\end{align}
 We pick $R \geq \frac{3({\ell}+1)}{\lambda}$ so that ${\ell}+1 -\lambda
(R+1) /3 \leq -\frac{\lambda}{3}$. As a result, we obtain
\begin{align}
 \textrm{R.H.S. of
(\ref{olp_an3ar})} &  =  o(1) \nonumber
\end{align}
 and thus
$\sum_{r=R+1}^{  r_n } f_{n,{\ell},r} =  o(1).$
We now obtain (\ref{eq:StillToShow1}). \hfill $\blacksquare$

\subsection{Establishing (\ref{eq:StillToShow2})}
\label{sec:Last_Parts_3}

Positive scalars $\lambda, \mu$ are given in the statement of
Proposition \ref{prop:OneLawAfterReductionPart2}. We need consider
only the case where $r_n \leq \lfloor \frac{n-{\ell}}{2} \rfloor$
for \textit{infinitely many} $n$, as otherwise
(\ref{eq:StillToShow2}) would hold trivially. From ${n \choose
{\ell}} \leq n^{{\ell} }$, ${ n-{\ell} \choose r} \leq { n \choose
r}$ and property (b) of Lemma \ref{olp_lem1}, we get for $r= r_n +1,
\ldots, \lfloor \frac{n-{\ell}}{2}\rfloor $,
\begin{align}\nonumber
 f_{n,{\ell},r}  &\leq n^{{\ell} } { n \choose r} \left( e^{-p_e r \lambda } +
e^{-K_n\mu}\right) ^{\frac{n-{\ell}}{2}}.
\end{align}

We will establish
(\ref{eq:StillToShow2}) in two steps. First set
$
\hat{r}_n = \left \lceil \frac{3 }{\lambda p_e} \right \rceil.
$
Obviously, the range $r= r_n+1, \ldots, \lfloor \frac{n-{\ell}}{2}
\rfloor $ is intersecting the range $r=\hat{r}_n, \ldots, \lfloor
\frac{n-{\ell}}{2} \rfloor $. We first consider the latter range
below.
 For $r= \hat{r}_n, \ldots,
\lfloor \frac{n-{\ell}}{2} \rfloor $, it follows that $e^{- p_e r
\lambda } \leq e^{-3} $. From Lemma \ref{lem:usefulcons2} (Appendix
\ref{sec:additional_lemmas_oy}), $K_n =
\Omega \left(\sqrt{\ln n}\right)$ holds. Then $e^{-K_n\mu} = o(1) <
\frac{1}{9} - e^{-3}$. Therefore,
\begin{align}\nonumber
 \left( e^{-p_e r \lambda } +
e^{-K_n\mu}\right) ^{\frac{n-{\ell}}{2}}  &\leq \left( \frac{1}{9}
\right) ^{\frac{n-{\ell}}{2}} = 3^{{\ell}-n}.
\end{align}
Then, we get
\begin{equation}
  \sum_{r=\hat{r}_n}^{\lfloor \frac{n-{\ell}}{2}\rfloor}
f_{n,{\ell},r}  \leq 3^{{\ell}-n} n^ {\ell}
\sum_{r=\hat{r}_n}^{\lfloor \frac{n-{\ell}}{2}\rfloor} { n \choose
r} \leq 3^{{\ell}-n} n^ {\ell} \cdot 2^n =o(1)
\label{eq:last_step_3b}
\end{equation}
upon using the binomial formula
$\sum_{r= \hat{r}_n }^{\lfloor \frac{n-{\ell}}{2} \rfloor} {n \choose
r} \leq \sum_{r= 0 }^{n} {n \choose
r} = 2^n$ and the fact that $\ell$ is constant.

If $\hat{r}_n \leq r_n+1$ for all $n$ sufficiently large, then the
desired condition (\ref{eq:StillToShow2}) is automatically satisfied
via (\ref{eq:last_step_3b}). On the other hand, if $ r_n+1 <
\hat{r}_n $, we should still consider the range $r= r_n +1, \ldots,
\hat{r}_n-1$. On that range, we use arguments similar to those
leading to (\ref{olpewrwfr2tae}) and
obtain
\begin{align}\label{eq:towards_end_oy}
 f_{n,{\ell},r}
 &\leq n^{{\ell}+r} e^{r} \left(p_e\right)^{r-1}  \left( e^{-p_e r \lambda } + e^{-K_n\mu}\right)
 ^{n-r-{\ell}}
\end{align}
upon using also property (b) of Lemma \ref{olp_lem1}.

 On the range $r= r_n +1, \ldots,
\hat{r}_n-1$, we have
\begin{align}\nonumber
 r & \geq r_n + 1 = \min \left ( {\left \lfloor \frac{P_n}{K_n} \right \rfloor},
\left \lfloor \frac{n}{2} \right \rfloor \right ) + 1 \geq
\min\left\{ \frac{P_n}{K_n},\frac{n}{2} \right\},
\end{align}
 and thus
%
 \begin{eqnarray}
\frac{e^{-\mu K_n}}{ p_e r \lambda} &\leq& \frac{e^{-\mu K_n}}{ p_e
 \lambda \cdot \min\{ \frac{P_n}{K_n},\frac{n}{2} \}}
\nonumber \\
&\leq& \max \left\{ \frac{K_n e^{-\mu K_n}}{ \sigma \lambda}~,~
\frac{2 e^{-\mu K_n}}{ \lambda} \right\}. \nonumber
\end{eqnarray}
since $P_n \geq \sigma n$ and $p_e n \geq 1$ for all $n$
sufficiently large.

Given $K_n = \Omega \left(\sqrt{\ln n}\right)$, \fo
\[
\lim_{n \to \infty} K_n e^{-\mu K_n} = 0 \quad \mbox{and} \quad
\lim_{n \to \infty} e^{-\mu K_n} = 0,
\]
whence we get
\begin{align}
 \lim_{n \to \infty} \frac{e^{-\mu K_n }}{ p_e r \lambda}&  = 0.
\nonumber
 \end{align}
Then for any given $0<\eta < 1$, there exists a finite integer
$n^\star(\eta)$ such that for all $n \geq n^\star(\eta)$, we have
\begin{align}
e^{-\mu K_n} &  \leq e^{-3}\eta  \cdot p_e r \lambda \leq e^{-3}
\cdot (e^{\eta p_e r \lambda} - 1) . \label{olp_x23}
\end{align}
From $r \leq \hat{r}_n-1 \leq \frac{3 }{\lambda p_e}$, \fo $p_e r
\lambda \leq 3$ and
\begin{align}
 e^{ - p_e r \lambda} & \geq e^{ -3}. \label{olp_x23a}
 \end{align}
Given (\ref{olp_x23}) and  (\ref{olp_x23a}), we obtain for all $n \geq
n^\star(\eta)$,
\begin{align}
e^{-\mu K_n} &  \leq e^{ - p_e r \lambda} \cdot (e^{\eta p_e r
\lambda} - 1) = e^{ - p_e r \lambda (1 - \eta)} - e^{ - p_e r
\lambda} \nonumber
\end{align}%
and thus
\begin{equation}
e^{- {p_e} r \lambda}  + e^{-\mu K_n} \leq e^{- {p_e} r \lambda
(1-\eta)}.
\label{eq:towards_end2_oy}
\end{equation}

Recalling (\ref{eq:DeviationCondition})
and the fact that   $n-\ell -r  \geq n/3$, we get
\begin{align}\label{eq:towards_end3_oy}
\lefteqn{e^{-{p_e} r \lambda(1-\eta) (n-r-{\ell}) }}
\\ \nonumber & \leq n^{-r
\lambda (1-\eta) /3 } (\ln n)^{-r \lambda {\ell} (1-\eta) /3}
e^{-r \lambda \beta_{{\ell},n} (1-\eta) / 3}.
\end{align}
Using (\ref{eq:towards_end2_oy}) and (\ref{eq:towards_end3_oy}) in (\ref{eq:towards_end_oy}),
and noting $p_e \leq 2 \frac{\ln n}{n}$,
we get
\begin{align}
  f_{n,{\ell},r} & \leq n^{{\ell}+r} e^{r}
\left(\frac{2\ln n}{n}\right)^{r-1}  \nonumber \\ & \quad  \times
n^{-r \lambda (1-\eta) /3 } (\ln n)^{-r \lambda {\ell} (1-\eta)
/3} e^{-r
\lambda \beta_{{\ell},n} (1-\eta) / 3} \nonumber \\
 & \leq n^{{\ell}+1-r \lambda (1-\eta) /3} \cdot (2e\ln n)^r
 \nonumber \\
 & = n^{{\ell}+1}
 \cdot (2e n^{-\lambda (1-\eta) /3} \ln n)^r.  \label{olp_an3arts}
\end{align}
Given $\lim_{n \to \infty} r_n = + \infty$, then
for any arbitrarily large integer $\hat{R}$,
we have $r_n \geq \hat{R}$
for all $n$ sufficiently large.
 From $ 2e n^{-\lambda (1-\eta)
/3} \ln n = o(1)$ and (\ref{olp_an3arts}), we have
\begin{align}
 \sum_{ r_n+1}^{\hat{r}_n-1} f_{n,{\ell},r}  & \leq
 \sum_{ \hat{R}+1}^{\infty}  n^{{\ell}+1}
 \cdot (2e n^{-\lambda (1-\eta) /3} \ln n)^r \nonumber \\   & \sim   n^{{\ell}+1} \cdot \frac{(2en^{-\lambda (1-\eta) /3} \ln n)^{\hat{R}+1}}{1-
 2en^{-\lambda (1-\eta) /3} \ln n } \nonumber \\
  &  \sim n^{{\ell}+1 -\lambda (1-\eta)(\hat{R}+1) /3 }
  (2e \ln n)^{\hat{R}+1}. \label{olp_an3arrst}
\end{align}
Since $\hat{R}$ was arbitrary, we pick $\hat{R} \geq \frac{3({\ell}+1)}{\lambda (1-\eta)}$. Then
 \begin{align}
{\ell}+1 -\lambda  (1-\eta)(\hat{R}+1) /3 \leq - \lambda
(1-\eta)/3. \nonumber
\end{align}
As a result, we have
$ \textrm{R.H.S. of
(\ref{olp_an3arrst})}  =  o(1)$, whence
\begin{align}
 \sum_{ r_n+1}^{\hat{r}_n-1} f_{n,{\ell},r} & =  o(1). \nonumber
\end{align}
%
%
 The desired conclusion
(\ref{eq:StillToShow2}) is now established. \hfill $\blacksquare$

Having established (\ref{eq:StillToShow0}), (\ref{eq:StillToShow1}) and
(\ref{eq:StillToShow2}), we now get (\ref{eq:OneLawToShow}) and this
completes the proof of Proposition \ref{prop:OneLawAfterReductionPart2}.
\myendpf

\section{Conclusion}
\label{sec:Conclusion}

We investigate random key graph with unreliable links
which amounts to the intersection of random key graphs with Erd\H{o}s-R\'enyi graphs. We derive
zero-one laws for  $k$-connectivity and minimum node degree being at lest $k$.
These zero-one laws are shown to improve the existing results on
$1$-connectivity of random key graphs with unreliable links
 as well as $k$-connectivity of random key graphs.

An extension of our work would be to consider a different unreliability model
than the independent on/off model used here. One possible candidate is
the so-called {\em disk model} \cite{penrose} where two nodes have
to be within a certain distance to each other to have a link in between; this induces
a {\em random geometric graph}.
Intersection of random key graphs with random geometric graphs has already received
some interest \cite{RybarczykGeom,KrishnanGaneshManjunath}, but the model is proven to be difficult to analyze with
 results obtained thus far for its connectivity \cite{ZhaoAllerton,ISIT_RKGRGG,Krzywdzi}, not for $k$-connectivity.

\section*{Acknowledgements}
This research was supported in part by CyLab at Carnegie Mellon University under grant
DAAD19-02-1-0389 and by MURI grant W 911 NF 0710287 from the US Army
Research Office. The views and conclusions contained in this document are those of the
authors and should not be interpreted as representing the official policies, either
expressed or implied, of any sponsoring institution, the U.S. government, or any other
entity.


 \begin{appendices}
\section{Additional Facts and Lemmas} \label{sec:addi:f:l}

\subsection{Facts} \label{apdfact}
We introduce additional facts below. Proofs
 of Facts \ref{fact1xynew} and \ref{1_x_y} are fairly standard and omitted here; interested reader is referred to
the full version \cite{ZhaoYaganGligorArxiv} for details. All other
 facts are established in Appendix
 \ref{appdexb}.



{\begin{fact} \label{fact1xynew} For $0\leq x<1$, the
following properties hold.\\
(a) If $0 < y < 1$, then
$ (1-x)^y   \leq 1 - xy.
$\\
(b) If $y = 0, 1, 2, \ldots$, then
\begin{align}
1 - xy & \leq (1-x)^y \leq 1 - xy + \frac{1}{2} x^2 y ^2. \nonumber
\end{align}
\end{fact}}
Fact \ref{fact1xynew} is used in proving the one-law (\ref{eq:OneLaw+kConnectivity_olp})
of Theorem \ref{thm:unreliableq1} as well as in proving Fact \ref{PncKn}, Fact \ref{fact:lemyagan2}, Lemma \ref{lem:secprob},
and Lemma \ref{lem:bounding_expectation}.

{\begin{fact}\label{1_x_y} Let $x$ and $y$ be both positive
functions of $n$. If $ x = o(1)$, then for any given constant
$\varepsilon > 0$, there
exists $N \in \mathbb{N}$ such that for any $n > N $, the following properties hold.\\
\noindent (a)
\begin{align}
e^{-xy-(\frac{1}{2}+\varepsilon)x^2 y}  \leq (1-x)^y  \leq
e^{-xy-\frac{1}{2}x^2 y}. \label{factsimleq}
\end{align}
(b) If $ x^2 y = o(1)$ further holds, then
\begin{align}
 (1-x)^y  \sim e^{-xy}. \label{factsim}
\end{align}
\end{fact}}
\noindent Fact \ref{1_x_y} is used in the proofs of Lemma \ref{lsn1}
and Lemma \ref{lem:event_f_result}.

{\begin{fact}\label{PncKn} Let integers $x$ and $y$ be both positive
functions of $n$, where $y \geq 2x$. For $z = 0, 1, \ldots, x$, we have
\begin{align}
\frac{\binom{y- z}{x} } {\binom{y}{x}} & \geq 1 - \frac{ z x }{y - z
}, \label{fone}
\end{align}
and
\begin{align}
 \frac{\binom{y- z}{x} } {\binom{y}{x}}
 & = 1 - \frac{xz}{y}
 \pm O\bigg(\frac{x^4}{y^2}\bigg). \label{fone2}
\end{align}
\end{fact}
}
\noindent Fact \ref{PncKn} is used in the proof of Lemma \ref{jzremfb}.

\begin{fact}\label{fact:lemyagan2}
{
Let $a, x$ and $y$ be positive integers satisfying $y \geq (2a+1)x$.
Then
\begin{align}\label{eq:P_aK_leq_q_a_oy}
\frac{{y- a x  \choose x}}{{y \choose x}} \geq { \left
[\frac{\binom{y- x}{x} } {\binom{y}{x}}\right ]} ^ {2a}
\end{align}
 }
\end{fact}
\noindent Fact \ref{fact:lemyagan2} is used in the proof of the one-law (\ref{eq:OneLaw+kConnectivity_olp})
of Theorem \ref{thm:unreliableq1}.

%
%
%

\subsection{Lemmas}
\label{sec:additional_lemmas_oy}

 We introduce additional lemmas below. The proofs
 of all the following lemmas are deferred to Appendix
 \ref{secprflem}.

 \begin{lem}\label{lem:usefulcons2}
Let $\ell$ be a non-negative constant integer. If $P_n = \Omega(n)$
and
(\ref{eq:DeviationCondition}) holds with $\beta_{\ell,n}
> 0$, then $K_n = \Omega \left(\sqrt{\ln n}\right)$.
\end{lem}
\noindent Lemma \ref{lem:usefulcons2} is used in proving the one-law (\ref{eq:OneLaw+kConnectivity_olp})
of Theorem \ref{thm:unreliableq1}.

\begin{lem}\label{jzremfb}
In $\mathbb{G}_{{on}}$, given $P_n \geq 2 K_n$, then the following properties
hold.

(a) $p_{s} =  \frac{K_n^2}{P_n} \pm O\left(
 \frac{K_n^4}{P_n^2} \right)$.

(b) (\cite[Lemma 7.4.3, pp. 118]{YaganThesis}) $p_{s} \leq
\frac{K_n^2}{P_n - K_n} $.


(c) $p_s = o(1)$ if and only if $\frac{K_n^2}{P_n} = o(1)$.

(d) If $p_s = o(1)$ or $\frac{K_n^2}{P_n} = o(1)$, then
$\frac{K_n^2}{P_n} =  p_{s} \pm O\left( p_{s} ^2 \right)$.
 \end{lem}
\noindent  Lemma \ref{jzremfb} is used in the proof of the zero-law (\ref{mnd_zero}) of
Theorem \ref{thm:unreliableq1},
as well as in the
proofs of Lemma \ref{lem:usefulcons2} and
Lemma \ref{lem:secprob}.

{
\begin{lem}\label{lem:secprob}
Consider $K_n, P_n$ with $K_n \leq P_n$. The following
properties hold for any three distinct nodes
 $v_{x}, v_{y} $ and $ v_j$.

 (a) We have
\begin{align}
 \mathbb{P}\left[\left(K_{{x} j} \cap K_{{y}
j}\right) \boldsymbol{\mid}\overline{K_{{x}{y}}} \right] &
\leq p_s ^ 2 .\label{eqn:lem:secprob1}
\end{align}

(b) If $p_s = o(1)$,
then for any $u = 0, 1, 2, \ldots, K_n$, we have
\begin{align}
&\mathbb{P}\left[\left( {K_{{x} j}} \cap
 {K_{{y} j}}\right) \boldsymbol{\mid}
\left(|S_{xy}|=u\right)\right]  =  \frac{u }{K_n} p_s  \pm
O\left(p_s^2 \right), \label{eqn:lem:secprob2xb}
 \\ & \mathbb{P}[ E_{ {{x} j}\cup  {{{y} j}}}
\boldsymbol{\mid}(|S_{xy}|=u)]  = 2p_e  - \frac{
{p_n\iffalse_{on}\fi}u}{ K_n} \cdot {p_e } \pm O({p_e }^2)
.\label{zjhaloxa}
\end{align}
\end{lem}}
\noindent Lemma \ref{lem:secprob} is used in the proof of
the zero-law (\ref{mnd_zero}) of Theorem \ref{thm:unreliableq1} as well as in the proof
of Lemma \ref{lem:event_f_result}.

\begin{lem} \label{lemb}
If $P_n \geq 2 K_n$, then we have
\begin{align}
 \mathbb{P}[|S_{xy}|=u ]
&\leq \frac{1}{u!} \bigg(\frac{K_n^2}{P_n-K_n}\bigg)^u . \nonumber
\end{align}
\end{lem}
\noindent  Lemma \ref{lemb} helps in proving the zero-law (\ref{mnd_zero}) of Theorem \ref{thm:unreliableq1}.

\begin{lem}[\textrm{\cite[Lemma 10.2]{yagan_onoff}} via the argument of \textrm{\cite[Lemma 7.4.5, pp. 124]{YaganThesis}}] {
For each $r=2, \ldots , n$, we have
\begin{equation}
\bP{ \mathcal{C}_{r} } \leq r^{r-2}  \left ( p_e \right)^{r-1}.
\label{eq:ProbabilityOfC}
\end{equation}
} \label{lem:ProbabilityOfC}
\end{lem}
\vspace{-4mm}
\noindent Lemma \ref{lem:ProbabilityOfC} is used in proving the one-law (\ref{eq:OneLaw+kConnectivity_olp})
of Theorem \ref{thm:unreliableq1}.

\begin{lem} \label{lem:{eq:crucial_bound_expectation}}

With $\boldsymbol{J} $ defined in (\ref{olp_xjdef}) for some
$\epsilon$, $\lambda$ and $\mu$ in $(0, \frac{1}{2})$, if
$\frac{K_n}{P_n} = o(1)$ and $p_e = o(1)$, then we have
\begin{align}\nonumber
 \lefteqn{\bE {
{ {P_n- L( \nu_{r} )} \choose K_n }\over{{P_n \choose K_n}}}}
\\
 &\leq
\min \left\{ e^{-p_e(1+\epsilon/2)},  e^{-p_e\lambda r}  +
e^{-K_n\mu} \1{r>r_n}
 \right\}
 \label{eq:crucial_bound_expectation}
\end{align}
for all $n$ sufficiently large and for each $r=2, 3,
\ldots, n$.
 \label{lem:bounding_expectation}
\end{lem}
\label{sec:addi:f:l:end}
\noindent Lemma \ref{lem:{eq:crucial_bound_expectation}} helps in proving the one-law (\ref{eq:OneLaw+kConnectivity_olp})
of Theorem \ref{thm:unreliableq1}.

\section{Proofs of Facts}\label{appdexb}

{\subsection{Proof of Fact
\ref{lem:graph_degreegeneral} (Section \ref{sec:Method_of_Moments_oy})}
\label{prf:lem:graph_degreegeneral}

\subsubsection{Proof of property (a)}

Clearly,  $[\delta=\ell]$ implies  $\left[X_{\ell} \geq 1
\right]$, whence $\mathbb{P}[\delta=\ell]  \leq\mathbb{P}\left[X_{\ell} \geq 1
\right]$.
 Since $X_{\ell}$ is a non-negative integer, we have
 $\mathbb{E}[X_{\ell}]  \geq
\mathbb{P}\left[X_{\ell} \geq 1 \right]$, leading  to
$\mathbb{P}[\delta=\ell] \leq \mathbb{E}[X_{\ell}]$. Then for
$\ell=0,1,\ldots, {k-1}$, given condition $\mathbb{E}[X_{\ell}] =
o(1)$, we obtain $\mathbb{P}[\delta=\ell] = o(1)$.

\subsubsection{Proof of property (b)}

For constant $k$, given $\mathbb{P}[\delta=\ell] = o(1)$ for
$\ell=0,1,\ldots, {k-1}$, we obtain
\begin{align}
 \mathbb{P}[\delta\geq k]   & = 1-\sum_{\ell=0}^{k-1}
\mathbb{P}[\delta=\ell] \to 1,\textrm{~as~}n \to +\infty. \nonumber
\end{align}

\subsubsection{Proof of property (c)}
Fix $\ell = 0,1, \ldots, k-1$.
From the method of second moment \cite[Remark 3.1, p. 54]{janson2011random},
we have
\begin{align}
\mathbb{P}[X_{\ell} = 0] \leq 1 -
\frac{\left\{\mathbb{E}\left[X_{\ell}\right]\right\}^2}{\mathbb{E}\big[{\big(X_{\ell}\big)}^2\big] }.
\end{align}
Then, from $\mathbb{E}[X_{\ell}] \neq 0$, and ${\mathbb{E}\big[{\big(X_{\ell}\big)}^2\big] } \sim {\left\{\mathbb{E}\left[X_{\ell}\right]\right\}^2}$, we
get
\begin{align}
\mathbb{P}[X_{\ell} = 0] = o(1). \nonumber
\end{align}
Therefore, we get $\lim_{n \to \infty} \mathbb{P}[\delta > \ell] = 0$.
The desired result $\lim_{n \to \infty} \mathbb{P}[\delta \geq k] = 0$ also
follows since
$\ell \leq k-1$. }

{\subsection{Proof of Fact \ref{PncKn}}

From $\binom{y - z}{x} =
 \frac{(y - z)!}{x! (y - z - x)!}$ and $\binom{y }{x} =
 \frac{(y )!}{x! (y - x)!}$, we get
 \begin{align}
 \frac{\binom{y - z}{x}}{\binom{y}{x}} &   = \frac{(y - z)!}{y !}
\cdot \frac{(y - x)!}{(y - z - x)!} = \prod_{t=0}^{z-1} \frac{y - x
- t}{y - t}. \nonumber
\end{align}
We define $g(t) = \frac{y - x - t}{y - t} = 1 - \frac{ x }{y - t}$,
where $t=0,1,2, \ldots, z $. Clearly, $g(t)$ decreases as $t$ increases for $t=0,1,2, \ldots,
z $, so $g(z ) \leq g(t) \leq g(0)$. As a result, we have
\begin{align}
 \left( 1 - \frac{ x }{y - z } \right)^z & \leq \frac{\binom{y - z}{x}}{\binom{y}{x}} \leq
 \left( 1 - \frac{ x }{y }  \right)^z.
 \label{xa2}
\end{align}
Given the above
expressions, we use Fact
\ref{fact1xynew} and obtain
\begin{align}
 \left( 1 - \frac{ x }{y - z } \right)^z  &  \geq 1 - \frac{ z x }{y - z } \label{xbxt}
\\  \left( 1 - \frac{ x }{y }  \right)^z &  \leq 1 -  \frac{z x }{y } + \frac{1}{2}\left( \frac{z
x }{y } \right)^2. \label{xa}
\end{align}
From (\ref{xa2}) and (\ref{xbxt}), we get (\ref{fone}).

Using $ 0 \leq z \leq x$ in the R.H.S. of (\ref{xa}),
we also have
\begin{align}
 \left( 1 - \frac{ x }{y }  \right)^z &  \leq 1 -  \frac{z x }{y } +
 O\left(\frac{x^4}{y^2}\right) .\label{xtr}
\end{align}
To evaluate R.H.S. of (\ref{xbxt}), we have
\begin{align}
 \textrm{R.H.S. of (\ref{xbxt})} - \left(1 -  \frac{z x }{y
}\right) & = - \frac{  z^2 x }{ y \left(y - z \right) }. \label{non1}
\end{align}
Given $y > 2x$ and $ 0 \leq z \leq x$, it follows that $z \leq
\frac{y}{2}$ and thus $y - z \geq y / 2$. Note that $x \geq 1$. Then,
we have
\begin{align}
 \frac{  z^2 x }{ y \left(y - z \right) } & \leq \frac{x ^3 }
 {y ^2 / 2} = \frac{2}{x} \cdot \frac{x^4}{y^2} =
   O\left(\frac{x^4}{y^2}\right). \label{non2}
\end{align}
Applying (\ref{non1}) and (\ref{non2}) into (\ref{xbxt}),  we get
\begin{align}
 \left( 1 - \frac{ x }{y - z } \right)^z &  \geq  1 -  \frac{z x}{y} -
O\left(\frac{x^4}{y^2}\right). \label{xb}
\end{align}
 Using (\ref{xtr}) and (\ref{xb}) in (\ref{xa2}), we obtain (\ref{fone2}).}

\subsection{Proof of Fact
\ref{fact:lemyagan2}}

The proof is similar to that of Lemma 5.1 in Ya\u{g}an
\cite{yagan_onoff}. First, given positive integer $a $, it holds
that
\begin{align}
\frac{{y-  a x  \choose x}}{{y \choose x}} & =
 \frac{\prod_{{\ell}=0}^{x-1}\left (y - a x -{\ell}\right )}{\prod_{{\ell}=0}^{x-1}\left (y-{\ell}\right )} =
\prod_{{\ell}=0}^{x-1} \left ( 1 - \frac{ a x }{y-{\ell}} \right ).
\label{eq:TowardsNewBounds}
\end{align}
Letting $a=1$ in (\ref{eq:TowardsNewBounds}), we obtain
\begin{align}
 \frac{{y- x \choose x}}{{y \choose x}} & = \prod_{{\ell}=0}^{x-1}
\left ( 1 - \frac{ x }{y-{\ell}} \right ). \label{eq:prelim_2}
\end{align}
From property (b) of Fact \ref{fact1xynew}, it follows that
\begin{align}
 \left ( 1 - \frac{ x }{y-{\ell}}
\right ) ^ {2a} & \leq 1 - \frac{2a x}{y-{\ell}} + \frac{1}{2} \left
(\frac{2a x}{y-{\ell}}\right )^2  \leq 1- \frac{a x}{y-{\ell}},
\label{eq:prelim_2factlem}
\end{align}
where, in the last step we used the fact that $a \leq \frac{y-x}{2x}$
since $y \geq (2 a +1)x$ by assumption.

From (\ref{eq:TowardsNewBounds}), (\ref{eq:prelim_2}) and
(\ref{eq:prelim_2factlem}), we get (\ref{eq:P_aK_leq_q_a_oy}).

%
%
%
%

\section{Proofs of Lemmas} \label{secprflem}

\subsection{Proof of Lemma \ref{lsn1} (Section \ref{sec:proof_of_zero_law_oy})} \label{prf:lsn1}

The events
$E_{1i}, E_{2i}, \ldots, E_{i-1,i}, E_{i+1,i} \ldots, E_{ni}$ are
mutually independent for any node $v_i$.
Thus, for each $i=1,2,\ldots,n$, the degree of node $v_i$ follows a Binomial
distribution $\textrm{Bin}(n-1,{p_e})$; i.e.,
\begin{align}
\mathbb{P}\left[D_{i,\ell}\right] & =
\binom{n-1}{\ell}{{p_e}}^{\ell} (1-{{p_e}})^{n-\ell-1}.
\label{eqn:eivis1}
\end{align}
Given $  {p_e} =  o\left(\frac{1}{\sqrt{ n}}\right)$ and constant
$\ell$, it follows that ${p_e} = o(1)$ and ${p_e}^2 (n-\ell-1) =
o(1)$. Then from property (b) of Fact \ref{1_x_y}, $(1-{{p_e}
})^{n-\ell-1} \sim e^{-{{p_e} }(n-\ell-1)}$ holds. Then given ${p_e}
= o(1)$ and constant $\ell$, we further get
$(1-{{p_e} })^{n-\ell-1}  \sim e^{-{p_e}n}$.
Using this and $\binom{n-1}{\ell} \sim (\ell!)^{-1}
n^{\ell}$ in (\ref{eqn:eivis1}), we obtain
\begin{align}
\mathbb{P}\left[D_{i,\ell}\right] & \sim
 \left( \ell! \right)^{-1}  \left({p_e} n\right)^{\ell}e^{- {p_e}n} .\nonumber
\end{align}

\subsection{Proof of Lemma \ref{lem:event_f_result} (Section \ref{proof_proposition1.2})} \label{proof:lem:event_f_result}

In graph $\mathbb{G}_{on}$, besides $v_{x} $ and $ v_{y}$, there are $(n-2)$
nodes, denoted by $v_{j_1},v_{j_2}, \ldots, v_{j_{n-2}}$ below. The
$(n-2)$ nodes are split into the four sets $ N_{x y} $, $N_{x
\overline{y}}$, $ N_{\overline{x}y } $ and $ N_{\overline{x}
\hspace{1.5pt} \overline{y}}$ as defined in Section \ref{sec:othernotation}.
According to the definition (\ref{def_event_f}), under event $\mathcal {F}$ we have
$|N_{x y}| =m_1 $, $|N_{x \overline{y}}| = m_2$,
 $ |N_{\overline{x}y }|=m_3 $, so that $
|N_{\overline{x} \hspace{1.5pt} \overline{y}}|= (n - m_1 - m_2 - m_3 - 2)$. Therefore,
given non-negative constant integers $m_1, m_2$ and $m_3$, the
constraints $0 \leq |N_{x y}| , |N_{x \overline{y}}|,
|N_{\overline{x}y }|
  , |N_{\overline{x} \hspace{1.5pt}
\overline{y}} | \leq n-2$ are satisfied. In this setting, it is clear that the number
of possible instances for realizing the event $\mathcal {F}$ is given by
 \begin{align}
\binom{n-2}{m_1} \cdot \binom{n-m_1-2}{m_2} \cdot \binom{n - m_1 -
m_2   - 2}{m_3}. \label{eq:number_of_instances_oy}
\end{align}
The event $\mathcal{J}$ defined below is an instance of $\mathcal
{F}$.
\begin{align}
 \mathcal {J}  & : =\Big(N_{x y} = \left\{v_{j_1},v_{j_2}, \ldots,
v_{j_{m_1}}\right\}\Big) \nonumber
\\ & \quad \textstyle\bigcap \Big(  N_{x \overline{y}}
 =  \left\{v_{j_{m_1+1}},v_{j_{m_1+2}}, \ldots, v_{j_{m_1 + m_2}}\right\} \Big)\nonumber
\\   &  \quad \textstyle\bigcap\Big( N_{\overline{x}y }  =
\left\{v_{j_{m_1 + m_2+1}},v_{j_{m_1 + m_2+2}}, \ldots, v_{j_{m_1 +
m_2+m_3}}\right\} \Big) \nonumber
\\   &  \quad \textstyle\bigcap \Big(N_{\overline{x} \hspace{1.5pt} \overline{y}} =
\left\{v_{j_{m_1 + m_2+m_3+1}},v_{j_{m_1 + m_2+m_3+2}}, \ldots,
v_{j_{n-2}}\right\} \Big) .\label{defe}
\end{align}
 It is clear that all instances of $\mathcal {F}$ happen with the same
probability. 
 Let node $v_j$ be any given node other than $v_{x} $ and $ v_{y}$ in
graph $\mathbb{G}_{on}$. Then
\begin{align}
 E_{{x} j \cap {y} j}&\Leftrightarrow \left( v_j \in N_{x y} \right);~ E_{{x} j\cap \overline{ {{y} j}}}
   \Leftrightarrow \left( v_j \in N_{x \overline{y}}\right);
   \label{equiv1}
\\  E_{\overline{{x} j}\cap {y} j}& \Leftrightarrow \left( v_j \in N_{\overline{x}y }
\right);\textrm{ and }
     E_{\overline{{x} j}\cap \overline{{{y} j}}}
  \Leftrightarrow
  \left( v_j \in N_{\overline{x} \hspace{1.5pt} \overline{y}} \right). \label{equiv2}
 \end{align}
 Applying the above equivalences (\ref{equiv1}) and (\ref{equiv2})
 to the definition of $\mathcal {J}$ in (\ref{defe}), we
obtain
 \begin{align}
\mathcal {J} &  =  \left(\bigcap_{i=1}^{m_1}  E_{{x} j_i \cap {y}
j_i}\right) \hspace{-2pt} {\textstyle\bigcap} \hspace{-2pt}
\left(\bigcap_{i=m_1+1}^{m_1+m_2} E_{{x} j_i\cap \overline{{y} j_i}}
\right) \nonumber \\  & \hspace{-2pt} \quad {\textstyle\bigcap}
\hspace{-2pt} \left(\bigcap_{i=m_1+m_2+1}^{m_1+m_2+m_3}
E_{\overline{{x} j_i}\cap {y} j_i} \right) \hspace{-2pt}
{\textstyle\bigcap} \hspace{-2pt}
\left(\bigcap_{i=m_1+m_2+m_3+1}^{n-2} E_{\overline{{x} j_i}\cap
\overline{{y} j_i}} \right) . \label{expr_math_e}
\end{align}

 Given $E_{xj}  = C_{xj} \cap K_{xj}~\textrm{and}~E_{yj} =
C_{yj} \cap K_{yj}, \label{exj_eyj}$
we have
\begin{align}
 E_{{x} j  \cap {y} j } & =\left( C_{xj} \cap
C_{yj}\right)  \cap  \left( K_{xj} \cap
K_{yj}\right).  \label{exj_eyj_exjyj}
\end{align}

For any node $v_j$ distinct from $v_{x} $ and $ v_{y}$, we have the
following observations: (a) events $C_{xj},
C_{yj}, C_{xj} \cap C_{yj},
K_{xj}, K_{yj}$ and thus $E_{xj},
E_{yj}$ given by (\ref{exj_eyj}) do not depend on any nodes
other than $v_x, v_y$ and $v_j$; (b) given $(|S_{xy}| = u)$, event
$K_{xj} \cap K_{yj}$ does not depend on any nodes
other than $v_x, v_y$ and $v_j$; (c) from (\ref{exj_eyj_exjyj}), and
observations (a) and (b) above, event $E_{{x} j \cap {y} j}$ does
not depend on any nodes other than $v_x, v_y$ and $v_j$
given that $(|S_{xy}| = u)$; (d) since
the relative complement of event $E_{{x} j \cap {y} j}$ with respect
to event $E_{{x} j}$ is event $E_{{x} j \cap \overline{{y} j }}$,
given observations (a) and (c) above, event $E_{{x} j \cap
\overline{{y} j }}$ and then similarly, events $E_{\overline{{x} j}
\cap {y} j } $ and $E_{\overline{{x} j} \cap \overline{{y} j }}$ do
not depend on any nodes other than $v_x, v_y$ and $v_j$.

From observations (c) and (d) above, we conclude that
\begin{align}
& E_{{x} j_1 \cap {y} j_1},  \ldots, E_{{x} j_{m_1} \cap {y}
j_{m_1}}, \nonumber \\ &  E_{{x} j_{m_1+1}\cap \overline{{y} j_{
m_1+1}}},
\ldots, E_{{x} j_{m_1+m_2}\cap \overline{{y} j_{m_1+m_2}}},  \nonumber \\
& E_{\overline{{x} j_{m_1+m_2+1}}\cap {y} j_{ m_1+m_2+1}},  \ldots,
E_{\overline{{x} j_{m_1+m_2+m_3}}\cap {y} j_{m_1+m_2+m_3}} , \nonumber \\
& E_{\overline{{x} j_{m_1+m_2+m_3+1}}\cap \overline{{y} j_{
m_1+m_2+m_3+1}}}, \ldots, E_{\overline{{x} j_{n-2}}\cap
\overline{{y} j_{n-2}}} \nonumber
\end{align}
 are mutually independent given that
$(|S_{xy}| = u)$.

Then from (\ref{eq:number_of_instances_oy}) and
(\ref{expr_math_e}),
we finally get
 \begin{align}
\lefteqn{ \mathbb{P}\left[\mathcal {F} \boldsymbol{\mid}
|S_{xy}| =u \right] }\nonumber\\  &=
\binom{n-2}{m_1} \binom{n-m_1-2}{m_2} \binom{n - m_1 - m_2   - 2}{
m_3 }
\nonumber\\
&\quad \times \{\mathbb{P}[E_{{x} j \cap {y} j} \boldsymbol{\mid}
(|S_{xy}| = u)]\}^{m_1} \nonumber\\
&\quad \times  \{\mathbb{P}[E_{{x} j \cap {\overline{y} j}}
\boldsymbol{\mid} (|S_{xy}| = u)]\}^{m_2}\nonumber\\  & \quad \times
\{\mathbb{P}[E_{\overline{{x} j}\cap {y} j}\boldsymbol{\mid}
(|S_{xy}| = u)]\}^{m_3} \nonumber\\
&\quad \times  \{\mathbb{P}[E_{\overline{{x} j}\cap \overline{{{y}
j}}} \boldsymbol{\mid} (|S_{xy}| = u)]\}^{n - m_1 - m_2 - m_3  - 2}.
\label{eqn:unre:f1h}
\end{align}
upon using exchangeability.

For any constants $m_1, m_2$ and $m_3$, we have
 \begin{align}
\lefteqn{\binom{n-2}{m_1} \binom{n-m_1-2}{m_2} \binom{n - m_1 - m_2
- 2}{ m_3 }} \nonumber\\& \sim   \frac{n^{m_1}}{{m_1}!} \cdot
\frac{n^{m_2}}{{m_2}!} \cdot \frac{n^{m_3}}{{m_3}!} =
\frac{n^{m_1+m_2+m_3}}{m_1! m_2! m_3! } . \label{eqn:unre:f1hfrac}
\end{align}
Now, we evaluate the probability
\begin{align}
\{\mathbb{P}[ E_{\overline{{x} j}\cap \overline{{{y} j}}}
\boldsymbol{\mid}(|S_{xy}|=u)]\}^{n - m_1 - m_2 - m_3  - 2}.
\label{pr_e_xj_yj}
\end{align}
It is clear that
\begin{align}
 (\ref{pr_e_xj_yj})  & = \left(1- \mathbb{P} [ E_{ {{x} j}\cup {{{y} j}}}
\boldsymbol{\mid}(|S_{xy}|=u)] \right) ^{n - m_1 - m_2 - m_3  - 2}.
\label{eqn:proe}
\end{align}
From Lemma \ref{lem:secprob} and the fact that $ {p_e} \leq \frac{\ln n + (k-1) \ln \ln n}{n}$
for all $n$ sufficiently large, we find
\begin{align}
\mathbb{P} [ E_{ {{x} j}\cup {{{y} j}}}
\boldsymbol{\mid}(|S_{xy}|=u)]  & =  2p_e  - \frac{
{p_n\iffalse_{on}\fi}u}{ K_n} \cdot {p_e } \pm O({p_e }^2) \nonumber
\\ & =  2p_e  - \frac{ {p_n\iffalse_{on}\fi}u}{ K_n} \cdot {p_e } \pm
o\left(\frac{1}{n}\right) \label{sa1}
\\ & = O\left(\frac{ \ln n}{n} \right)  = o(1).
\label{p1aba1655}
\end{align}
%
%
%
%
%
Then using the above relation, given constants $m_1, m_2$ and $m_3$,
we obtain
\begin{align}
 \lefteqn{(n - m_1 - m_2 - m_3  - 2)\{\mathbb{P} [ E_{ {{x} j}\cup {{{y} j}}}
\boldsymbol{\mid}(|S_{xy}|=u)]\}^2}  \nonumber \\   & = (n - m_1 -
m_2 - m_3  - 2) \cdot \left[ O\left(\frac{ \ln n}{n}
\right)\right]^2 = o(1). \label{p1aba1655ast}
\end{align}
Given (\ref{p1aba1655}) and (\ref{p1aba1655ast}), we use property
(b) of Fact \ref{1_x_y} to evaluate R.H.S. of (\ref{eqn:proe})
(i.e., (\ref{pr_e_xj_yj})).
We get
 \begin{align}
\textrm{(\ref{pr_e_xj_yj})}  & \sim  e^{ - (n - m_1 - m_2 - m_3  -
2) \mathbb{P} [ E_{ {{x} j}\cup {{{y} j}}}
\boldsymbol{\mid}(|S_{xy}|=u)] }.
\label{p1aba1655aa12}
\end{align}
Substituting (\ref{sa1}) and (\ref{p1aba1655}) into
(\ref{p1aba1655aa12}), given constants $m_1, m_2$ and $m_3$,
we find
 \begin{align}
\textrm{(\ref{pr_e_xj_yj})}  & \sim  e^{ - n [  2p_e   - \frac{
{p_n\iffalse_{on}\fi}u}{ K_n} \cdot {p_e  } \pm
o\left(\frac{1}{n}\right)]} \cdot e^{(m_1 + m _2 + m_3 +2)\cdot
o(1)} \nonumber
\\  & \sim e^{-2p_en + \frac{ {p_n\iffalse_{on}\fi}u}{ K_n} \cdot {p_e n }}.
\label{tehab}
\end{align}
Applying (\ref{eqn:unre:f1hfrac}) and (\ref{tehab})
into
(\ref{eqn:unre:f1h}), we obtain (\ref{event_f_result}) and this establishes
Lemma
\ref{lem:event_f_result}.

\subsection{Proof of Lemma \ref{lem:usefulcons2}}

The proof is similar to  \cite[Lemma 5.3]{yagan_onoff}.
Given $\ell, \beta_{\ell,n}
> 0$ and (\ref{eq:DeviationCondition}), we obtain
$p_e = p_n \cdot p_s \geq \frac{\ln n}{n}$. Since
$p_n \leq 1$, we get $p_s \geq \frac{\ln n}{n} $. Then using $p_{s} \leq
\frac{K_n^2}{P_n - K_n} $ given in property (b) of Lemma
\ref{jzremfb}, $\frac{K_n^2}{P_n - K_n} \geq \frac{\ln n}{n} $
holds. Using this, we find
\begin{equation}
K_n^2  = \frac{K_n^2}{P_n - K_n} \cdot \left(P_n - K_n \right)
 \geq  \frac{\ln n}{n} \cdot P_n - \frac{K_n \ln n}{n} . \label{olp5}
 \end{equation}
Given $K_n \geq 1$, we have $\frac{K_n \ln n}{n}  < \frac{K_n ^2}{2}$ for all $n$ sufficiently large. From
(\ref{olp5}) and  $P_n = \Omega \left( n \right)$, we now get
\begin{align}
K_n^2 & >  \frac{1}{2} \cdot \frac{\ln n}{n} \cdot P_n = \Omega
\left( \ln n \right)  \nonumber
 \end{align}
 The desired result $K_n= \Omega \left(\sqrt{\ln n}\right)$ is now immediate.

{\subsection{Proof of Lemma \ref{jzremfb}}

\subsubsection{Proof of property (a)}

Recall from (\ref{hh2psps}) that given $P_n \geq 2 K_n$, we have
\begin{align}
p_{s} &= 1- \mathbb{P}[S_i \cap S_j = \emptyset] = 1-
\frac{\binom{P_n- K_n}{K_n} } {\binom{P_n}{K_n}}. \label{hh2}
\end{align}
We use Fact \ref{PncKn} (in particular (\ref{fone2}))
to evaluate R.H.S. of (\ref{hh2}) and obtain
\begin{align}
p_{s} & =  \frac{K_n^2}{P_n} \pm O
\left(\left(\frac{K_n^2}{P_n}\right)^2\right). \label{hh3}
\end{align}%

\subsubsection{Proof of property (b)}

Property (b) is proved in \cite[Lemma 7.4.3, pp. 118]{YaganThesis}.

\subsubsection{Proof of property (c)}

 From (\ref{hh3}), $p_s = o(1)$ if
and only if $\frac{K_n^2}{P_n} = o(1)$; namely, property (b) holds.

\subsubsection{Proof of property (d)}

From property (c), given $p_s = o(1)$ or $\frac{K_n^2}{P_n} = o(1)$,
we use property (b) and have $\frac{K_n^2}{P_n} = o(1)$. From
(\ref{hh3}) and $\frac{K_n^2}{P_n} = o(1)$, it follows that $p_{s}
\sim \frac{K_n^2}{P_n}$. Therefore,
\begin{align}
p_{s} - \frac{K_n^2}{P_n} & =   \pm O
\left(\left(\frac{K_n^2}{P_n}\right)^2\right) =   \pm O
\left(\left(p_{s}\right)^2\right). \nonumber
\end{align}
Then, we get $\frac{K_n^2}{P_n} =  p_{s} \pm O\left( \left(p_{s}\right)^2
\right)$.}

{\subsection{Proof of Lemma \ref{lem:secprob}}

\subsubsection{Proof of property (a)}
\label{proofaa}

We start by computing the probability
$\mathbb{P}\left[\left( {K_{{x} j}} \cap
 {K_{{y} j}}\right) \boldsymbol{\mid}
\left(|S_{xy}|=u\right)\right]$ for each $u = 0, 1, 2, \ldots, K_n$.
First, note that
\begin{align}
 \lefteqn { \mathbb{P}\left[\left(K_{{x}
j} \cap K_{{y} j}\right) \boldsymbol{\mid} \left(|S_{xy}|=u\right)
\right] } \nonumber \\
  & =  1 - \mathbb{P}\left[\left(\overline{K_{{x} j}} \cup \overline{K_{{y} j}}\right)
\boldsymbol{\mid} \left(|S_{xy}|=u\right) \right]. \label{jzsec:2}
\end{align}
From the inclusion-exclusion principle,  this yields
\begin{align}
 \lefteqn{ \mathbb{P}\left[\left(K_{{x}
j} \cap K_{{y} j}\right) \boldsymbol{\mid} \left(|S_{xy}|=u\right)
\right]}
\nonumber \\
 & = 1 - \mathbb{P}\left[
 \overline{K_{{x} j}}
\boldsymbol{\mid} \left(|S_{xy}|=u\right) \right] -
 \mathbb{P}\left[
 \overline{K_{{y} j}}
\boldsymbol{\mid} \left(|S_{xy}|=u\right) \right]  \nonumber \\
 &  ~~~~~ +
\mathbb{P}\left[\left(\overline{K_{{x} j}} \cap \overline{K_{{y}
j}}\right) \boldsymbol{\mid} \left(|S_{xy}|=u\right) \right]
. \label{se4se4a}
\end{align}
Note that for each $u = 0, 1, 2, \ldots, K_n$, events
$\overline{K_{{x} j}}$ and $\overline{K_{{y} j}}$ are both
independent of $\left(|S_{xy}|=u\right)$; however, $\overline{K_{{x}
j}} \cap \overline{K_{{y} j}}$ is not independent of
$\left(|S_{xy}|=u\right)$. Thus, we get
\begin{align}
\mathbb{P}\left[
 \overline{K_{{x} j}}
\boldsymbol{\mid} |S_{xy}|=u \right]  & =
\mathbb{P}\left[
 \overline{K_{{x} j}} \right] = 1-p_s \label{se1} \\
 \mathbb{P}\left[
 \overline{K_{{y} j}}
\boldsymbol{\mid} |S_{xy}|=u \right]  & =
\mathbb{P}\left[
 \overline{K_{{y} j}} \right] = 1-p_s .\label{se2}
\end{align}
Substituting (\ref{se1}) and (\ref{se2}) into (\ref{se4se4a}), it
follows that
\begin{align}
 \lefteqn{ \mathbb{P}\left[\left(K_{{x}
j} \cap K_{{y} j}\right) \boldsymbol{\mid} \left(|S_{xy}|=u\right)
\right]}
\nonumber \\
 & = 2p_s - 1 + \mathbb{P}\left[\left(\overline{K_{{x} j}} \cap \overline{K_{{y}
j}}\right) \boldsymbol{\mid} \left(|S_{xy}|=u\right) \right].
\label{se4u_K_n}
\end{align}
Given that the events $\overline{{K}_{{x}{y}}}$ and
$\left(|S_{xy}|=0\right)$ are equivalent, letting $u = 0$ in
(\ref{se4u_K_n}), we obtain
\begin{align}
  \mathbb{P}\left[\left(K_{{x}
j} \cap K_{{y} j}\right) \boldsymbol{\mid}
\overline{K_{{x}{y}}} \right]
  & = 2p_s - 1 + \mathbb{P}\left[\left(\overline{K_{{x} j}} \cap \overline{K_{{y}
j}}\right) \boldsymbol{\mid} \overline{{K}_{{x}{y}}}\right].
\label{se4}
\end{align}
Since events $\overline{K_{{x} j}} $ and $\overline{K_{{y} j}} $ are
equivalent to $ [( S_{x}\cap S_j ) = \emptyset ]$ and $[( S_{y}\cap
S_j)
  = \emptyset]$, respectively, we have
\begin{align}
(\overline{K\iffalse\textrm{SE}\fi_{{x} j}} \cap
\overline{K\iffalse\textrm{SE}\fi_{{y} j}}) & \Leftrightarrow
\Big\{ S_j
\subseteq [\mathcal {P}_n\setminus(S_{x} \cup S_{y})] \Big\}.
\label{event_equiv}
\end{align}
Therefore, from (\ref{event_equiv}),
$(\overline{K\iffalse\textrm{SE}\fi_{{x} j}} \cap
\overline{K\iffalse\textrm{SE}\fi_{{y} j}})$ equals the event that
the $K_n$ keys forming $S_j$ are all from $\left[\mathcal
{P}_n\setminus(S_{x} \cup S_{y})\right]$. From $|\mathcal {P}_n| =
P_n$, $|S_{x}|=K_n$ and $|S_{y}|=K_n$, we get
\begin{align}
\left | \mathcal {P}_n \setminus \left(S_{x} \cup S_{y}\right)
\right |
   & =  P_n-2K_n+ |S_{xy}|. \label{pnknsxy}
\end{align}

Under $\overline{K_{{x}{y}}}$ we have $|S_{xy}|= 0$ so that
$
\left| \mathcal {P}_n \setminus \left(S_{x} \cup S_{y}\right)
\right|
  =  P_n-2K_n$.
  Clearly, if $P_n < 3
K_n$, then $\mathbb{P}\left[\left(\overline{K_{{x} j}} \cap
\overline{K_{{y} j}}\right)
\boldsymbol{\mid}\overline{K_{{x}{y}}} \right] = 0 \leq (1-p_s)
^2 $. Below we consider the case of $P_n \geq 3 K_n$. We have
\begin{align}
\mathbb{P}\left[\left(\overline{K_{{x} j}} \cap
\overline{K_{{y} j}}\right)
\boldsymbol{\mid}\overline{K_{{x}{y}}} \right] & =  \frac{
\binom{P_n-2K_n }{K_n }}{\binom{P_n}{K_n}} .\label{jzsec:1}
\end{align}
Applying \cite[Lemma 5.1]{yagan_onoff} to R.H.S. of
(\ref{jzsec:1}), we get
\begin{align}
\mathbb{P}\left[\left(\overline{K_{{x} j}} \cap
\overline{K_{{y} j}}\right)
\boldsymbol{\mid}\overline{K_{{x}{y}}} \right] & \leq
(1-p_s)^2. \label{se3}
\end{align}

Using (\ref{se3}) in (\ref{se4}),
we obtain
\begin{align}
  \mathbb{P}\left[\left(K_{{x}
j} \cap K_{{y} j}\right)
\boldsymbol{\mid}\overline{K_{{x}{y}}} \right]   \leq
1 - 2(1-p_s ) + (1-p_s)^2  = p_s ^2 .\nonumber
\end{align}

\subsubsection{Proof of property (b)}

We first establish (\ref{eqn:lem:secprob2xb}).
Given $p_s=o(1)$, from property (c) of Lemma \ref{jzremfb},
$\frac{K_n^2}{P_n }=o(1)$ follows. Then $P_n  > 3K_n $ holds for all
$n$ sufficiently large. We first compute
$\mathbb{P}[ (\overline{K\iffalse\textrm{SE}\fi_{{x} j}}
\cap \overline{K\iffalse\textrm{SE}\fi_{{y} j}})
 \boldsymbol{\mid}(|S_{xy}|=u)]$ to derive $\mathbb{P}[
( {K\iffalse\textrm{SE}\fi_{{x} j}} \cap
 {K\iffalse\textrm{SE}\fi_{{y} j}})
 \boldsymbol{\mid}(|S_{xy}|=u)]$ from (\ref{se4u_K_n}).
 As presented in (\ref{event_equiv}), event $(\overline{K\iffalse\textrm{SE}\fi_{{x}
j}} \cap \overline{K\iffalse\textrm{SE}\fi_{{y} j}})$ is equivalent
to event $\Big\{ S_j \subseteq [\mathcal {P}_n\setminus(S_{x} \cup
S_{y})] \Big\}$. Given
 $|S_{xy}|=u$ and (\ref{pnknsxy}), it follows that
$
\left| \mathcal {P}_n \setminus \left(S_{x} \cup S_{y}\right)
\right|
=  P_n-2K_n + u$. Also, for $0 \leq u \leq K_n$, it holds that $ P_n-2K_n + u  \geq  K_n $
since $P_n  > 3K_n $.
%
%
%
Then for all $n$
sufficiently large, we have
\begin{align}
\hspace{-2mm}\mathbb{P}[ (\overline{K\iffalse\textrm{SE}\fi_{{x} j}}
\cap \overline{K\iffalse\textrm{SE}\fi_{{y} j}})
\boldsymbol{\mid} |S_{xy}|=u] =&  \frac{
\binom{P_n-2K_n+u}{K_n }}{\binom{P_n}{K_n}} \nonumber
\\  =&  \prod_{t=0}^{K_n-1} \left(1- \frac{2K_n-u}{P_n- t}\right).
 \label{pn2knpn}
\end{align}
Now, it is a simple matter to check that
\begin{equation}
\mathbb{P}[ (\overline{K\iffalse\textrm{SE}\fi_{{x} j}} \cap
\overline{K\iffalse\textrm{SE}\fi_{{y} j}})
\boldsymbol{\mid} |S_{xy}|=u ]  \leq \left(1-\frac{2K_n-u}{P_n }\right)^{K_n}
\label{up}
\end{equation}
and
\begin{align}
 \mathbb{P}[
(\overline{K\iffalse\textrm{SE}\fi_{{x} j}} \cap
\overline{K\iffalse\textrm{SE}\fi_{{y} j}})
\boldsymbol{\mid} |S_{xy}|=u] \geq
\left(1-\frac{2K_n-u}{P_n-K_n}\right)^{K_n}.
\label{lo}
\end{align}
We first evaluate R.H.S. of (\ref{up}).
It is clear that  $0<\frac{2K_n-u}{P_n}<1$
for all sufficiently large since $P_n > 3K_n$ and $u \leq K_n$.
We utilize Fact
\ref{fact1xynew} to get
\begin{align}
\lefteqn{\text{R.H.S. of
(\ref{up})}} \nonumber\\
 & \leq  1-\frac{ K_n \left( 2K_n-u \right) }{P_n}  +
 \frac{1}{2} \left[\frac{ K_n \left( 2K_n-u \right) }{P_n}\right]^2.
 \label{upppertcf}
\end{align}
Applying (\ref{upppertcf}) to (\ref{up}), we obtain
\begin{align}
\lefteqn{\mathbb{P}[ (\overline{K\iffalse\textrm{SE}\fi_{{x} j}}
\cap \overline{K\iffalse\textrm{SE}\fi_{{y} j}})
\boldsymbol{\mid} |S_{xy}|=u]} \nonumber \\ & \leq
1-\frac{2K_n^2}{P_n} + \frac{u K_n}{P_n}
+O\left(\frac{K_n^4}{P_n^2}\right). \label{uppper}
\end{align}

Then we evaluate R.H.S. of (\ref{lo}). With $0 \leq u \leq K_n$
and $P_n > 3K_n$,
 it follows that
$ 0<\frac{2K_n-u}{P_n-K_n} <1$
for all $n$ sufficiently large.
We utilize Fact
\ref{fact1xynew} and (\ref{lo}) to get
\begin{align}
 \mathbb{P}[ (\overline{K\iffalse\textrm{SE}\fi_{{x} j}} \cap
\overline{K\iffalse\textrm{SE}\fi_{{y} j}})
\boldsymbol{\mid} |S_{xy}|=u]
 & \geq   1-\frac{ K_n\left( 2K_n-u \right) }{P_n-K_n}.  \label{uppperlobg}
\end{align}
It is easy to see that
\begin{align}
\frac{  K_n\left( 2K_n-u \right)}{P_n-K_n} -  \frac{ K_n \left(
2K_n-u \right) }{P_n} & = O\left(\frac{K_n^4}{P_n^2}\right).
\label{uppperloska2}
\end{align}
Applying (\ref{uppperloska2}) to (\ref{uppperlobg})
and using (\ref{uppper}) it follows that
\begin{align}
\mathbb{P}[ (\overline{K\iffalse\textrm{SE}\fi_{{x} j}}
\cap \overline{K\iffalse\textrm{SE}\fi_{{y} j}})
 \boldsymbol{\mid} |S_{xy}|=u ] = 1-\frac{2K_n^2}{P_n}  + \frac{u K_n}{P_n} \pm
O\left(\frac{K_n^4}{P_n^2}\right). \nonumber 
\end{align}
Given $p_s=o(1)$, from property (d) of Lemma \ref{jzremfb}, we have
that
$ \frac{K_n^2}{P_n}  =  p_{s} \pm O\left( p_{s} ^2 \right) \sim p_s$.
Given $0 \leq u \leq K_n$, this yields
\begin{align}
\lefteqn{\mathbb{P}[ (\overline{K\iffalse\textrm{SE}\fi_{{x} j}}
\cap \overline{K\iffalse\textrm{SE}\fi_{{y} j}})
 \boldsymbol{\mid} |S_{xy}|=u]}  \nonumber\\ & =
1 - 2\left[p_{s} \pm O\left( p_{s} ^2 \right)\right] +  \frac{u
}{K_n} \left[p_{s} \pm O\left( p_{s} ^2 \right)\right] \pm O\left(
p_{s} ^2 \right) \nonumber\\ & = 1 -  2p_s  + \frac{ {
\iffalse_{on}\fi}u}{ K_n} \cdot {p_s } \pm O({p_s }^2) .
\label{pkse_yj}
\end{align}
Applying (\ref{pkse_yj}) to (\ref{se4u_K_n}), we obtain
\begin{align}
\mathbb{P}\left[\left(K_{{x}
j} \cap K_{{y} j}\right) \boldsymbol{\mid} \left(|S_{xy}|=u\right)
\right]  = \frac{ { \iffalse_{on}\fi}u}{ K_n} \cdot {p_s } \pm
O({p_s }^2) \label{eq:auxiliary_for_last_lemma_oy}
\end{align}
and this establishes (\ref{eqn:lem:secprob2xb}).

We now turn to the proof of (\ref{zjhaloxa}). From (\ref{eqn:lem:secprob2xb}), we obtain
\begin{align}
\lefteqn{ \mathbb{P} [ E_{ {{x} j}\cup  {{{y} j}}}
\boldsymbol{\mid}(|S_{xy}|=u)] } \nonumber \\& = \mathbb{P} [
E_{xj} \boldsymbol{\mid}(|S_{xy}|=u)] + \mathbb{P} [ E_{yj}
\boldsymbol{\mid}(|S_{xy}|=u)]\nonumber \\  & \quad - \mathbb{P} [
E_{ {{x} j}\cap {{{y} j}}} \boldsymbol{\mid}(|S_{xy}|=u)].
 \nonumber \\& = 2p_e - \mathbb{P} [
C_{xj}] \cdot  \mathbb{P} [C_{yj}] \cdot
\mathbb{P}\left[\left( {K_{{x} j}} \cap
 {K_{{y} j}}\right) \boldsymbol{\mid}
  \left(|S_{xy}|=u\right)\right] \nonumber \\& =
{p_n\iffalse_{on}\fi}^2 \cdot \left[ \frac{u }{K_n} p_s \pm
O\left({p_s }^2  \right) \right]
\nonumber \\
& = \frac{ {p_n\iffalse_{on}\fi}u}{ K_n} \cdot {p_e } \pm O({p_e
}^2). \nonumber
\end{align}
The desired result (\ref{zjhaloxa}) is now established.} 

\subsection{Proof of Lemma \ref{lemb}}

It is not difficult to see that
\begin{align}\nonumber
\lefteqn{\mathbb{P}[|S_{xy}|=u ]}  \\
\nonumber
 & =
 \frac{\binom{K_n}{u} \cdot \binom{P_n-K_n}{K_n-u} }
 {\binom{P_n}{K_n} }.
 \\ \nonumber
 & =\frac{1}{u!} \cdot \left[ \frac{K_n!}{ (K_n-u)!} \right]^2
  \cdot \frac{(P_n-K_n)!}{ (P_n-2K_n+u)!} \cdot
  \frac{(P_n-K_n)!}{P_n!}
  \\ \nonumber
  & \leq \frac{1}{u!} \cdot K_n^{2 u}  \cdot (P_n-K_n)^{K_n-u}   \cdot(P_n-K_n)^{-K_n}
  \\ \nonumber
  & =\frac{1}{u!} \bigg(\frac{K_n^2}{P_n-K_n}\bigg)^u.
\end{align}


\subsection{Proof of Lemma
\ref{lem:bounding_expectation}}

Recall $J_{i}$ defined in (\ref{olp_xjdef}). Here we still use $Y_{i}$
defined in (\ref{eq:X_S_thetaa}) for $j \geq 2$. Then
(\ref{olpxja}) follows. We define $M(|\nu_r|) $ and $Q(|\nu_r|) $ as
follows:
\begin{align}
M(\nu_{r})& = \1{|\nu_{r}|>0} \cdot \max \{ K_n, Y_{n,|\nu_{r}|} +1\} \label{olp:deflvrmr} \\
Q(\nu_{r}) & = K_n \1{|\nu_{r}| = 1} + ( \lfloor (1+\varepsilon)K_n \rfloor
+1) \1{|\nu_{r}|>1}  \label{olp:deflvrqr}
\end{align}
Lemma \ref{lem:bounding_expectation} is an extension of a similar
result established in \cite[Lemma 10.1, pp. 11]{yagan_onoff}. There,
it was shown that for $r=1, 2, \ldots, \lfloor \frac{n}{2} \rfloor$,
\begin{align}
\bE { { {P_n- M(\nu_{r})} \choose K_n }\over{{P_n \choose K_n}}} &
\leq e^{-{p_e}\lambda r}+ e^{-K_n\mu} \1{r>r_n}. \label{olp:pmvr}
\end{align}
 Recalling the definition of $L(\nu_{r}) $ in (\ref{olp:deflvr}) and
using the definitions of $M(\nu_{r})$ and $Q(\nu_{r})$ in
(\ref{olp:deflvrmr}) and (\ref{olp:deflvrqr}), we have the following
cases.

(a) If $|\nu_{r}| = 0$, then $L(\nu_{r}) = M(\nu_{r}) = Q(\nu_{r}) = 0$.

(b) If $|\nu_{r}| = 1$, then $L(\nu_{r}) = M(\nu_{r}) = Q(\nu_{r})  = K_n$.

(c) If $|\nu_{r}| \geq 2$, then
\begin{align}
L(\nu_{r}) & =  \max \left\{ K_n  , J_{n,|\nu_{r }|} +1 \right\}
\label{olp_lvr} \\ M(\nu_{r}) & =  \max \left\{ K_n  , Y_{n,|\nu_{r }|}
+1 \right\} \label{olp_mvr} \\ Q(\nu_{r}) & = \lfloor (1+\varepsilon)K_n
\rfloor +1.  \label{olp_qvr}
 \end{align}
Then for case (c), we further have the following two subcases.

(c1) If $|\nu_{r}| = 2, 3, \ldots, r_n$, given (\ref{olp_lvr}),
(\ref{olp_mvr}) and $J_{ |\nu_{r}|} = \max\{(1+\varepsilon) K_n , Y_{
|\nu_{r}|}\}$ from (\ref{olpxja}),
\fo 
\begin{align}
L(\nu_{r}) & =  \max \left\{ \lfloor (1+\varepsilon)K_n \rfloor + 1 ,
Y_{n,|\nu_{r }|} +1 \right\} \label{olp_lvr2}
 \end{align}
resulting in $L(\nu_{r}) = \max\left\{ M(\nu_{r}) , Q(\nu_{r}) \right\}$
from (\ref{olp_mvr}) and (\ref{olp_qvr}).

(c2) If $|\nu_{r}| = r_n +1, r_n +2, \ldots, n$, given (\ref{olp_lvr}),
(\ref{olp_mvr}) and $J_{ |\nu_{r}|} = Y_{ |\nu_{r}|}$ from (\ref{olpxja}), \fo 
\begin{align}
L(\nu_{r}) &  = M(\nu_{r}) =  \max \left\{ K_n, \lfloor \mu P_n \rfloor
+1 \right\}. \label{olp_lvr3}
 \end{align}
Given $\frac{K_n}{P_n} = o(1)$, then $\lfloor \mu P_n \rfloor \geq
\lfloor ( 1+\varepsilon)K_n \rfloor $ for all $n$ sufficiently large.
 Consequently, from (\ref{olp_qvr}) and (\ref{olp_lvr3}), \fo
$L(\nu_{r}) = \max\left\{ M(\nu_{r}) , Q(\nu_{r}) \right\}$.

Summarizing cases (a), (b), and (c1)-(c2) above, given any
$|\nu_{r}|$, we have $L(\nu_{r}) = \max\left\{ M(\nu_{r}) , Q(\nu_{r})
\right\}$ for all $n$ sufficiently large. This yields
%
\begin{align}
\lefteqn{\bE { { {P_n- L(\nu_{r})} \choose K_n }\over{{P_n \choose
K_n}}} }
 \nonumber \\& \leq \min \left\{  \bE { { {P_n- M(\nu_{r})} \choose K_n
}\over{{P_n \choose K_n}}}, \bE { { {P_n- Q(\nu_{r})} \choose K_n
}\over{{P_n \choose K_n}}} \right \}. \label{eq:olplvrqvr}
\end{align}
We will show the following result: for all $n$ sufficiently large
and for any $r=2, 3, \ldots, n$,
\begin{align}
 \bE { { {P_n- Q(\nu_r)} \choose K_n }\over{{P_n \choose K_n}}} & \leq
 e^{-{p_e}(1+\varepsilon/2)}.
 \label{eq:to_show_crucial_bound}
\end{align}
Clearly, if (\ref{eq:to_show_crucial_bound}) holds, we can substitute
(\ref{olp:pmvr}) and (\ref{eq:to_show_crucial_bound}) into
(\ref{eq:olplvrqvr}) and obtain
(\ref{eq:crucial_bound_expectation}), which establishes Lemma
\ref{lem:{eq:crucial_bound_expectation}}.

For any given $n$ and any given $r$, from (\ref{olp:deflvrqr}), we
get
\begin{align}
  \bE { { {P_n- Q(\nu_r)} \choose K_n }\over{{P_n \choose
K_n}}} \leq 
\bE { { {P_n- \lceil K_n  \{\1{ |\nu_{r}| = 1} + (1+\varepsilon) \1{
|\nu_{r}|
> 1}  \} \rceil} \choose K_n }\over{{P_n \choose K_n}}}.  \label{olp_jzra01}
\end{align}
From Lemma 5.1 in Ya\u{g}an \cite{yagan_onoff}, \fo
\begin{equation}
 \textrm{R.H.S. of (\ref{olp_jzra01})}
  \leq \bE{{(1-p_s)}^{\1{ |\nu_{r}| = 1} + (1+\varepsilon)
\1{ |\nu_{r}| > 1}}}. \label{eq:int_1}
\end{equation}
Then from (\ref{eq:v_r_alpha}), we obtain
\begin{align}
 \lefteqn{\textrm{R.H.S. of (\ref{eq:int_1})}}
 \nonumber \\   &= \mathbb{P} [|\nu_{r}| = 0]
 + (1-p_s) \mathbb{P} [|\nu_{r}| = 1] \nonumber \\
& \quad   + {(1-p_s)}^{1+\varepsilon} \mathbb{P} [|\nu_{r}| \geq 2]
\nonumber \\  &=  (1-p_n)^r + r p_n(1-p_n)^{r-1}{(1-p_s)}
\nonumber \\
& \quad  + [1-(1-p_n)^r -r p_n (1-p_n)^{r-1}] {(1-p_s)}^{1+\varepsilon}.
\label{fgam_1}
\end{align}

We introduce a continuous variable $\gamma$ and define $f(\gamma, p_n,
p_s)$ as follows, where $\gamma \geq 1$.
\begin{align}\nonumber
f(\gamma, p_n, p_s) &=  (1-p_n)^{\gamma} + {\gamma}
p_n(1-p_n)^{{\gamma}-1}{(1-p_s)} \\
& ~  + [1-(1-p_n)^{\gamma} -{\gamma}p_n (1-p_n)^{{\gamma}-1}]
{(1-p_s)}^{1+\varepsilon}. \label{fgam_2}
\end{align}
From (\ref{fgam_1}) and (\ref{fgam_2}), we obtain
\begin{align}
 \textrm{R.H.S. of
(\ref{eq:int_1})} = f(r, p_n, p_s). \label{rhs:eq:int_1}
\end{align}
Note that since $r$ is an integer, we cannot take the partial
derivative of $f(r, p_n, p_s)$ with respect to $r$. We have introduced
continuous variable $\gamma$ and hence can take the partial
derivative of $f(\gamma, p_n, p_s)$ with respect to $\gamma$. We get
\begin{align}
 \lefteqn{\frac{\partial f({\gamma}, p_n, p_s)}{\partial {\gamma}} }
 \nonumber \\  & = (1-p_n)^{\gamma}
[1-(1-p_s)^{1+\varepsilon}] \ln (1-p_n) \nonumber \\ &  \quad + p_n
(1-p_n)^{{\gamma}-1} [1-p_s-{(1-p_s)}^{1+\varepsilon}] [1+{\gamma} \ln
(1-p_n)] \nonumber \\
& \leq (1-p_n)^{\gamma}
[1-p_s - (1-p_s)^{1+\varepsilon}] \ln (1-p_n) \nonumber \\ &  \quad + p_n
(1-p_n)^{{\gamma}-1} [1-p_s-{(1-p_s)}^{1+\varepsilon}] [1+{\gamma} \ln
(1-p_n)], \nonumber
\end{align}
where, in the last step, we used the fact that $\ln (1-p_n) \leq 0$.
Therefore, it's clear that
\begin{align}
\lefteqn{ \frac{1}{(1-p_n)^{{\gamma}-1} [1-p_s-(1-p_s)^{1+\varepsilon}]}
\frac{\partial f({\gamma}, p_n, p_s)}{\partial {\gamma}} }\nonumber \\
& \leq (1-p_n) \ln (1-p_n) + p_n [1+{\gamma} \ln (1-p_n)]\nonumber \\
& = (1-p_n + p_n \gamma)
  \ln (1-p_n) + p_n
  \nonumber
\end{align}
with ${(1-p_n)^{{\gamma}-1} [1-p_s-(1-p_s)^{1+\varepsilon}]} \geq 0$.
Using $\ln (1-p_n) \leq - p_n < 0$ and $\gamma \geq 1$, we get
\begin{align}
\lefteqn{ \frac{1}{(1-p_n)^{{\gamma}-1} [1-p_s-(1-p_s)^{1+\varepsilon}]}
\frac{\partial f({\gamma}, p_n, p_s)}{\partial {\gamma}} }\nonumber \\
&  \leq - p_n (1-p_n + p_n \gamma) + p_n
\nonumber \\
& = p_n^2 (1 - {\gamma}) \leq 0.\hspace{4cm}
\end{align}

Given $p_n$ and $p_s$, then $f({\gamma}, p_n, p_s)$ is decreasing with
respect to ${\gamma}$ for $\gamma \geq 1$. Then given $r \geq 2$,
(\ref{eq:int_1}) and (\ref{rhs:eq:int_1}), we have
 \begin{align}
\lefteqn{\textrm{R.H.S. of (\ref{olp_jzra01})}} \nonumber \\ & \leq
f(2, p_n, p_s) \nonumber \\ & = (1-p_n)^2 + 2p_n(1-p_n){(1-p_s)} + p_n^2
{(1-p_s)}^{ 1+\varepsilon } \label{eq:int_2}
\\
&\leq (1-p_n)^2 + 2p_n(1-p_n){(1-p_s)} + p_n^2 {(1-p_s)} (1-\varepsilon p_s)
\label{eq:int_2b}
\\
&= 1-{p_e} [2 -  \varepsilon p_e -( 1 - \varepsilon) p_n ] \label{eq:int_2a}
\\
&\leq \exp\left\{-{p_e} [2 -  \varepsilon p_e -( 1 - \varepsilon) p_n ]
\right\} \label{eq:int_35}
\end{align}
where in (\ref{eq:int_2}) we use $0<p_s<1 $, $0<\varepsilon<1 $ and
Fact \ref{fact1xynew} to obtain ${(1-p_s)} ^ \varepsilon \leq 1-
\varepsilon p_s$; and in (\ref{eq:int_2b}) we use $p_e = p_n p_s$; and in
(\ref{eq:int_2a}) we use the  $ 1- x \leq
e^{-x}$ that holds for any $x \geq 0$.

Given $p_e = o(1)$, then $p_e \leq \frac{1}{2}$ for all $n$ sufficiently
large. Using this and $0< p_n \leq 1$, we obtain
\begin{align}
2 -  \varepsilon p_e -( 1 - \varepsilon) p_n & \geq 2 - \frac{\varepsilon}{2}
-( 1 - \varepsilon) = 1 + \frac{\varepsilon}{2} \nonumber
\end{align}
for all $n$ sufficiently large.
Applying the above result to (\ref{eq:int_35}), we obtain
\begin{align}
 \textrm{R.H.S. of (\ref{olp_jzra01})} & \leq  e^{-{p_e}(1+\varepsilon/2)}.
 \label{eq:to_show2_crucial_bound_2}
 \end{align}
 Applying
(\ref{eq:to_show2_crucial_bound_2}) to (\ref{olp_jzra01}), we get
(\ref{eq:to_show_crucial_bound}) and Lemma
\ref{lem:bounding_expectation} is now established.
 \myendpf

\end{appendices}

\newpage

\begin{IEEEbiographynophoto}
{Jun Zhao} (S'10) received the B.S.
degree in Electrical Engineering from Shanghai Jiao Tong University (China) in 2010. Currently, he is a Ph.D. candidate in the Department
of Electrical and Computer Engineering at
Carnegie Mellon University, Pittsburgh, PA. He has served as a session chair in Allerton Conference on Communication, Control, and Computing 2014 and
a session co-chair in Information Theory and Applications Workshop 2015.
His research
interests include network science, wireless security,
 graph theory and algorithms.
\end{IEEEbiographynophoto}

\begin{IEEEbiographynophoto}
{Osman Ya\u{g}an} (S'07-M'12) received the B.S.
degree in Electrical and Electronics Engineering from the Middle
East Technical University, Ankara (Turkey) in 2007, and the Ph.D.
degree in Electrical and Computer Engineering from the University
of Maryland, College Park, MD in 2011.
He is an Assistant Research Professor of Electrical and Computer Engineering (ECE) at Carnegie Mellon University (CMU) with an appointment in the Silicon Valley Campus. Prior to joining the faculty of the ECE department in August 2013, he was a Postdoctoral Research Fellow in CyLab at CMU. He has also held a visiting Postdoctoral Scholar position at Arizona State University during Fall 2011.
He
has served as a Technical Program Committee member of several international conferences including SECRYPT 2012, IEEE GLOBECOM 2013-2015, IEEE GlobalSIP 2013, IEEE PIMRC 2014.
His research interests
include wireless network security,
dynamical processes in complex networks,
percolation theory, random
graphs and their applications.
\end{IEEEbiographynophoto}

\begin{IEEEbiographynophoto}
{Virgil Gligor} (M'76-SM'11) received the B.S. and Ph.D. degrees from the University of California at Berkeley in 1972 and 1976, respectively. He taught at the University of Maryland between 1976 and 2007, and is currently a Professor of Electrical and Computer Engineering at Carnegie Mellon University and co-Director of CyLab. For nearly four decades,  his research interests have ranged from access control mechanisms, penetration analysis, and denial-of-service protection to cryptographic protocols and applied cryptography. Gligor was an editorial board member of several IEEE and ACM journals, and the Editor in Chief of the IEEE Transactions on Dependable and Secure Computing. He received the 2006 National Information Systems Security Award jointly given by NIST and NSA in the US, and the 2011 Outstanding Innovation Award given by the ACM Special Interest Group on Security, Audit and Control, and the 2013 IEEE Computer Society Technical Achievement Award. He also served as the chair of the Association for Computing Machinery Special Interest Group on Security Audit and Control (ACM SIGSAC) between 2005 and 2009.
\end{IEEEbiographynophoto}

\end{document}




\IEEEpeerreviewmaketitle